\documentclass[fleqn,usenatbib]{mnras}

\usepackage[T1]{fontenc}
\usepackage{soul}
\usepackage{graphicx}
\usepackage{amsmath}
\usepackage{amssymb}
\usepackage{comment}
\usepackage{newtxtext,newtxmath}
\usepackage{tabularx}

\newcommand{\MSUN}{\rm{M}_{\odot}}

\title[Assembly histories from integral galaxy properties]{ERGO-ML I: Inferring the assembly histories of IllustrisTNG galaxies from integral observable properties via invertible neural networks}

\author[L. Eisert et al.]{Lukas Eisert,$^{1}$\thanks{E-mail: eisert@mpia.de}
Annalisa Pillepich,$^{1}$
Dylan Nelson,$^{2}$ Ralf S. Klessen,$^{2}$
Marc Huertas-Company$^{3}$ and\newauthor
Vicente Rodriguez-Gomez$^{4}$
\\\\
$^{1}$Max-Planck-Institut f{\"u}r Astronomie, K{\"o}nigstuhl 17, 69117 Heidelberg, Germany\\
$^{2}$Universit\"{a}t Heidelberg, Zentrum f\"{u}r Astronomie, Institut f\"{u}r theoretische Astrophysik, Albert-Ueberle-Str. 2, 69120 Heidelberg, Germany\\
$^{3}$ Departamento de Astrof\'{i}sica, Instituto de Astrof\'{i}sica de Canarias, Universidad de La Laguna, E-38200 La Laguna, Spain \\
$^{4}$ Instituto de Radioastronom\'{i}ıa y Astrof\'{i}ısica, Universidad Nacional Aut\'{o}noma de M\'{e}xico, Apdo. Postal 72-3, 58089 Morelia, Mexico.
}

\date{}
\pubyear{2021}

\begin{document}
\label{firstpage}
\pagerange{\pageref{firstpage}--\pageref{lastpage}}
\maketitle

\begin{abstract}
A fundamental prediction of the $\Lambda$CDM cosmology is the hierarchical build-up of structure and therefore the successive merging of galaxies into more massive ones.
As one can only observe galaxies at one specific time in cosmic history, this merger history remains in principle unobservable.
By using the TNG100 simulation of the IllustrisTNG project, we show that it is possible to infer the unobservable stellar assembly and merger history of central galaxies from their observable properties by using machine learning techniques. In particular, in this first paper of ERGO-ML (Extracting Reality from Galaxy Observables with Machine Learning), we choose a set of 7 observable integral properties of galaxies (i.e. total stellar mass, redshift, color, stellar size, morphology, metallicity, and age) to infer, from those, the stellar ex-situ fraction, the average merger lookback times and mass ratios, and the lookback time and stellar mass of the last major merger. To do so, we use and compare a Multilayer Perceptron Neural Network and a conditional Invertible Neural Network (cINN): thanks to the latter we are also able to infer the posterior distribution for these parameters and hence estimate the uncertainties in the predictions. We find that the stellar ex-situ fraction and the time of the last major merger are well determined by the selected set of observables, that the mass-weighted merger mass ratio is unconstrained, and that, beyond stellar mass, stellar morphology and stellar age are the most informative properties. Finally, we show that the cINN recovers the remaining unexplained scatter and secondary cross-correlations. Our tools can be applied to large galaxy surveys in order to infer unobservable properties of galaxies' past, enabling empirical studies of galaxy evolution enriched by cosmological simulations.
\end{abstract}


%%%%%%%%%%%%%%%%% INTRODUCTION %%%%%%%%%%%%%%%%%%

\section{Introduction}

% hierarchical growth 
One of the main predictions of the $\Lambda$CDM paradigm is the hierarchical formation of structure whereby galaxies merge into progressively more massive systems \citep[e.g.][]{Lacey&Cole_1997, Springel_2005, Genel_2009}. 

% galaxy evolution
This hierarchical growth is a fundamental physical principle for the evolution of galaxies, both within populations and for individual objects. For example, it determines that more massive and luminous galaxies, which are thought to reside in more massive dark matter haloes \citep{Wechsler&Tinker_2018}, are rarer than less massive and luminous ones \citep{Schechter_1976, Blanton&Moustakas_2009}. Moreover, already a few decades ago, it had been suggested that the (dry) mergers of massive disc galaxies produce galaxies with elliptical morphologies \citep[e.g.][]{Quinn_1993}, i.e. with most of their stars in hot orbits, such that massive elliptical galaxies have often been associated to a recent violent merger history.

Over the last years, cosmological simulations have been able to quantitatively follow the evolution and interactions of representative populations of dark-matter haloes \citep[e.g.][]{BoylanKolchin_2009, Potter_2017, Ishiyama_2021} and galaxies \citep{Vogelsberger.2014Nature, Schaye_2015, nelson2019illustristng} within ever-increasing volumes. They have hence demonstrated that mergers between galaxies are more frequent at higher redshifts \citep{Fakhouri&Ma_2008, Rodriguez_Gomez_2015} and that the number of massive mergers is a steep function of galaxy and halo mass \citep{Rodriguez_Gomez_2015}. Moreover, it has been shown, via cosmological hydrodynamical galaxy simulations, that at high stellar masses ($> 10^{10-11}\,\MSUN$) mergers become an important source for the stellar inventory of a galaxy in addition to in-situ star formation. As a result,  massive galaxies are made of larger fractions of stellar material that is accreted from mergers and orbiting satellites \citep[e.g.][]{rodriguezgomez2016exsitu, Pillepich_2018}.

% past history, unobservable, galactic archaelogy
Astronomical observations provide snapshots of the state of galaxies at distinct points in time, such that their actual merger and accretion histories remain in principle unobservable. Whereas it is not uncommon to observationally capture galaxies in the act of merging with one another \citep{HST_Merger} and whereas recent mergers can leave clear signatures in the stellar mass maps of observed galaxies, via e.g. stellar shells \citep{Turnbull_1999}, tidal tails \citep{martinezdelgado2010stellar}, and disturbed stellar morphologies \citep{Ellison_2013}, the inference of more ancient events and interactions is substantially more challenging. 

Recent developments in both theory and observations have enabled progress in this direction, namely in inferring aspects of the past histories of galaxies given their properties at the time of inspection. For example, the combination of models with 6D + metallicity data for millions of individual stars from the {\it Gaia} mission \citep{GaiaDR22018} and other surveys \citep[e.g. LAMOST, GALAH, and APOGEE:][]{Deng2012, DeSilva2015, Majewski2017} have allowed the reconstruction of discrete merger events of the assembly history of our own Galaxy \citep{Helmi2018, Belokurov2018}. It is now believed that the Milky Way has undergone its last major merger sometime at $z=1-2$ \citep{Bonaca2020}, even if the merging galaxy, dubbed Gaia-Sausage-Enceladus and with estimated $5-6\times 10^8 \MSUN$ in stars \citep{Naidu2021}, has long been destroyed. 

In a similar fashion, but going beyond the Local Volume, \cite{Zhu2021} and \textcolor{blue}{Zhu et al. (2022, submitted)} have uncovered that two early-type galaxies in the Fornax cluster each underwent a merger as early as 8 and 10 billion years ago, respectively, and quantified their stellar mass and merger time. To achieve this, they have combined population-orbit super-position methods \citep{Zhu2020}, the results of the cosmological simulation TNG50 \citep{Pillepich_2019, Nelson_2019}, and the observed luminosity density, stellar kinematic, age and metallicity maps from the MUSE IFU instrument on the VLT \citep{Sarzi2018}.

% galactic archaelogy, still too costly observations and too few galaxies
Although spectacular, these results are destined to be limited to small samples of galaxies in the nearby Universe, mostly because of the scope of the required observational techniques. 
%
% This paper: why
In this paper, we investigate whether, and to what degree, less-costly observational galaxy data can be queried to successfully infer aspects of the past stellar assembly and merger history of large samples galaxies. In particular, we aim to see whether scalar features that summarize galaxy integral properties, such as galaxy stellar masses, galaxy colors or labels of stellar morphology, possess sufficient information content to determine, e.g., the amount of accreted stellar mass or the mass and time of the last major merger. This possibility would open up a powerful new avenue to quantify the merger history of thousands of observed galaxies across cosmic epochs, because it would be achievable with data from large photometric surveys such as SDSS\footnote{Sloan Digital Sky Survey, \url{https://www.sdss.org/}}, GAMA\footnote{Galaxy And Mass Assembly, \url{http://www.gama-survey.org/}}, DES\footnote{Dark Energy Survey, \url{https://www.darkenergysurvey.org/}} and HSC-SSP\footnote{Hyper Suprime-Cam Subaru Strategic Program, \url{https://hsc.mtk.nao.ac.jp/ssp/}}, but also future ones such as LSST\footnote{Large Synoptic Survey Telescope/Vera C. Rubin Observatory, \url{https://www.lsst.org/}}.

% This paper: how
To do so, we use data from the IllustrisTNG project{\footnote{\url{www.tng-project.org}}}, which provides a series of cosmological, gravity + magnetohydrodynamics simulations of galaxies in representative portions of synthetic universes \citep{nelson2019illustristng}. As such simulations predict the physical state of thousands of individual and realistic galaxies across cosmic time, they allow us to understand and quantify how unobservable dynamical processes of galaxy evolution, including the merger history, influence the observable appearance of galaxies. The ``by hand'' investigation of these connections has been addressed in the past \citep[e.g.][]{rodriguezgomez2016exsitu, rodriguezgomez2016role}, also using as observational inputs the properties of the stellar haloes \citep{Pillepich_2014, Cook_2016, Pop_2018, Merritt_2020}. Here we take a leap forward and utilize modern machine learning (ML)  methods, training from the IllustrisTNG galaxies to derive a continuous mapping function between their observable and their unobservable properties.

% ERGO-ML
This is the first paper of a wider project (ERGO-ML) where we aim at Extracting Reality from Galaxy Observables with Machine Learning, across a wide range of physical properties of galaxies and utilizing diverse sets of observations, e.g. from photometry to spatially-resolved spectroscopy, from stellar light to gaseous signatures, from the innermost regions of galaxies to their dim stellar and gaseous haloes, from scalar features to maps and multi-dimensional data cubes. We will use and combine state-of-the-art cosmological simulations of galaxies to test methods for direct application to observational data in order to ``train on simulations and apply to observations''. All this is now possible thanks to the breadth of scope of current cosmological galaxy simulations \citep{Vogelberger_2020}, the large numbers of the galaxies simulated therein \citep{nelson2015illustris, nelson2019illustristng} and their increasing realism, which since a few years has also being quantified with ML methods \citep[e.g.][]{Huertas_Company_2019, Zanisi_2020}.

% Other past ML works for unobservable galaxy properties
The potential of combining ML and galaxy (simulation) data for similar scientific goals has been shown over the past couple of years in a wide array of works.
A large fraction of these have focused on determining the properties of the underlying dark-matter haloes \citep[e.g.][]{Rios_2021, Marttens_2021} and on identifying merging galaxies or merger remnants from images \citep[e.g.][]{Bottrell_2019, Ferreiras_2020, Bottrell_2020, Ciprianovic_2021}. Very recently, \cite{Rui_2021} have also used data from IllustrisTNG to determine, via a Random Forest, the accreted stellar mass fraction.

% This paper: why better/different
In this paper, we build further by focusing on summary statistics of the entire merger and assembly history of galaxies. We also go beyond the methodologies adopted thus far in the field by obtaining not only point predictions, but also by quantifying the full posterior distributions, and hence uncertainties, of the model predictions. 

% This paper is organized as follows
Our paper is organized as follows. In Section~\ref{sec:methods}, we provide the details for the rationale of this work and for the galaxy simulations and the ML algorithms adopted throughout. The architecture and training of the latter are described in Section~\ref{sec:training}. We show that the unobservable past history of galaxies can be decoded from a handful of galaxy features in Section~\ref{sec:results}, where we compare and quantify the results from two complementary models, built respectively from a Multilayer Perceptron Neural Network and a Conditional Invertible Neural Network. We comment on the technical aspects of the inference and on the scientific implications of our findings in Section~\ref{sec:discussion} and conclude and summarize in Section~\ref{sec:conclusions}.

\begin{figure}
	\centering
	\includegraphics[trim={1cm 7cm 0cm 0cm},width=9cm]{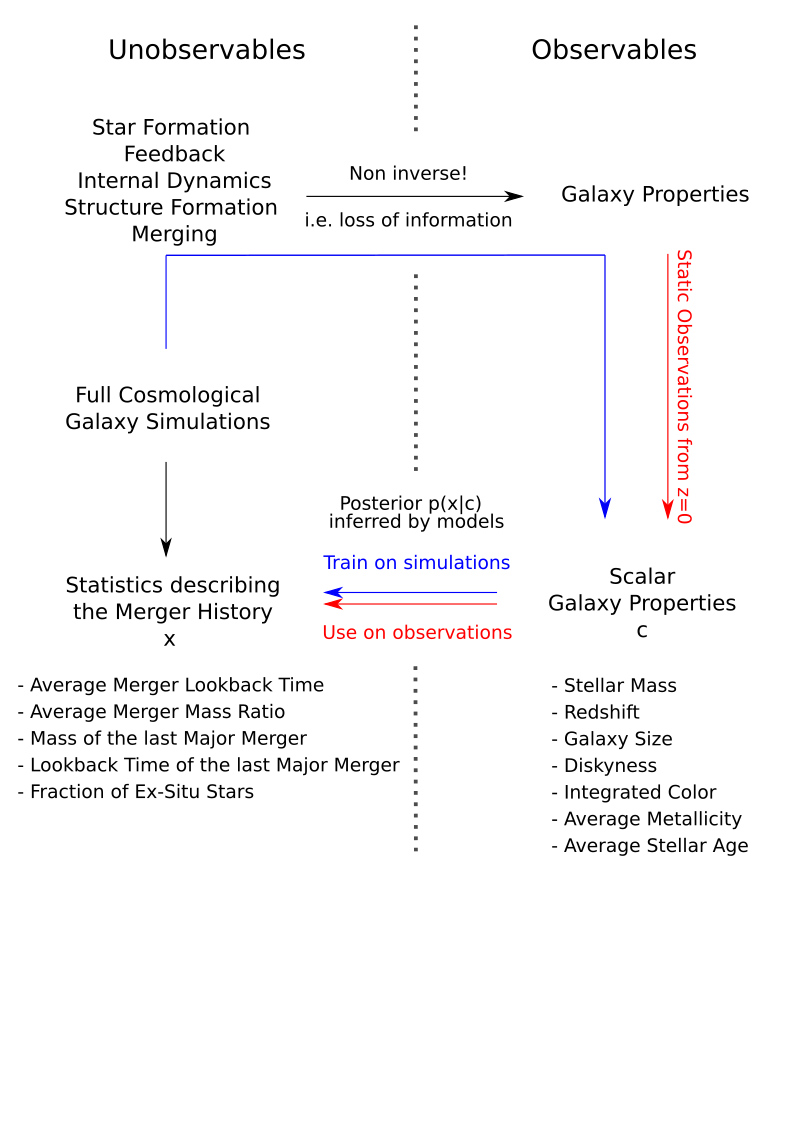}
	\caption{{\bf Visual scheme of the setup and goals of this paper, ERGO-ML I.} We can predict the observable properties of thousands of galaxies given their individual histories and evolution using a cosmological hydrodynamical galaxy simulation, here TNG100. The goal of this paper is to model the inverse of this, i.e. the mapping from the observables to the unobservable stellar mass assembly and merger history (see also Table~\ref{tab:properties}). Because of the possible loss of information in the forward step -- that is, the observables are not sufficient to fully constrain the history --, we are especially looking for the posterior distribution, which is the probability that a certain merger history $x$ is compatible with a set of observations $c$. In this work, we focus on utilizing scalar integral properties, i.e. galaxy features, and test this scheme on simulated galaxies. However, in subsequent work, we will go beyond features and use maps of galaxies and of their stellar haloes as inputs, and we will also apply the models developed here to actual observational data.}
	\label{fig:methods/scheme}
\end{figure}

%%%%%%%%%%%%%%%%% METHODS %%%%%%%%%%%%%%%%%%

\section{Setup, simulation data and methods}
\label{sec:methods}
\subsection{Rationale}

We aim to uncover and investigate the connections between observable (input or feature) and unobservable (output or desiderata or target) properties of galaxies, with a focus on their unobservable past merger history. To do so, we use data from the IllustrisTNG simulations to train, and hence compare the results of, two ML algorithms: a Multilayer Perceptron (MLP) Neural Network and a Conditional Invertible Neural Network (cINN). 

In both approaches, the known parameters are 7 galaxy integral galaxy properties (i.e. features) that describe the state of the galaxy at a certain point in time. We focus on galaxy properties that are in principle directly observable or derivable from observations, namely: galaxy stellar mass, galaxy redshift i.e. lookback time, stellar half-light radius, galaxy stellar morphology (i.e. the discyness, measured via the disc-to-total mass ratio ), integrated galaxy color, average stellar metallicity, average stellar age.

The output target quantities we elect to investigate are 5 summary statistics of the galaxies' assembly and merger histories that are directly predicted by the cosmological simulations and are extractable from the simulation output data. These are the total stellar ex-situ mass fraction, the lookback time and stellar mass of the last merger with a stellar mass ratio $\ge \frac{1}{4}$, the mass-weighted average merger mass ratio, and the mass-weighted average merger lookback time.

The rationale of our study is summarized in Fig.~\ref{fig:methods/scheme} and a succinct overview of the observable and unobservable galaxy properties we focus on is in Table~\ref{tab:properties}. In the following, we briefly describe the simulation data we use to uncover the connections between observables and past history of galaxies, the galaxy properties, and the functioning of the MLP and cINN neural networks and the way we use them.

\subsection{The IllustrisTNG simulations for galaxy formation and evolution}

We utilize the publicly-available results \citep{nelson2019illustristng} of the large-scale cosmological magneto-hydrodynamical simulation project IllustrisTNG \citep[TNG hereafter, ][]{Springel_2017, naiman2017results, Marinacci_2018, nelson2017results, Nelson_2019, Pillepich_2018,Pillepich_2019}. These consist of multiple simulation runs  for galaxy formation and evolution based on the moving mesh code \textsc{AREPO} \citep{Springel_2010} and cover a broad range of resolutions, simulation volumes and additional dark matter only runs to trace the influence of baryons on the cosmological evolution. 

The full-physics runs of TNG follow the evolution of not only dark-matter particles but also of gas cells, and stars and black holes particles (collectively referred to as resolution elements), from redshift $z=127$ to $z=0$. The initial conditions defined in periodic-boundary cubic domains are given by the Zeldovich approximation, via the N-GenIC code \citep{2005Natur.435..629S}. The underlying cosmology is given by Planck \citep{Planck_2015}.

The TNG simulations are hereby the direct successor of the Illustris project \citep{Vogelsberger.2014Nature}, but introduce magneto-hydrodynamics and several updates to the physical model \citep[see details in the method papers:][]{Weinberger_2016, Pillepich_2018a}. In this work, we will use the outcome of the run called TNG100, as further elaborated in Section \ref{sec:sample}.\\

\subsection{Galaxy identification and past history}

The TNG simulations return tens of thousands of well-resolved galaxies across 14 billion years of cosmic evolution, for which integral properties can be measured from the constituents resolution elements. To contain the amount of saved data to a manageable size, the resolution elements for each of the simulated ingredients have been stored only at $100$ equally distributed points in (simulated) cosmic time, called snapshots. We have at our disposal 50 snapshots at $z\leq1$ and 50 at $z>1$ \citep[see][for more details]{nelson2019illustristng}.

Within the simulation output, galaxies -- i.e. sets of resolution elements that together constitute individual galaxies -- are identified by the  \textsc{Friend-of-Friends} \citep[FoF,][]{doi:10.1111/j.1365-2966.2009.15034.x} and \textsc{Subfind} \citep{2001MNRAS.328..726S} algorithms. Based on these two methods, 
a hierarchical set of two catalogues are available:
\begin{enumerate}
	\item The \textsc{Subfind} catalogue, which lists the subhaloes within each snapshot and  provides a number of integral statistics for each object, such as e.g. the total mass of each subhalo for each particle type.
	\item The FoF catalogue, which groups these subhaloes into FoF haloes.
\end{enumerate}
Typically, the most massive subhalo in a FoF halo is called {\it central} or primary subhalo, while all other subhaloes in that group are {\it satellites}. Any subhalo with non-vanishing stellar mass is a galaxy, whether central or satellite.

As we want to investigate the formation history of the galaxies, we need the connection of identified subhaloes along their cosmological evolution. In this paper, we hence utilize merger trees created by \textsc{Sublink} \citep{Rodriguez_Gomez_2015}. This method constructs a tree-like structure out of the previously identified subhaloes, such that for each root subhalo there is a link to one or more progenitors in the previous snapshot.
If there is more than one progenitor, the progenitor along the most massive branch is chosen as the main or first progenitor. Throughout this work, the time evolution of the galaxies of interest is given by such main progenitor branches, whereas the secondary progenitors constitute {\it merging galaxies}, in short: mergers. 

In particular, in this paper we adopt the so-called \textsc{SublinkGal} code, a version of \textsc{Sublink} whereby star particles and star-forming gas cells -- instead of dark-matter particles -- are tracked in time to construct the trees of galaxies. In comparison to the default functioning of the \textsc{SublinkGal} code, we apply additional measures to obtain a cleaner catalog of merger events.
Namely, we consider only galaxies with cosmological origin\footnote{Namely, throughout this analysis, we exclude so-called spurious subhaloes, i.e. gravitationally-bound sets of stars that formed out of gas that in turn fragmented in already-established, parent galaxies. To do so, we impose SubhaloFlag $\equiv 1$ \citep{nelson2019illustristng}.}, with at least 50 stellar particles (galaxy stellar mass $\gtrsim 5\times10^{6} \rm{M}_{\odot}$), and whose main progenitor history is recorded for at least three snapshots.

\subsection{Galaxy sample}
\label{sec:sample}

The TNG suite consists of three flagship simulations of different resolutions and volumes: TNG50, TNG100 and TNG300. These evolve cubic volumes of approximately 50, 100, and 300 comoving Mpc a side, with the best resolution achieved in the smaller-volume TNG50 \citep{Nelson_2019, Pillepich_2019}. However, a larger volume allows for more simulated galaxies, importantly especially at the high-mass end. With about ten thousand galaxies above $10^{9}\, \MSUN$ at every snapshot, TNG100 provides a galaxy sample that is sufficient for our ML approach and is therefore our choice for the following study. 

From TNG100, we consider only central galaxies (i.e. the first galaxy within a FoF group and hence the galaxy at the deepest minimum of the FoF potential) with cosmological origin \citep[i.e. non-spurious subhaloes as per][]{nelson2019illustristng}, 
and in the $z=0-1$ range and with total stellar masses between $10^{10}$ and $10^{12}\, \MSUN$. We choose this limit as the highest mass end is not well sampled and would require a dedicated work.

The lower limit in stellar mass is set for a number of reasons:
\begin{itemize}
	\item to ensure that the galaxy properties are well captured not only at the time of inspection, with about $\gtrsim10^4$ stellar particle per galaxy, but also at earlier times when the inspected galaxies and their potential merging companions have lower mass and are hence less resolved;
	\item to focus on a mass regime with a richer merger history, given that the merger rates of galaxies decline towards low masses \citep{Rodriguez_Gomez_2015}; 
	\item to focus on a mass regime that is accessible by current large photometric surveys of  galaxies, such as SDSS, that are complete down to $\gtrsim 5-7\times 10^9\, \MSUN$.
\end{itemize}
With $49$ discrete snapshots between $z=0$ and $z=1$, our TNG100 sample includes 182'625 galaxies with total stellar mass in the $10^{10-12}\,\MSUN$ range.

\subsection{Galaxy observables and unobservables}
\label{sec:properties}

\begin{table*}
	\centering
	\begin{tabularx}{\linewidth}{{>{\hsize=0.4\hsize}X>
	                               {\hsize=0.5\hsize}X>                            {\hsize=1.1\hsize}X}}
	\textbf{Galaxy Observables} & &\\
	\hline
	\hline
	
	Name & Definition & Note\\
	\hline
	&&\\
    Stellar Mass & Stellar mass within twice the radius that contains half of the total galaxy mass i.e. within two times the stellar half-mass radius. & With this mass definition, which we use throughout, the minimum and maximum mass of our sample fall slightly below the nominal limits of $10^{10}$ and $10^{12}\,\MSUN$ used for the sample selection.\\
    %\hline
    &&\\
    Lookback Time & Time at which the galaxy is considered. & For plotting and when utilizing ML methods, we use the lookback time rather than redshift, as it represents a linear measure of time and the TNG snapshots are also approximately uniformly distributed in time.  \\
    %cosmological age of the galaxy in terms of lookback time as seen from redshift $z=0$\\
    %\hline
	&&\\
	Half Light Radius & Stellar half-light radius in the SDSS-r band. & This radius is measured relative to the center of the galaxy, which is defined as the position of the resolution element with the lowest potential energy in the respective subhalo. We measure this quantity using the dust-attenuated SSDS magnitudes from \cite{nelson2017results}.\\
	%\hline
	&&\\
	Fraction of Disc Stars & Disc-to-total ratio, obtained via the fractional mass of disc stars: $[D/T] = M^{\star}_{\rm circ} / M^{\star}_{\rm tot}$, i.e. of stars on circular orbits. &  To identify stars on circular orbits within the simulation, we use the star-wise epsilon parameter defined by e.g. \cite{Marinacci_2013} as $\epsilon = J_z / J(E)$,
whereby $J_z$ is the angular momentum around the symmetry axis and $J(E)$ is the maximum momentum $J$ for a certain binding energy $E$.
We define a stellar particle to be on a circular orbit when $\epsilon > 0.7$. As galaxy-wide statistics, we use the fraction of stellar particles that are identified as circular within an aperture of $10 \times$ the stellar half-mass radius around the galaxy center, from \cite{2015ApJ...804L..40G}.\\
	%\hline
	&&\\
	$g-r$ Color & $g-r$ in SDSS bands & These are dust-attenuated and measured within twice the stellar half-mass radius; from \cite{nelson2017results}.  \\
	%\hline
	&&\\
	Stellar Metallicity & SDSS-r-band luminosity-weighted average stellar metallicity. & The average is taken across gravitationally-bound stellar particles within twice the stellar half-mass radius. In the simulation, the stellar metallicity is the fraction of all elements heavier than Helium divided by the total stellar-particle mass. \\
	%\hline
	&&\\
	Stellar age & SDSS-r-band luminosity-weighted average stellar age. & The average is taken across gravitationally-bound stellar particles within twice the stellar half-mass radius.\\
	\vspace{0.5cm}
	\textbf{Galaxy Unobservables} &&\\
	\hline
	\hline
	Name & Description & Note\\
	\hline
	&&\\
	Mean Merger Time & Mass-weighted average lookback time of all past merger events in a galaxy history. & Adopted from \cite{rodriguezgomez2016exsitu}. The lookback time is equal to the last snapshot the galaxies were both identified as unique subhaloes by \textsc{SublinkGal}. ``Mass weighted'' in this context means that the mass of the secondary progenitor is used as weight in the averaging. This makes both this and the following statistics more sensitive to {\it massive} merger events.\\
	&&\\
	Mean Merger Mass Ratio & Mass-weighted average stellar mass-ratios of all  past merger events in a galaxy history & Adopted from \cite{rodriguezgomez2016exsitu}. The mass ratio is measured at $t_{\rm max}$ i.e. the time the secondary reached its maximum stellar mass. \\
	&&\\
	Last Major Merger Mass & Total stellar mass of the last merger with a stellar mass ratio $\ge 1/4$. & The mass is defined as the maximum total stellar mass the secondary had during its lifetime. If there is no major merger, the parameter is set to the value of $10^6\,\MSUN$, i.e. below the resolution limit of TNG100.\\
	&&\\
	Last Major Merger Time & Lookback time of the last merger with a stellar mass ratio $\ge 1/4$. & The lookback time is relative to the redshift of the galaxy under consideration. If there is no major merger, the parameter is set to the unphysical value of $15$ Gyr.\\
	&&\\
	Stellar Ex-Situ Fraction & Ratio between ex-situ to total stellar mass, where totoal = ex-situ + in-situ. & We can track individual stellar particles through cosmic time using their unique particle ID and call in-situ the stars of a galaxy that have formed within its main progenitor branch and call ex-situ those that formed in galaxies on secondary branches of the galaxy merger tree, following the definitions and results of \cite{rodriguezgomez2016exsitu, Pillepich_2018}. This mass ratio includes all stellar particles identified by \textsc{Subfind} to belong to the respective subhalo i.e. all gravitationally-bound stellar particles with no galactocentric distance cut.\\
	\end{tabularx}
	\caption{{\bf Observable and unobservable galaxy properties used and connected in this work.} At the top, we list and define the observable integral properties of galaxies (i.e. features) that are are used in this work to infer unobservable properties of galaxies' past assembly and merger history, bottom.}
	\label{tab:properties}
\end{table*}

%Here the observable
We characterize the simulated galaxies by the fundamental properties listed and described in the top panel of Table~\ref{tab:properties}. These include the galaxy stellar mass and the redshift, as from previous studies \citep[e.g.][]{Rodriguez_Gomez_2015, rodriguezgomez2016role}, we know that the merger activity of galaxies is highly dependent on these (see Introduction). 

Additionally, we also use as observable inputs information on the galaxy stellar extent and structure, such as the
size of the galaxies and their morphology (i.e. the stellar disk-to-total mass ratio).
While the stellar half-light radius gives information about the spatial distribution of the stars, the fraction of disc stars gives us a possibility to differentiate between disc-like and elliptical galaxies. Disc galaxies are thought to be mainly build by in-situ star formation, whereas a large fraction of non-circular orbits could be an indicator for past merger events. In observations, similar morphological estimators can be obtained via photometric morphological decompositions or by using integral field unit spectroscopy \citep[e.g. ManGA,][]{Bundy_2014}.

Finally, as further information that can be gained via photometric observations of galaxies, we also characterize the simulated galaxies by their integrated galaxy color, average stellar metallicity and average stellar age.
These may especially contain information about the formation history of the stellar populations.

%And here now the unobservable
We elect five summary statistics to characterize
the past history of a galaxy: these are listed in Table~\ref{tab:properties}, bottom,  and we will further comment on our
choices in the Discussion. In fact, finding a set of variables that fully represent the merger history of any individual galaxy is not a trivial task. From the simulations, we can construct this history from the corresponding \textsc{Sublink} tree of progenitor subhaloes/galaxies which have previously merged with the galaxy of question. Although it might be possible to use the whole list of progenitors as a whole, we streamline our analysis by deriving a set of meaningful numbers for each galaxy that concatenate the information of the merger tree into a set of scalar statistics. 

As anticipated in the Introduction, cosmological simulations have demonstrated that galaxies can assemble their stars via two channels: by in-situ star formation, whereby gas is converted into stars, and by the stripping and accretion of stars that formed in other galaxies that in turn became satellites of, or merged with, others \citep[e.g.][]{Oser_2010, Pillepich_2014}. The second stellar mass assembly channel is referred to as ``ex-situ'', and accreted or ex-situ stars can be found throughout galaxy bodies, from their innermost regions to the stellar haloes \citep[e.g.][]{Pillepich_2015, rodriguezgomez2016exsitu}.
It is therefore a natural choice to use the ex-situ stellar mass fraction as a primary indicator for summarizing the past merger history of a galaxy. 

For any galaxy, we also record the mass-weighted average mean lookback time and the mass-weighted average stellar mass ratio of all past merger events. Previous analyses of previous cosmological simulations have shown that there exist a strong correlation between these statistics and the stellar ex-situ fractions \citep{rodriguezgomez2016exsitu}: we will further investigate on these within the TNG framework in the next Sections.

Finally, as recent major mergers may have a strong impact on the properties of galaxies, we also consider the lookback time and the stellar mass of the last major merger a galaxy underwent \citep{rodriguezgomez2016exsitu}. By major merger, we mean one with stellar mass fraction larger than 1/4. However, not all galaxies undergo such massive mergers ever: for galaxies with no major merger in their lifetime, we set these mass and time to unphysical values, so that we can later on distinguish these no-major-merger galaxies from the remaining population.

\begin{figure*}
	\centering
	\includegraphics[trim={51cm 0  0 0},clip,width=17cm]{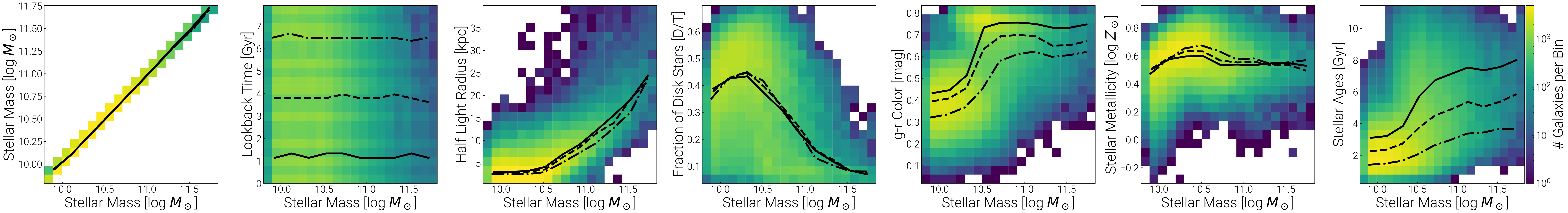}
	\caption{{\bf Overview of the observable statistics used in this work, as a function of galaxy stellar mass and redshift, and predicted by TNG100}: in each row we plot the whole simulated sample (182'625 galaxies in total) introduced in Section \ref{sec:sample} as 2D histograms between one of the observable quantities and the stellar mass (within $2 \times r^\star_{1/2}$).
	Curves denote the medians as function of the stellar mass at redshift $z < 0.2$ (solid), $0.2 < z < 0.5$ (dashed) and $0.5 < z < 1$ (dot-dashed). Note that there is in fact a bi-modality in the data between disc-blue and elliptical-red galaxies, which is not well represented by the median lines.}
	\label{fig:methods/observables}
\end{figure*}

\begin{figure*}
	\centering
	\includegraphics[width=17cm]{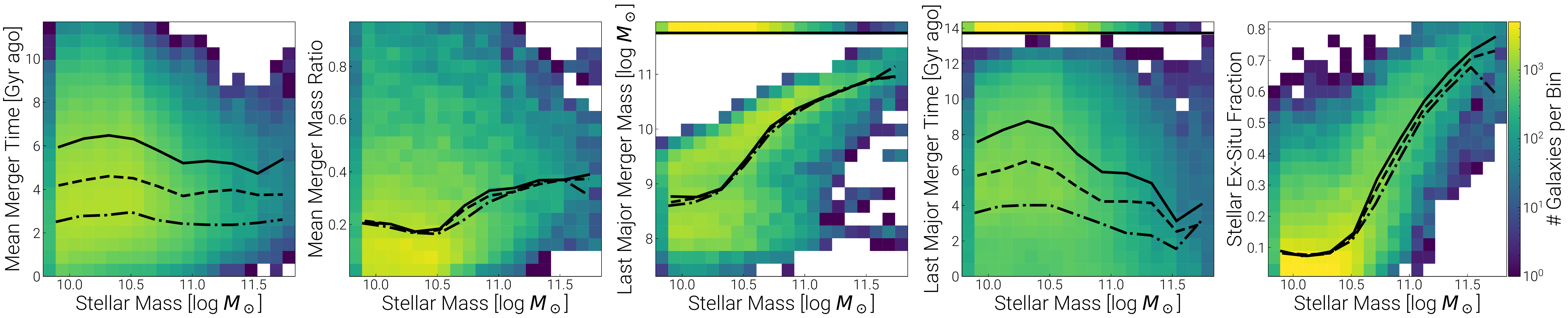}
	\caption{{\bf Overview of the unobservable statistics used in this work and predicted by TNG100}: in each row we plot the whole simulated sample introduced in Section \ref{sec:sample} as 2D histograms between one of the unobservable quantities and the stellar mass (within $2 \times r^\star_{1/2}$).
	Curves denote the medians as function of the stellar mass for redshift $z < 0.2$ (solid), $0.2 < z < 0.5$ (dashed) and $0.5 < z < 1$ (dot-dashed). Here, for the mass and the lookback time of the last major merger, we plot the galaxies without any major merger in their history as a bar at the top. Especially for the stellar ex-situ fraction and the mass of the last major merger, the simulations predict a strong positive correlation with the stellar mass of the respective galaxy; on the other hand, these relationships depend only weakly on redshift. This is naturally different for the unobservables that are defined relative to the redshift of the galaxy at hand, namely the mean merger lookback time and the time of the last major merger.}
	\label{fig:methods/unobservables}
\end{figure*}

We show an overview of all observable and unobservable quantities and how they relate to galaxy mass in Fig.~\ref{fig:methods/observables} and Fig.~\ref{fig:methods/unobservables}.


\subsection{Inference methods}

\subsubsection{Multilayer Perceptron (MLP)}
\label{sec:methods_mlp}

As a first way to tackle the problem outlined in Fig.~\ref{fig:methods/scheme}, we use a Multilayer Perceptron (MLP) neural network, which consists of multiple layers of artificial neurons (i.e. linear transformations) and a nonlinear rectifier in between. The $7$ galaxy observables are given as input for the first linear layer, whereas the 5 merger-statistics unobservables are given by the output of the last linear layer. With this, we get a regression model that maps the observables to the unobservable parameters. The model is trained on the simulation-based dataset by minimizing the mean squared error loss 
\begin{align}
    L_{\rm MSE} = \frac{1}{N} \sum^N_i(\hat{x}_i - x_i)^2
\end{align}
with $x$ denoting the ground truth merger-history set of parameters from the TNG simulations, $\hat{x}$ being the MLP model predictions, and $N$ the number of samples in the training batch.
For the implementation and optimization of this network, we utilize the framework \textsc{KERAS} \citep{chollet2015keras}.

\subsubsection{Conditional Invertible Neural Network (cINN)} 
\label{sec:methods_cINN}

A regression model like the MLP will converge to the correct solution only if the underlying connection between the 7 observables and the 5 merger statistics is describable as a function and is therefore unambiguous. However, it is possible that the information contained in the chosen set of observables is insufficient and that galaxies with (nearly) identical observable parameters may actually have distinct assembly and merger histories. It is also possible that the observable properties of galaxies may not be uniquely determined solely by their past stellar assembly and merger history, but also by other physical processes and conditions, as it is certainly the case. It is hence important to not only perform a simple scalar regression with one single set of point predictions, but to also know the posterior distribution $p(x|y)$: this is the conditional probability that a set of merger statistics $x$ is compatible with the condition that a set of galaxy observables $y$ is given.

We achieve this by using the Conditional Invertible Neural Networks (cINNs) developed by \cite{2019arXiv190702392A}, which are a generalization of their Invertible Neural Networks \citep{2018arXiv180804730A}. The latter utilize as building blocks the invertible architecture proposed by \cite{2016arXiv160508803D} and an additional set of latent variables $z$, that store the information that would be otherwise lost. This network is therefore describable as an invertible function $f(x) = [y,z]$ from the merger-history statics $x$, to a tuple of observables $y$, and a set of latent variables $z$. It was shown by \cite{2018arXiv180804730A} that by training this network as proposed, the inverse $f^{-1}(y,z)$ will converge against the posterior $p(x|y)$ when sampling over the gaussian latent space $z$. However, a simple INN demands for the forward process $f(x) = [y,z]$ to be well defined, this not being the case for our physical problem. The cINNs circumvent this issue by allowing for the set of observables $y$, to be introduced in the network as conditions $c$, which are then concatenated into the conditional affine coupling layers. The resulting neural network is therefore equivalent to an invertible function $f(x,c) = z$ with latent variables $z$ whose dimension is equal to the dimension of $x$. By applying a negative log likelihood loss 
\begin{align}
    L_{\rm NLL} = \left|~ \frac{|z|^2}{2} - \log(|J|) ~\right|,
\end{align}
with the Jacobian $J$ of the invertible function $f$, the distribution of $z$ is enforced to take the form of a multivariate gaussian.

In practice, the cINN provides a mapping between the posterior $p(x|c)$ and a multivariate gaussian distribution conditioned by the galaxy observables $c$; by sampling the gaussian latent space $z$, one can therefore recover the posterior $f^{-1}(z,c) = p(x|c)$.

Conditional Invertible Neural Networks can be easily implemented by using the \textsc{FrEIA} framework \citep[Framework for Easily Invertible Architectures, ][]{FrEIA}, which is based on \textsc{pytorch} \citep{NEURIPS2019_9015}. They have been already successfully applied to similar problems in astronomy \citep[e.g.][]{10.1093/mnras/staa2931}.

%%%%%%%%%%%%%%%%% TRAINING %%%%%%%%%%%%%%%%%%

\section{Network Training}
\label{sec:training}
\subsection{Preprocessing of the sample}

Before tackling the training of the MLP and cINN, a couple of interventions need to be executed on the data.

\subsubsection{Feature Scaling}

As standard in  ML applications, we scale all quantities to a comparable range of values, for the following reasons:
\begin{enumerate}
	\item the MLP loss function we are using is not invariant regarding the magnitude of the unobservable quantities. A feature with large absolute variations would be thus more important in terms of loss than a feature with small absolute variations;
	\item the gradient descent algorithms that are used to optimize the neural networks are sensitive to the scale of the input -- the change in weights is proportional to the feature values;
	\item many default parameters of ML algorithms are given relative to standardized inputs. 
\end{enumerate} 
Therefore, we scale each of the quantities introduced in Section~\ref{sec:properties} by subtracting the mean and dividing by the standard deviation of each variable distribution. We hence get a  sample of inputs and outputs in scaled units, whereby the mean is $0$ and the standard deviation is $1$ for each quantity across the whole sample.
During the whole training and testing, we only use these scaled quantities and scale them back to physical units only when enunciating and visualizing the final results.

\subsubsection{Sample split, in the context of galaxy populations across cosmic epochs}
\label{sec:split}

We randomly split the TNG100 galaxy sample into a training, a validation and a test subset, in fractions of 80, 10, and 10 per cent, respectively. We fit the MLP and cINN models only to the training set, separately, whereas the validation set is used during the fitting process to evaluate the performance of the models. The test set is needed to test the generalization performance of the final models and consists of galaxies that are never seen before by the networks.

Because cosmological simulations follow the same galaxy populations across cosmic epochs, progenitors and descendants of the same galaxy (i.e. of the same main branch) may appear over multiple, consecutive snapshots. This special tree-like structure of the sample under consideration may require special precaution, as some information may unintentionally slip between the training and validation/test subsamples. For example, a galaxy in the validation/test set may have progenitor(s) and/or descendant(s) in the training sample with possibly similar observable and unobservable parameters: training a network under these circumstances may lead to a good optimization along the merger tree of each galaxy, but possibly to a bad generalization performance across galaxies throughout the TNG100 galaxy catalogues. 

The issue is more severe for more closely spaced snapshots: in the case of TNG, the time span between consecutive snapshots is about 150 Myr; moreover, this issue is progressively more problematic for smaller training sets. In the real Universe, the properties of the galaxy populations at different redshifts certainly reflect the underlying evolution of individual galaxies and their properties across time. However, in early tests, we observed that training our network(s) without reducing the merger tree-like information between training and validation/test sets causes overfitting and therefore can eventually lead to a bad performance on real observational data.
To be more specific: without paying attention to the tree-like structure of the data, our ML model(s) performed worse when trained (and tested) on larger galaxy samples (i.e. available from the TNG300) in contrast to the expectation that a larger training sample should yield more information  that a network could learn from. We believe that this is caused by the fact that a larger sample (where galaxies are in general closer to each other in parameter-space) counteracts the overfitting caused by the information that is present in the tree-like structure. 

To overcome this problem we split the TNG100 galaxy sample into training, validation and test subsets according to the merger trees, such that all progenitors and descendants of each galaxy are contained in the same subset. 

In practice, we perform the following steps:
\begin{enumerate}
    \item get the root descendant for each galaxy (at each snapshot); i.e. the ID of the galaxy it will eventually result in according to {\sc SublinkGal};
    \item split the list of these root descendants in training, validation and test sets;
    \item sort each galaxy according to the associated root descendant into the $3$ subsets.
\end{enumerate}
With this procedure it is ensured that the progenitors and descendants of each test and validation galaxy are not contained in the training set. We have verified that, after splitting according to root descendants, training and validation galaxies indeed lie much further apart in parameter space than with the default random splitting.

\subsection{Network configurations and training}

For both the MLP and cINN architectures, we want to find the optimal architecture and optimal set of training parameters. To do so, we perform for each network type a 
parameter search i.e. we train each network multiple times on the same set of TNG100 training galaxies but alternating the architecture and training parameters at each step.
Each training is stopped as soon as the validation loss (i.e. the model loss when applied to the validation galaxy subsample) does not decrease anymore.
From all resulting models, we choose then the network configuration that yields the lowest validation loss.
In the following, we describe the architectures and training parameters we have found to suit best for the particular problem at hand. 

\begin{table}
	\centering
	\begin{tabular}{c|c|c}
		\textbf{Layer} & \textbf{Output Shape} & \textbf{Parameters} \\
		\hline
		Input & 7 & 0 \\
		\hline
		Dense & 256 & 2048 \\
		\hline
		Batch Normalization & 256 & 1024 \\
		\hline
		ReLU Activation & 256 & 0 \\
		\hline
		Dense & 256 & 65792 \\
		\hline
		Batch Normalization & 256 & 1024 \\
		\hline
		ReLU Activation & 256 & 0 \\
		\hline
		Dense & 256 & 65792 \\
		\hline
		Batch Normalization & 256 & 1024 \\
		\hline
		ReLU Activation & 256 & 0 \\
		\hline
		Dense & 256 & 65792 \\
		\hline
		Batch Normalization & 256 & 1024 \\
		\hline
		ReLU Activation & 256 & 0 \\
		\hline
		Dense & 5 & 1285 \\
	\end{tabular}
	\caption{{\bf Layer configuration used for the Multilayer Perceptron neural network adopted in this work} (MLP, Sections~\ref{sec:methods_mlp} and \ref{sec:training_mlp}). We give the layer type (left column), the output shape of the respective layer (middle column), and the number of trainable parameters per layer (right column). Within our physical problem at hand, the 7 input parameters of the MLP are the 7 observable galaxy properties, whereas the target/output quantities are 5 summary statistics of  the galaxies' assembly and merger history; both chosen and defined in Section~\ref{sec:properties}. In total, this model has $202'757$ trainable parameters.}
	\label{tab:mlparchitecture}
\end{table}

\subsubsection{MLP}
\label{sec:training_mlp}

The basic architecture of an MLP is a series of dense neuron layers (i.e. each neuron is connected to all neurons of the previous and next layer) and of activation and regularization layers.
To do a recursion on $5$ unobservable quantities, we need an output layer with $5$ neurons (one for each unobservable quantity) and also an input layer with $7$ values. There is no straightforward way to determine the configuration of the layers between the input and the output layers. To reduce the number of possible architectures to explore, we use some basic constrains: a) We keep the number of neurons constant for all dense layers. Thus we only have to find the number of neurons per layer (the width of the MLP) and the numbers of layers (the depth of the MLP). b) Between each dense layer, we use one activation layer and one regularization layer. And c) we use a rectified linear unit as activation function.

The optimizer \textsc{Adam} \citep[]{kingma2017adam} proves to be a stable and (sufficiently) fast converging method and performs well with the default values given by the \textsc{Keras} framework implementation. We therefore use this optimizer in all of the following training procedures. If not stated otherwise, the learning rate is set to the default value $\eta = 10^{-3}$.

We use the mean squared error (MSE) as loss function, which leads to a faster and more stable convergence than the mean absolute error when starting from random weights.

As regularization method, we include a Batch Normalization \citep{DBLP:journals/corr/IoffeS15} between each dense neuron layer and activation function. 
Because we deal with a very unevenly distributed sample (especially as a function of galaxy stellar mass, with much fewer galaxies at the high-mass end), we have to take special care of an appropriate batch size. 
%A too small batch size would cause the gradient updates to be calculated from an unrepresentative batch of our sample (i.e. the gradient becomes too noisy). A too large batch size tends to slow down the training and might cause the optimization to miss the global optimum. We ultimately choose to use a batch size of $256$.

Regarding the node configuration, we choose to use $3$ hidden layers (in addition to the in- and output layer) with 256 nodes each. 

The resulting configuration of the MLP adopted in this work is summarized in Table \ref{tab:mlparchitecture}. After the validation loss has converged with this training procedure, an additional training step is performed: we change the loss function to the mean absolute error (MAE) and halve the learning rate to $\eta = 0.5 \times 10^{-3}$. We do so because the MSE loss function leads to a better convergence when starting from a set of randomly initialized weights. However, we expect that our data set is contaminated by ambiguous outliers, i.e. of case galaxies whose merger statistics are not properly described by the set of $7$ input quantities. We therefore focus more on the optimization regarding the more ``well-behaved'' galaxies in the sample.

\subsubsection{cINN}
\label{sec:training_cINN}

For the cINN, we rely on conditional affine coupling blocks \citep[in so-called GLOW configuration,][]{2018arXiv180703039K} that utilise two standard (and therefore non-invertible) MLP sub-networks. Additionally, a further ``conditioning'' MLP network is introduced between the conditions (i.e. the 7 galaxy observables of Section~\ref{sec:properties}) and the coupling layers. In practice, this conditioning network works like the point-prediction model of Sections~\ref{sec:methods_mlp} and \ref{sec:training_mlp}, whereas the coupling layers provide a model for the posterior around the point predictions.

For the conditioning network we use the architecture shown in Table~\ref{tab:conditional_architecture}; the sub-networks are summarized in Table~\ref{tab:subnetwork_architecture}. For the cINN itself, we use in total 12 coupling layers with a random perturbation layer between each. With this configuration, we have in total 1'262'584 parameters to train.

\begin{table}
	\centering
	\begin{tabular}{c|c|c}
		\textbf{Layer} & \textbf{Output Shape} & \textbf{Parameters} \\
		\hline
		Input & 7 & 0 \\
		\hline
		Dense & 128 & 1024 \\
		\hline
		ReLU Activation & 128 & 0 \\
		\hline
		Dense & 128 & 16512 \\
		\hline
		ReLU Activation & 128 & 0 \\
		\hline
		Dense & 128 & 16512 \\
		\hline
		ReLU Activation & 128 & 0 \\
		\hline
		Dense & 128 & 16512 \\
	\end{tabular}
	\caption{{\bf Layer configuration used for the {\it conditioning network} of the conditional Invertible Neural Network adopted in this work} (cINN, Sections~\ref{sec:methods_cINN} and \ref{sec:training_mlp}). From left to right, we give the layer type, the output shape of the respective layer, and the number of trainable parameters per layer.}
	\label{tab:conditional_architecture}
\end{table}

\begin{table}
	\centering
	\begin{tabular}{c|c|c}
		\textbf{Layer} & \textbf{Output Shape} & \textbf{Parameters} \\
		\hline
		Input & 130 & 0 \\
		\hline
		Dense & 128 & 16768 \\
		\hline
		ReLU Activation & 128 & 0 \\
		\hline
		Dense & 128 & 16512 \\
		\hline
		ReLU Activation & 128 & 0 \\
		\hline
		Dense & 128 & 16512 \\
		\hline
		ReLU Activation & 128 & 0 \\
		\hline
		Dense & 4-6 & 516-774 \\
	\end{tabular}
	\caption{{\bf Layer configuration used for the cINN {\it sub-networks}} (see Section~\ref{sec:training_cINN}). Columns as in Table~\ref{tab:conditional_architecture}.}
	\label{tab:subnetwork_architecture}
\end{table}

As already described in Section~\ref{sec:methods_cINN}, this network is trained by propagating the obervables and unobservables of the training subset through the network and then adjust the layer weights of the model in such a way that the negative log likelihood loss $L_{\rm NLL}$ is minimized.

During the training, we add a small amount of gaussian noise with $\sigma = 0.025$ to all inputs and outputs. This is done to lower the risk of overfitting, especially in the case of the parameters based on lookback time, which enter the network in a discrete fashion as the simulation only provides us with $50$ discrete snapshots. Again, we use \textsc{Adam} as optimizer and start with a learning rate of $2\times10^{-3}$, which is then gradually reduced (i.e. multiplied by $0.7$) after each $5$ runs through the whole training sample. We train the cINN in batches of $2048$ training galaxies. To avoid overfitting, we calculate the loss for the validation set and terminate the training if the validation loss has not decreased for $5$ following training epochs.

These training parameters (and model architecture) were determined by a parameter search: on the one hand we choose the parameters such that the validation loss is minimized; on the other hand, we also check that the latent space $z$ indeed becomes a gaussian and we also test if the prior distribution of the validation set is successfully recovered (both utilizing a Maximum Mean Discrepancy between the two distributions).

\subsubsection{Ensemble Training}
\label{sec:ensembling}

As the network's initial weights as well as the gaussian noise augmentation are randomly chosen, the training of deep learning models is a stochastic process. Therefore, even when using the same data, the same network architecture and the same training parameters, the results of a network may differ, albeit slightly, from training to training: some will perform better or worse than others for specific galaxies. To take these network-to-network variations into account, we train multiple networks and average their results. In this way, we ensure a globally-optimal result over the whole sample while reducing the risk of overfitting as we can keep the single networks simpler and profit from the combined prediction capability. Furthermore, we ensure that our results (Section~\ref{sec:results}) are less dependent on the random choice of initial parameters.

For the MLP, we train $7$ equivalent networks with random initial weights. The ensemble prediction is then given by the galaxy-wise median of the $7$ single network predictions. We choose $7$ models because the additional training and validation effort from this is still computationally feasible. 

For the cINN, we also train 7 networks with random initial weights. However, in this case we cannot simply take medians of scalar point predictions. Instead, when sampling the posterior, the posterior samples are drawn alternately from the 7 single networks. The resulting ensemble posterior for each galaxy is therefore the normalized sum of the 7 individual network posteriors. 

\begin{figure*}
	\centering
	\includegraphics[trim={3.5cm 3cm 1cm 3cm},width=19cm]{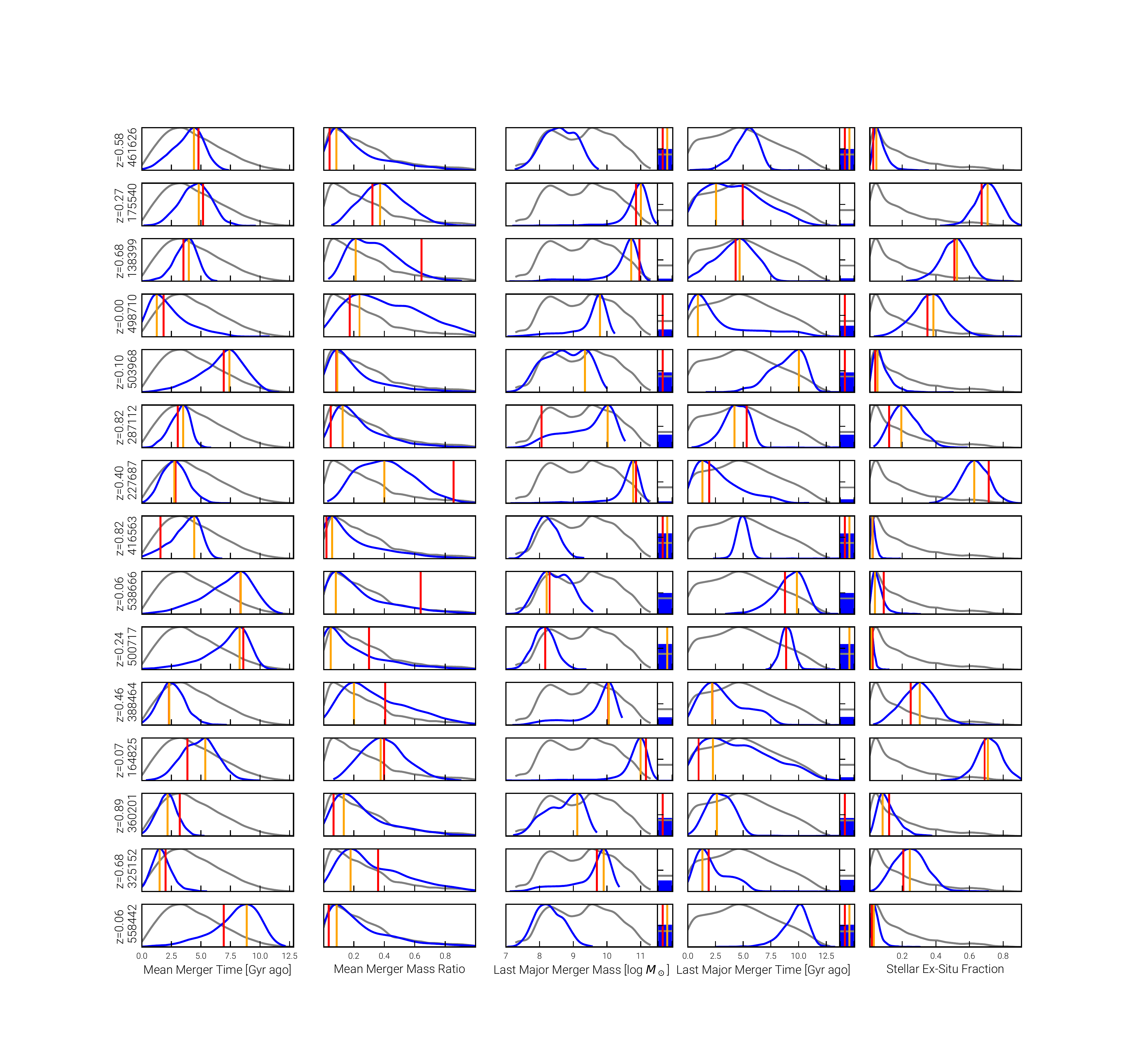}
	\caption{{\bf Posterior distributions obtained with our cINN} (Sections~\ref{sec:methods_cINN} and \ref{sec:training_cINN}) for 15 randomly-chosen galaxies from the TNG100 simulation test set (out of in total 17982 galaxies). Each row corresponds to one of the 15 galaxies (we show the redshift and the (snapshot-wise) unique {\sc Subfind} ID on the left) whereas each column shows results for the five inferred unobservable statistics of their assembly and merger history. We show the prior distributions of the overall sample of test galaxies (grey), the predicted posterior by our cINN approach (blue), the MAP prediction (Maximum A-Posteriori estimation, yellow), and the ground truth (red) -- i.e. the actual outcome of the TNG100 simulation.
	The distributions are normalized such that the maximum value is equal to 1. For the last major merger, we introduce an additional bin to show the fraction of posterior samples that fall outside the valid regime, i.e. the probability that the galaxy has never had a major merger in its merging history. This bin is chosen as the MAP for this galaxy only if the fraction of posterior samples falling into this bin is $> 0.5$ i.e. if the predicted probability for a major merger to have occurred is larger than 50 per cent.}
	\label{fig:results/TNG100/posterior_example}
\end{figure*}

%%%%%%%%%%%%%%%%% RESULTS %%%%%%%%%%%%%%%%%%

\section{Results: recovering the unobservable properties of galaxies' past}
\label{sec:results}

In this Section, we show the main scientific results of the paper, i.e. that it is possible to recover aspects of the past assembly and merger history of galaxies from a set of integral properties at a given time. In practice, in what follows, we apply the previously introduced and trained networks to the unseen test data set from the TNG100 simulation. We first focus on the overall performance of the cINN and then compare its predictions from those of the simpler MLP point-recursion model. We also show how the cINN recovers the cross-correlations among input and output galaxy properties and identify the most informative galaxy properties for the inference of the past history.

\subsection{Testing procedure}
\label{sec:test}

To test the performance of the algorithms and their prediction capabilities, we apply the trained networks to a previously-unseen test sample. To ensure full comparability, we use the same set of training, validation, and test TNG100 galaxies for both MLP and cINN.

Our main measure for the goodness of fit is the prediction error for each galaxy: this is the difference between the predicted values, inferred by each algorithm given the observables, and the ground-truth values given by the TNG100 simulation.

In the case of the simple recursion network (MLP) we get point predictions for each galaxy that can be directly used for the error measurement.

However, for the cINN the procedure is more articulate: we first sample the posterior for each galaxy (i.e. conditioned by the corresponding set of observables) by drawing $1000$ random normal-distributed points in the latent space $z$ and by propagating them backwards through the cINN.
The resulting posterior is then represented by the point density of the sampled point cloud and can be further postprocessed. To be able to compare this point cloud to the point predictions of the MLP, we identify as ``point prediction'' of the cINN the Maximum A-Posteriori estimation (MAP) i.e. the position of the highest point density in unobservable space.

Technically, to obtain the MAP for each unobservable quantity, we perform a gaussian kernel-density estimation on the set of posterior samples and evaluate the estimated density on a uniform grid of $512$ points along each dimension of the unobservable space and use the position of the highest density. Note that with this, we impose a strict upper limit on the prediction accuracy due to the discrete nature of this grid approach. However, with $512$ grid points we are well below the level of noise that was added during the training.

For the lookback time and mass of the last major merger, we define an additional bin for the galaxies or predictions that have no major merger in their history (we call them for convenience NMMs). The cINN MAP is set to this bin if at least 50 per cent of the posterior sample points fall into this bin i.e. have a predicted lookback time larger than $13.7$ Gyr -- the ground truth for these galaxies is set to $15$ Gyr; see Section \ref{sec:properties} -- or a predicted stellar mass below $10^7\, \MSUN$, as the ground truth for these galaxies is set to $10^6\, \MSUN$.

To take also the additional information of the posterior distribution into account, we calculate the standard deviation of the posterior point cloud along each dimension i.e. we take the root of the diagonal elements of the covariance matrix. So, whereas the prediction errors give an assessment on the accuracy of the predictions, we investigate below if the standard deviation of the posteriors serves as a meaningful error bar (i.e. level of uncertainty or precision) for the MAP predictions.

\subsection{Performance of the cINN}
\label{sec:performance}

\subsubsection{Posterior distributions from the cINNs}

A strength of cINN architectures is that they return a posterior distribution for each target quantity\footnote{It would in principle be possible to obtain a posterior distribution also for the MLP predictions: however, for the MLP, a parameterized distribution and therefore prior assumptions about the shape of the posterior would be needed, which is not the case for the cINN}. However, the performance evaluation is not as straightforward as e.g. the validation of a simple recursion method. To get a first impression of the performance of the cINN, we show in Fig.~\ref{fig:results/TNG100/posterior_example} a few examples of the cINN posteriors inferred for a random subset of TNG100 test galaxies. Each row shows results for each test galaxy, 15 in total; columns refer to the five summary statistics elected to summarize the assembly and merger history of galaxies (see Section~\ref{sec:properties}): from left to right, mean merger lookback time, mean meger mass ratio, stellar mass and time of the last major merger, fraction of ex-situ stars. Gray curves denote the prior distributions, i.e. the distributions of values encompassed by the whole sample of test galaxies; blue curves show the posterior distribution predicted by the cINN for each test galaxy and unobservable property. In each panel, we mark the MAP (the Maximum A-Posteriori estimation, Section~\ref{sec:test}) in orange and the simulation ground truth in red. It is important to note that here the posteriors (blue curves) are a convolution of the ``true'' data-driven posterior and the uncertainty introduced by the network-to-network variations dealt with in Section~\ref{sec:ensembling}.

Fig.~\ref{fig:results/TNG100/posterior_example} shows that the posteriors of the mean merger lookback time and particularly of the stellar ex-situ fraction are typically peaked and rather symmetrically distributed around this peak, denoting qualitatively a good prediction power of the cINN model in terms of point predictions. In contrast to that, the posterior of the mean merger mass ratio of the tested galaxies is often similar to the prior of this parameter. For the  mass and time of the last major merger, we can also see broader and more complex posteriors, at times with multiple peaks e.g. fifth galaxy from the top. Note that the additional bin, the one for galaxies with no major merger in their history, can be seen as a posterior peak as well. It is important to emphasize, however, that everywhere where the posterior probability is $>0$, a certain merger statistic configuration is in general predicted to be possible: i.e. we have to check if the inferred posteriors follow the distribution of the ground truth later on.

\begin{figure*}
	\centering
	\includegraphics[width=18cm]{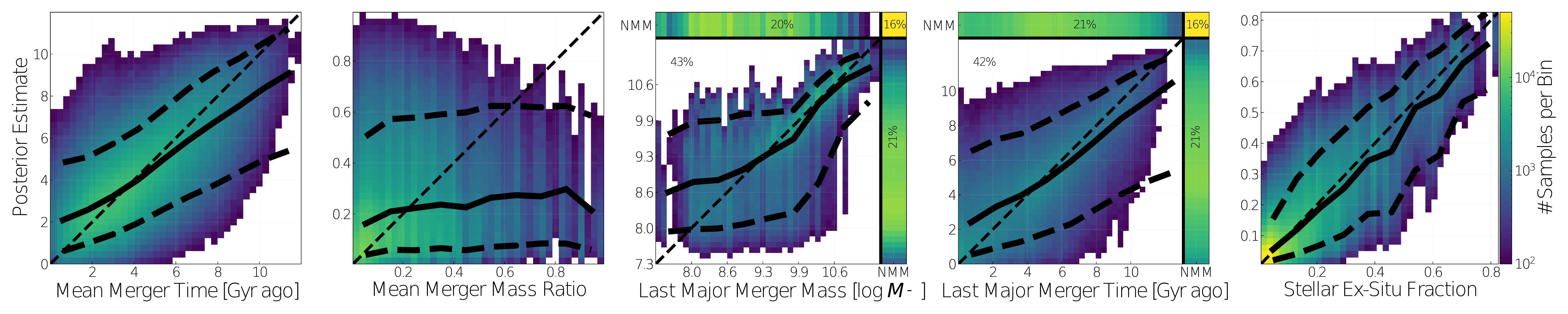}
	\includegraphics[width=18cm]{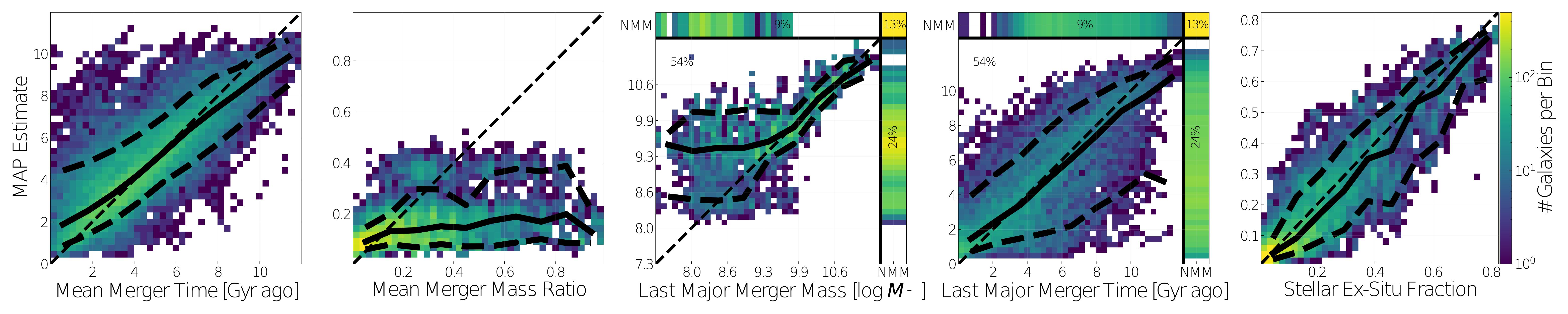}
	\includegraphics[width=18cm]{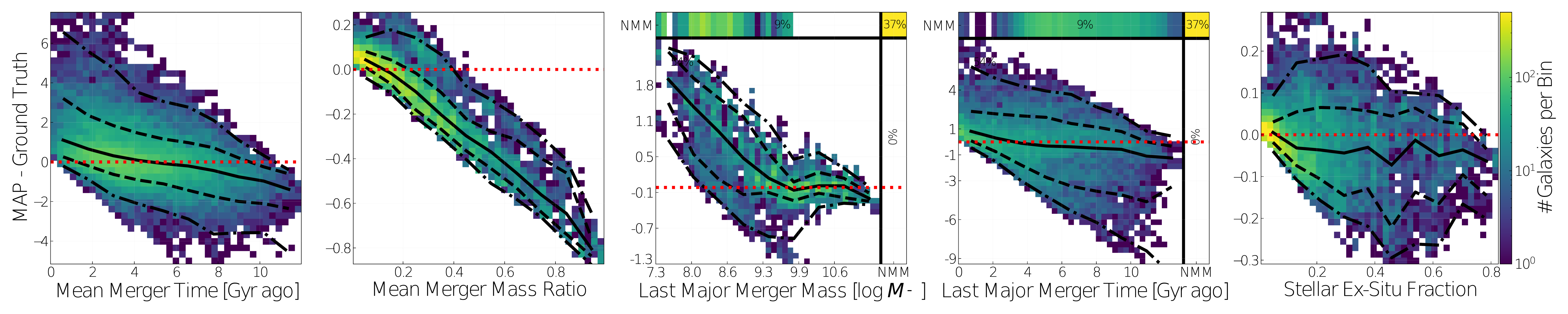}
	\caption{{\bf Prediction performance of our cINN based on training, validating and testing against the outcome of the TNG100 cosmological simulation.} The columns show results for the five inferred unobservable statistics of the assembly and merger history of galaxies, for 17`982 TNG100 test galaxies . {\bf Top:} 2D histograms of the cINN posteriors (y-axis) over simulation ground truth (x-axis) for each of the $5$ output quantities. Namely, for each test galaxy, we draw $100$ random posterior samples and stack them across all test galaxies. The measured median is given as solid black line while the area included within the black dashed lines contain 80 per cent of the datapoints. For comparison, we also show the ideal result (i.e. the case whereby MAP prediction and ground truth concur) as dotted black diagonal.
	{\bf Middle:} 2D histograms of the cINN MAPs (y-axis) vs. simulation ground truth (x-axis) for each of the $5$ output quantities. Meaning of the lines and annotations as in the top panels.
	{\bf Bottom:} 2D histograms of the cINN MAP errors, i.e the MAP - Ground Truth (y-axis) vs. simulation ground truth (x-axis) for each of the $5$ output quantities. The solid black line shows the median, the area within the dashed and dash-dotted lines contain $68 \%$ ($1\sigma$) and $95 \%$ ($2 \sigma$) of the datapoints in bins of ground truth values. We show for reference the ideal (zero error) baseline as red dotted line. For the time and mass of the last major merger (third and fourth panels from the left), we also show the galaxies that have no major merger (NMM) in their main progenitor branch according to the cINN (top), the ground truth (right) and both (upper right corner). The percentage denotes the overall fraction of galaxies in each of those 3 bins. Note that for the bottom plot, the right bin is empty as the error is undefined if the galaxy had no major merger according to the ground truth. 
	}
	\label{fig:results/TNG100/map}
\end{figure*}

\subsubsection{cINN predictions vs. ground truth}
\label{sec:cinn_vs_gt}

In Fig.~\ref{fig:results/TNG100/map}, we therefore systematically compare the cINN predictions with the ground truth from the TNG100 simulation. In particular, in the top panels of Fig.~\ref{fig:results/TNG100/map} we show the stacked posterior distributions predicted by the cINN for all galaxies of the test sample. In practice, we show the probability distribution functions of Fig.~\ref{fig:results/TNG100/posterior_example} for all test galaxies by stacking them in bins of true mean merger lookback time, mean meger mass ratio, mass and time of the last major merger, and fraction of ex-situ stars, from left to right, with colors denoting the normalized probability distribution function. In each panel, the black thin dashed line denotes the 1:1 relation; the solid black curve shows the median of the stacked posteriors and the dashed thick black curves encompass their 10th and 90th percentiles.

In the middle panels of Fig.~\ref{fig:results/TNG100/map}, we plot the MAP estimates for each test galaxy vs. the simulation ground-truth values, again different column for the five merger statistics. As we aim to quantify the outcome for 17`982 test galaxies, the results are given as color-coded numbers of galaxies per image pixel. In each panel, the black thin dashed line denotes the 1:1 relation; the solid black curve shows the running median of the galaxy test sample in the depicted plane; and the dashed black curves indicate the location of the 10th and 90th percentiles of the test galaxy samples in bins of true merger statistics. 

In the bottom panels of Fig.~\ref{fig:results/TNG100/map}, we replace the MAP estimates of the middle panel with the error on the prediction, i.e. the difference between the MAP estimate and the simulation ground truth for each test galaxy, for each inferred target quantity. The median prediction error is given by the solid black curves whereas the dashed (dot-dashed) curves encompass 68 (95) per cent of the test galaxies.

As also seen in Fig.~\ref{fig:results/TNG100/posterior_example}, Fig.~\ref{fig:results/TNG100/map} demonstrates that the cINN returns very good predictions for both the mean merger lookback time (leftmost panels) and the mass fraction of ex-situ stars (rightmost panels). The MAP estimates for the ex-situ fractions are on average slightly ($\approx 5$ per cent) below the ground truth, this being the case also from the stacked overall posteriors: this could be caused by the unevenly-distributed prior distribution of the ex-situ parameter, which strongly favours lower ex-situ fractions. Nevertheless, for 68 per cent of the galaxies, the prediction error (bottom panels) for the ex-situ fraction does not exceed $\pm 10-15$ percentage points from the ground truth: the model is therefore well able to discriminate between low and high ex-situ galaxies. The MAP error is especially low in the low ex-situ-fraction regime (i.e. ex-situ mass fractions lower than 10 per cent) where the predicted ex-situ fraction does not exceed $\approx \pm 5$ percentage points for $95$ per cent of the test galaxies. Similar promising results hold for the mass-weighted mean merger time, with errors on the MAP prediction smaller than $\pm 1.5$ lookback billion years for $\approx 69$ per cent of the test galaxies.

On the other hand, the cINN predictions for the time and mass of the last major merger, and especially for the mean merger mass ratio, are more complex. As noted before, the mean merger mass ratio posteriors are often similar to the prior of this parameter. We can see this also in terms of the MAPs, whose peaks are (independently from the ground truth) always close to the prior peak. In practice, our models cannot recover this particular summary statistics of the galaxies' history. We will further elaborate on this in Section~\ref{sec:mass_ratio}.

For the inference of the mass and time of the last major merger, it should be noted that 37 per cent of the TNG100 test galaxies have never had a major merger in their history according to the simulation (i.e. ground truth). The cINN predicts in $24$ per cent of the cases that galaxies have had a major merger even though they have experienced in truth no major merger (so, a wrong prediction). In contrast to that, $9$ per cent of galaxies with ground-truth major merger have been predicted to have no major merger, while $13$ per cent have been correctly identified to have no major merger in their history. These numbers are consistent for both the mass and time of the last major merger. 
%This means, that if the model outputs a No-Major-Merger MAP, there is roughy a $50$ per cent chance that the respective galaxy had indeed no major merger -- however, this corresponds to our choice to set the NMM MAP when more than $50$ per cent of the posterior samples fall into the NMM bin (more on this in Section~\ref{sec:nmm}). 

For the remaining galaxies that have undergone a major merger and that the cINN has correctly recognized as such ($\approx 54$ per cent of the overall test sample), the cINN performs well, particularly in timing the merger, with a good alignment to the ideal line. For the median galaxy, the cINN predicts an MAP value for the time of the last major merger that is close to the true one by less than $\pm 1$ billion years, also for very ancient mergers. Galaxies with a more recent major merger (within the last $2-3$ Gyr) have lower MAP errors throughout the galaxy population than ancient ones (and there is a much lower tendency to wrongly classify them as NMM). In contrast to that, there is a large fraction of galaxies with an early major merger ($> 8$ Gyr ago) that are wrongly predicted to have a much more recent major merger. For the stellar mass of the major merger, for the bulk (68 per cent) of the galaxy population, the cINN predicts the correct mass of the last major merger to within $\pm \,0.1-0.2\,$dex but only for mergers more massive than $\sim10^{9.5}\,\MSUN$ in stars. On the other hand, the prediction becomes ambiguous at lower masses. These two trends are naturally related to each other: because of the monotonically mass growth of galaxies, massive mergers will also happen at later cosmological times. The lower ambiguity for recent major mergers translates therefore directly to the better performance for more massive mergers and vice versa. In conclusion, the network performs especially well in identifying recent and/or massive major mergers, while the predictions become less accurate if the last major merger was longer ago (or not present at all). This supports the initial expectation and foundation of the whole exercise, namely that recent mergers have a more pronounced impact on the appearance of a galaxy.

It should be noted that the trends in the prediction error for the mean merger lookback time, the lookback time of the last major merger and (partially) the ex-situ fraction are also caused by the physical limits that our model learns: e.g. there are no negative lookback times or ex-situ fractions. Additionally, it is important to emphasize that choosing the MAP estimate is only one of the possible ways to condense the information contained in the posterior distributions. We have also evaluated the cINN performance based on the median and the mean values of the posteriors: the latter yields slightly better results regarding the under-prediction of the ex-situ fraction and the mass ratio but at the cost of returning worse performances for the other target statistics.

\begin{figure*}
	\centering
	\includegraphics[width=18cm]{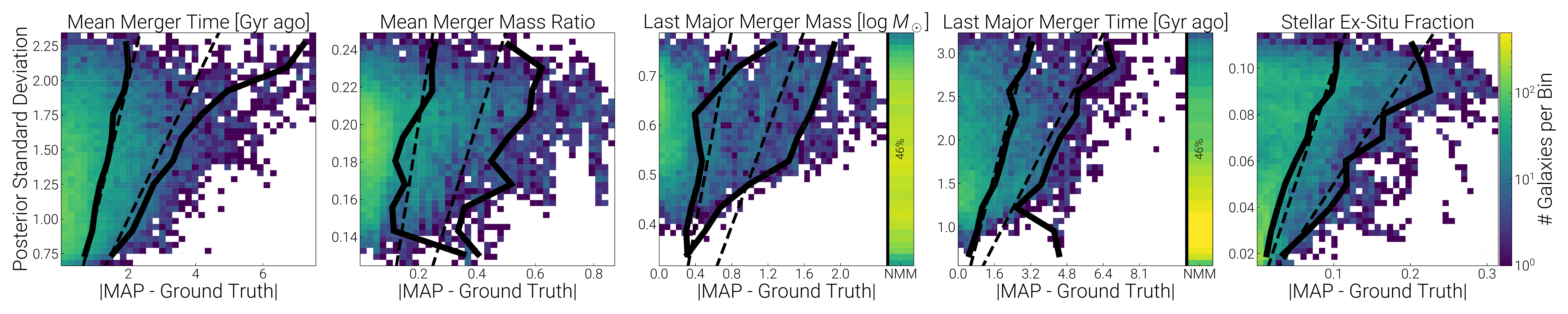}	\includegraphics[width=18cm]{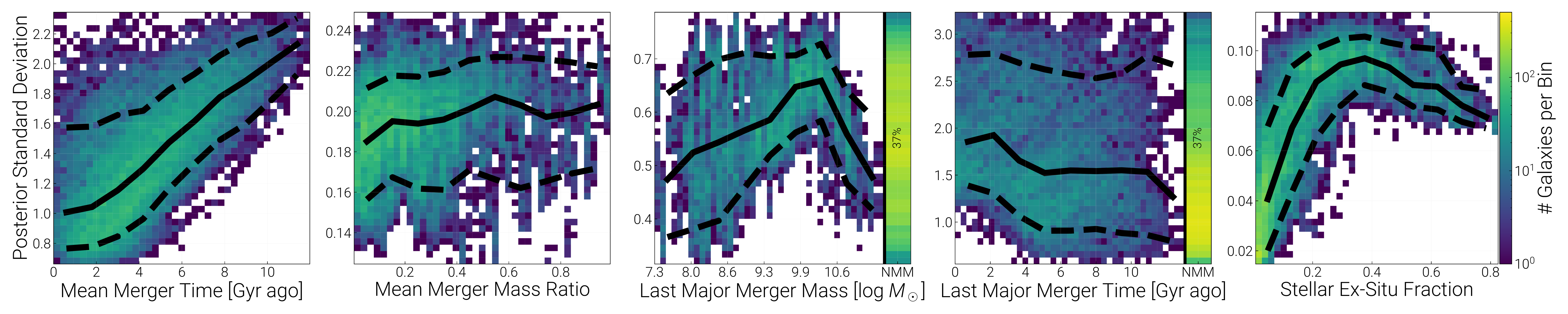}
	\caption{{\bf Uncertainties of the cINN predictions.} {\bf Top:} 2D histograms of the standard deviation of the cINN posterior distributions vs. the absolute value of the MAP prediction error for all TNG100 test galaxies. In bins of posterior standard deviation, we plot where the 68th and 95th percentiles of the test galaxies lie: thick black curves. The dashed lines correspond to the expected ``ideal'' result, assuming that the MAP prediction error is normal distributed.
	{\bf Bottom:} 2D histogram of the standard deviation of the cINN posterior distribution vs. the simulation ground truth. The black curves denote the median relationships, whereas 80 per cent of the test galaxies are contained within the area enclosed by dashed lines, in bins of predicted properties (x-axis).
	For the time and mass of the last major merger, we plot the galaxies that have never had a major merger according to the ground truth (i.e. the TNG100 simulation) in an extra bin (NMM) and show the fraction of galaxies that fall into this bin: in these cases, the standard deviations are only calculated from the fraction of posterior samples that fall into the physical range (i.e. lookback time $\le 13.7$ Gyr and stellar mass $\ge 10^7\, \MSUN$).
	}
	\label{fig:results/TNG100/sigma}
\end{figure*}

\subsubsection{Interpretation of the uncertainties of the cINN predictions}
\label{sec:uncertainties}

In a next step, we want to estimate the errorbars on the cINN point predictions, i.e. on the chosen MAP estimates, by evaluating the meaning of the standard deviation of each posterior. 

%It should be emphasized that 'error' in this context denotes the epistemic prediction uncertainty given by the limited information given by the observables. 

It should be emphasized that the cINN posteriors throughout this work include prediction uncertainties given by both the limited information given by the selected observables as well as by the network-to-network variations (see Section~\ref{sec:ensembling}). We proceed under the ansatz that the former is dominant.  

To use the standard deviation of the cINN posteriors as a meaningful measure for this uncertainty we need to answer the following questions:
\begin{itemize}
    \item Are the posteriors ``gaussian-like''? Namely, are the posteriors symmetric around one peak and hence with the well-known relations between the variance and the confidence intervals?
    \item Is the standard deviation of the posterior related to the prediction error of our test sample? In other words, is the width of the posteriors related to the difference between the MAP estimate and the ground truth?  
\end{itemize}
For example, from Fig.~\ref{fig:results/TNG100/posterior_example}, we already know that for the  mass and time  of the last major merger and the mean merger mass ratio the cINN often returns asymmetric posteriors. In such cases, the meaning of the posterior standard deviation is therefore in question. 

To investigate these  relations, in the top panels of Fig.~\ref{fig:results/TNG100/sigma} we plot the standard deviation of the posteriors vs. the prediction errors (here the absolute value of the difference between the MAP estimate and the ground truth) for the whole TNG100 test sample. To evaluate whether the error is indeed gaussian, we indicate with thin dashed lines the relationships we would expect if the cINN posteriors were normally distributed. Namely, in general terms, we would expect that $68$ per cent of the test galaxies with a posterior standard deviation equal to e.g. 1 have a prediction error  $\le 1$; similarly, we would expect that 95 per cent of the test galaxies with a posterior standard deviation equal to e.g. 1 have a prediction error  $\le 2$; and so on. We can then compare such idealized gaussian expectations with the locus of the galaxies in the plane of posterior standard deviation vs. prediction error: namely, in bins of the posterior standard deviation, we measure where the 68 and 95 percentiles of the galaxies lie: solid thick lines. 

For the ex-situ fraction we find that the empirical standard deviations (solid lines) of the cINN align well with the 'ideal' gaussian ones (dotted lines): we conclude that the ex-situ standard deviations can be well interpreted as a gaussian error. 
With caution, this can also be stated for the mean merger lookback time (especially at the 1-$\sigma$ level) and for the time of the last major merger. For the other two target quantities, we can see that more attention needs to be placed on the shape of the posteriors (as expected from the examples in Fig.~\ref{fig:results/TNG100/posterior_example}).

In the bottom panels Fig.~\ref{fig:results/TNG100/sigma}, we also quantify the connection between the standard deviation of the posteriors and the ground truth values. There are two strong correlations to highlight and comment upon. Firstly, the standard deviation of the mean merger lookback time increases with the true mean merger lookback time. However, we find that this linear dependency is build up by galaxies at different redshifts, with the galaxies at $z=0$ populating the top right corner and those at $z=1$ populating the lower left region. 

Secondly, the standard deviation of the ex-situ fraction increases strongly with the true ex-situ fraction, reaches a peak at around fractions of $\approx 0.3 - 0.4$ and then slightly decreases again at higher ex-situ fractions. The low ex-situ fractions can therefore be estimated with small uncertainties (percentages of a few) for a large part of the overall galaxy sample.
%(low mass, blue disks). 
However, as the fraction of ex-situ stars increases, also the uncertainty increases up to $\approx \pm 10$ per cent. We interpret this as a result of the diverse and complex evolution history that leads to an increased ambiguity between the chosen set of observable inputs and the ex-situ fraction. The downturn at the high ex-situ end could be explained with the fact that this region of the parameter space is constituted again by a simpler population of galaxies, namely massive red galaxies whose stellar mass growth is dominated by the contribution from mergers.

We further quantify the degree to which the posterior distribution itself is related to the distribution of ground truth galaxies in Appendix~\ref{sec:calibrationerrors}.

\begin{figure*}
	\includegraphics[trim={0 0 4.2cm 0},clip,width=15cm]{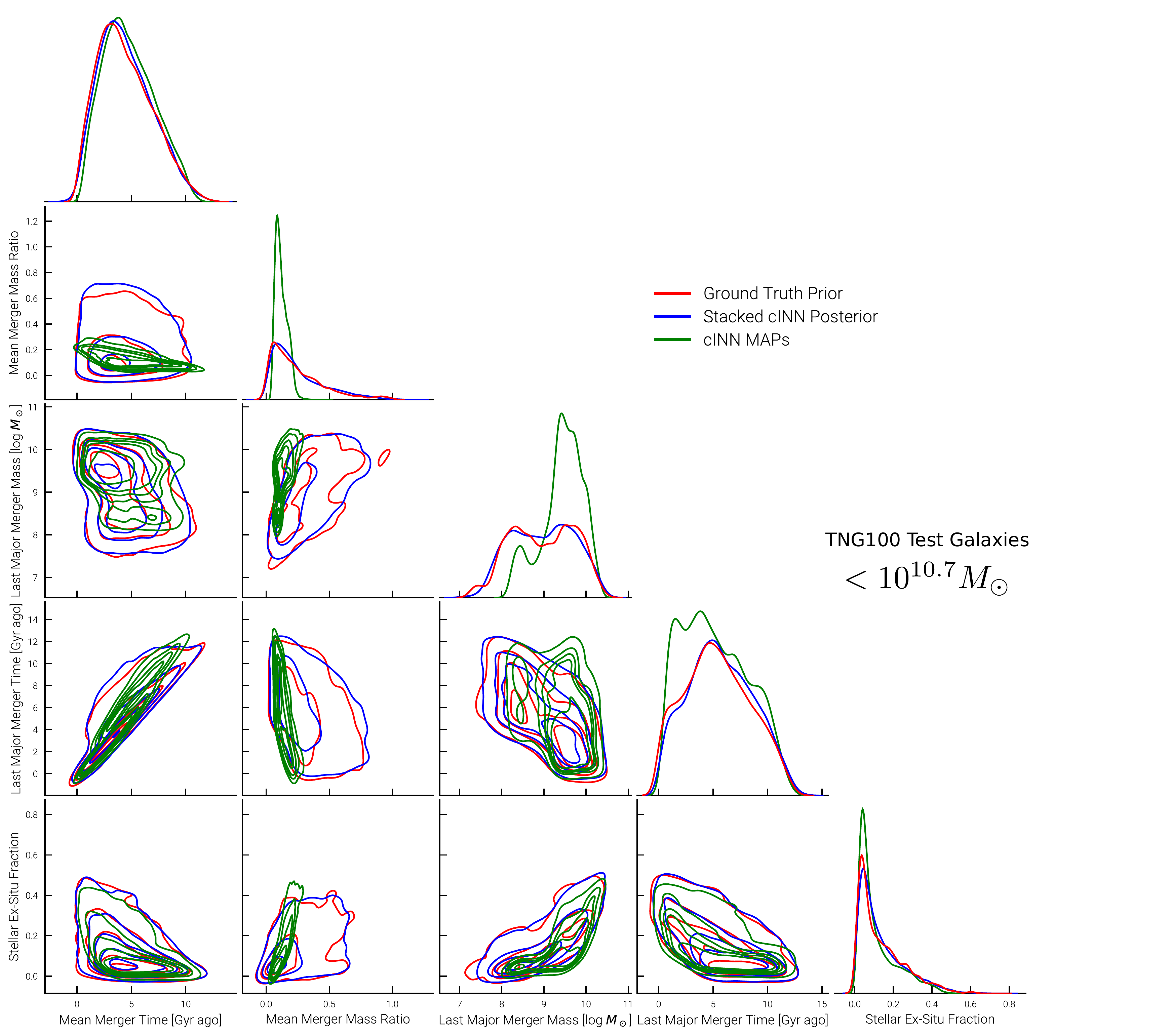}
	\caption{{\bf Recovery of the cross-correlations among the statistics of the stellar assembly and merger history.} We show how test galaxies occupy the parameter space among pairs of the cINN-predicted properties with contours at fixed number density: distributions of the ground truth test sample (ground truth priors, red), the posterior distribution predicted by the cINN trained and tested on the TNG100 galaxies (blue), and the MAP estimates based on these posteriors (green). For the posterior distribution, we include $10$ randomly drawn posterior samples per galaxy. Here we focus on galaxies with stellar mass in the $10^{10-10.7}\, \MSUN$ range. The prior distributions are in general well reconstructed by the cINN although a certain level of smoothing is visible especially in the less sampled higher-mass bin: see next Figure. The MAPs are in general heavily biased towards the peaks of the prior distributions; however, also the ``point-predictions'' from the cINN recover well the underlying true distributions (green vs. red contours) in the case of the stellar ex-situ fraction and the mean merger lookback time.}
	\label{fig:cinn_prior_lowmass}
\end{figure*}

\begin{figure*}
	\includegraphics[trim={0 0 4.2cm 0},clip,width=15cm]{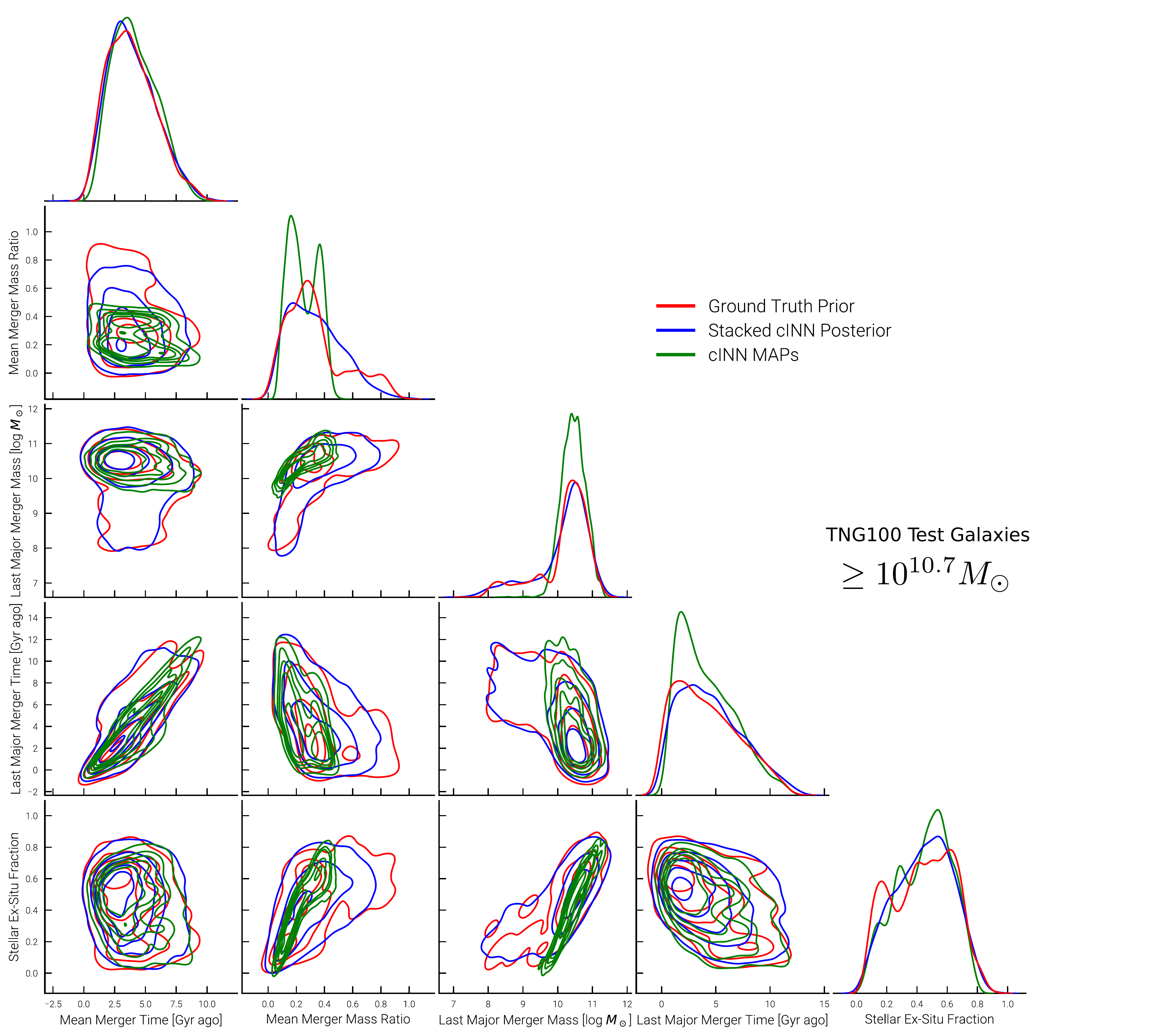}
	\caption{As in Fig.~\ref{fig:cinn_prior_lowmass} but for high-mass galaxies, with stellar mass in the $10^{10.7-12}\,\MSUN$ range.}
	\label{fig:cinn_prior_highmass}
\end{figure*}

\subsubsection{Recovery of secondary dependencies and cross correlations}
\label{sec:degeneracies}

In addition to the predictive power of the cINNs to return unobservable assembly and merger statistics from observed quantities, it is important also to check to what degree the cINN is capable to learn the cross correlations among the 5 statistics themselves. In fact, we know from the analyses of cosmological simulations that galaxies properties within both sets of observable and unobservable statistics are highly correlated with one another. So for example, at fixed galaxy stellar mass, galaxies with more massive mergers have larger ex-situ mass fractions \citep{rodriguezgomez2016exsitu}.  
In Figs.~\ref{fig:cinn_prior_lowmass} and \ref{fig:cinn_prior_highmass}, we compare a) the prior distributions of the ground truth, i.e. how the test galaxies populate the parameter space of the 5 target statistics (red); b) the distributions from the cINN by integrating its posterior samples (blue): namely, for each galaxy in the test sample we use the ground truth observables and randomly sample $10$ posterior points for the $5$ statistics; the overall distribution is then given by the sum of all points for all test galaxies; finally c) the distributions of the MAP values from the cINN (green). The resulting distributions are shown in Fig.~\ref{fig:cinn_prior_lowmass} for low-mass galaxies (stellar mass $<10^{10.7}\, \MSUN$) and in Fig.~\ref{fig:cinn_prior_highmass} for high-mass ones ($ >10^{10.7}\, \MSUN$).

The cross correlations among the target quantities are well reproduced by the cINN posteriors (blue vs. red contours), although a certain amount of smoothing is visible, particularly in the less-sampled high-mass end population. For example, the cINN posteriors recover the following trends: 
\begin{itemize}
    \item The lookback time of the last major merger and the mean merger lookback time are tightly correlated, with only a small fraction of galaxies having a significant amount of minor mergers after their last major merger event (galaxies with no major mergers are excluded here).
    \item Smaller mean merger lookback times, larger mean mass ratios, larger masses of the last major merger and smaller times since the last major merger imply larger ex-situ stellar mass fractions at the time of inspection. These relation steepen for higher-mass galaxies, with the exception of the mean merger lookback time that slightly decouple from the ex-situ fraction in high-mass galaxies.
    \item The mass and the time of the last major merger are connected, with more recent mergers being typically also more massive.
\end{itemize}

These correlations -- extracted from the TNG100 simulation and successfully recovered by our cINN -- are qualitatively in agreement with those predicted by the Illustris simulation and analyzed by \cite{rodriguezgomez2016role} and \cite{rodriguezgomez2016exsitu}, whose works have inspired our selection of stellar assembly and merger statistics.

However, whereas the integrated posteriors follow the general trends of the ground truth, this is not the case for their MAP values: these are somewhat biased and askew distributions, which naturally favour the peak of the posterior distributions (by construction). This is especially pronounced and critical for the mean merger mass ratio of low-mass galaxies, and less so for higher-mass galaxies, for which the MAPs seem to be able to roughly learn the correlation with e.g. the ex-situ fraction and to slightly grasp the trend towards higher-mass ratios. Also in the case of the cross-correlations and consistently with what seen in the cINN performance analysis so far, the MAPs work best for the stellar ex-situ fraction, the mean merger lookback time, and the time of the last major merger. For the other outputs, the MAPs have to be interpreted with caution, e.g. for the mass of the last major merger for massive galaxies with last major merger smaller than $10^{9.5}\,\MSUN$.

\begin{figure*}
	\centering
	\includegraphics[width=17cm]{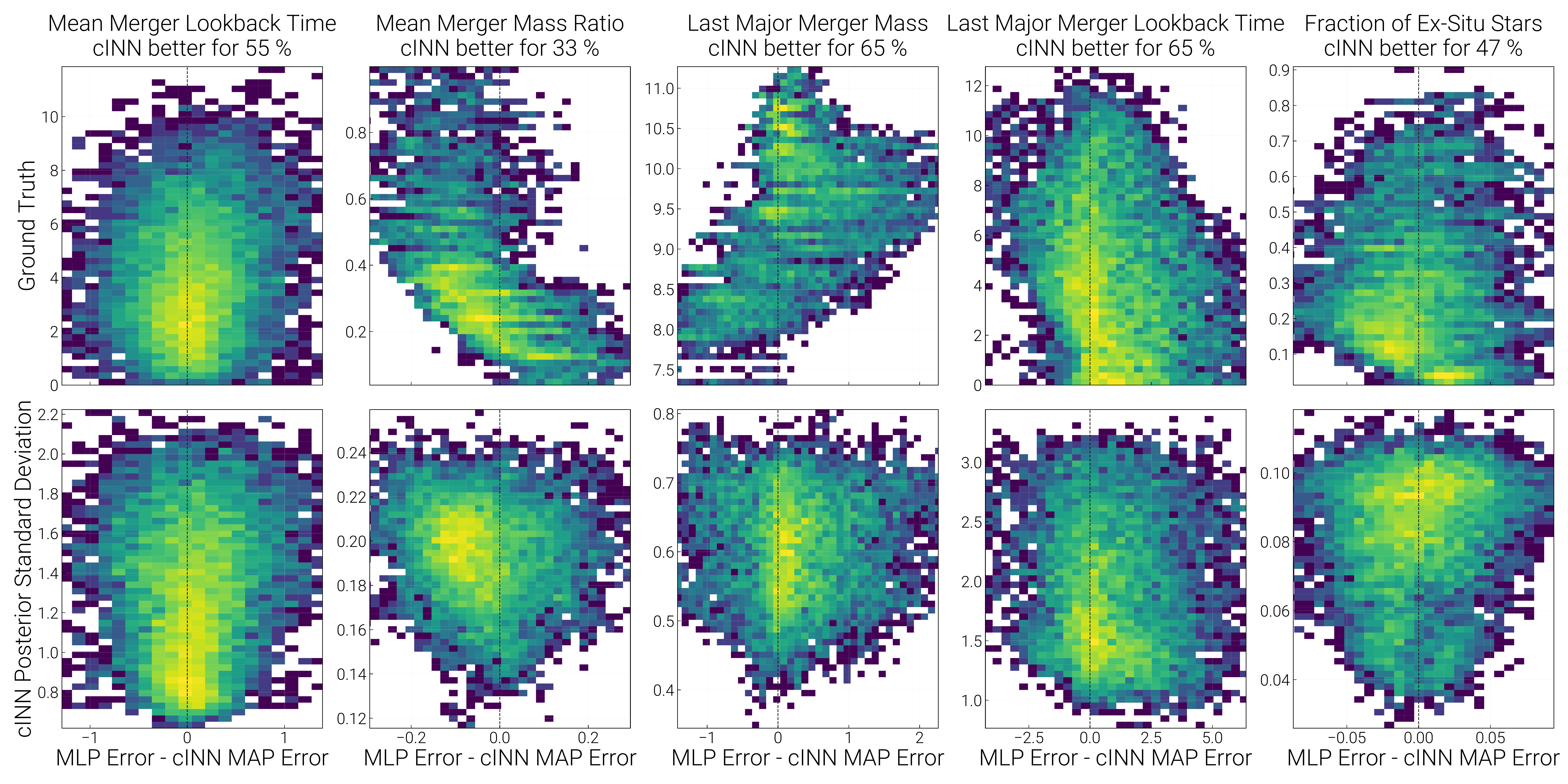}
	\caption{{\bf Comparison of the MLP with the cINN MAP predictions}. On the x-axis, we plot the difference between the prediction errors from the two methods: $|{\rm MLP} - {\rm Ground~Truth}| - |{\rm MAP} - {\rm Ground~Truth}|$, where here ${\rm MLP}$ represents the MLP point-prediction value and ${\rm MAP}$ is the cINN Maximum A-Posteriori estimation. Therefore, a positive value of this quantity means that the cINN performs better in the predition than the MLP.  {\bf Top:} ground truth vs. difference between the errors of the MLP and cINN methods. We observe that the distributions are in general symmetrical around $0$, but for the case of the mean merger mass ratio (second column from the left). The performance of the two models is therefore similar in terms of their point predictions, with a slightly better performance for the cINN. {\bf Bottom:} standard deviation of the cINN posterior vs. difference between the errors of the MLP and cINN methods. The galaxies where the MLP and cINN MAPs disagree more substantially are also predicted to have a broader posterior distribution. This is especially pronounced for the ex-situ fraction. We speculate that this is an effect caused by the intrinsic uncertainty in the prediction of the selected unobservables. In these plots, for the time and mass of the last major mergers, we exclude galaxies with no major merger in their history. 
	}
	\label{fig:results/MLP_cINN_comparison}
\end{figure*}

\subsection{Comparison between MLP and cINN: do they yield the same results?}
\label{sec:mlp}

How well does the much simpler MLP architecture (Sections~\ref{sec:methods_mlp} and \ref{sec:training_mlp}) perform in comparison to the cINN (Sections~\ref{sec:methods_cINN} and \ref{sec:training_cINN})? We quantify this in Fig.~\ref{fig:results/MLP_cINN_comparison}, by noting that exactly the same set of training/validation and test galaxies have been used for both MLP and cINN. We can hence directly compare the galaxy-wise point predictions by the MLP with the MAPs of the cINN (Section~\ref{sec:test}). 

In Fig.~\ref{fig:results/MLP_cINN_comparison}, we show the difference in absolute prediction error between the two models, namely $|{\rm MLP} - {\rm Ground~Truth}| - |{\rm MAP} - {\rm Ground~Truth}|$, where here ${\rm MLP}$ represents the MLP point-prediction value and ${\rm MAP}$ is the cINN Maximum A-Posteriori estimation, which we have already quantified and analyzed e.g. in Figs.~\ref{fig:results/TNG100/map} and \ref{fig:results/TNG100/sigma}. Therefore, the cINN works better on galaxies where this quantity is positive, the MLP works better where it is negative. If the error of both methods is comparable, this value is close to $0$.

In the top panels of Fig.~\ref{fig:results/MLP_cINN_comparison}, we compare the difference between the prediction errors against the true values of the target quantities. We see that this metric follows overall a rather symmetric distribution around $0$ for all the assembly and merger statistics, implying that the MLP and the cINN (evaluated via the MAPs) perform similarly well. 

We find larger deviations for the mean merger mass ratio, where the MLP performs slightly better than the cINN. The point-recursion with a mean squared error loss is able to track the dependence better than the MAP of the cINN posteriors, which instead tend to broaden towards higher mass ratios but keep their peak (and therefore the MAP) at a similar position. However, this over-performance of the MLP is on average not larger than $\pm 0.1$: therefore, not even the MLP is capable of inferring this parameter.

We also see that the lookback time of the last major merger has a significant fraction of galaxies that do not work well with the MLP model. This might be caused by our choice to set this parameter to 15 Gyr for those galaxies with no major mergers. This choice does not work well with point predictions, as the MLP is not able to express ambiguities and therefore fall in between the peaks of the potentially multi-modal posterior distributions. The same problem might be the case for the mass of the last major merger, where the cINN MAPs perform substantially better than the MLP predictions.

We hence conclude that both the MLP and the cINN methods are rather equivalent regarding their point prediction capability, particularly so for inferring the ex-situ-fractions and the mean merger lookback time and the time of the last major merger. This is a confirmation of our implementation of the two methods, but it is not too surprising, as the minimization of the mean squared error (which is used to train the MLP) corresponds to the squared term in the log-likelihood loss (used to train the cINN). 

Further deviations (in terms of scatter) between the two models might be explained by galaxies that have an intrinsically non-well defined point prediction i.e. a broad posterior distribution. To test this hypothesis, we also plot the difference between the prediction errors from the MLP and the cINN against the cINN posterior standard deviation for each merger statistic: lower panels of Fig.~\ref{fig:results/MLP_cINN_comparison}. We see that the difference in performance between the MLP and the cINN indeed increases with the width of the inferred posterior distributions,  especially for the stellar ex-situ fraction and the mean merger lookback time, i.e. the two statistics we identified to have gaussian like posteriors in the previous section. The intrinsic uncertainty of the prediction therefore plays an important role in the diverging results between the two methods.

\begin{figure}
	\centering
	\includegraphics[width=8.5cm]{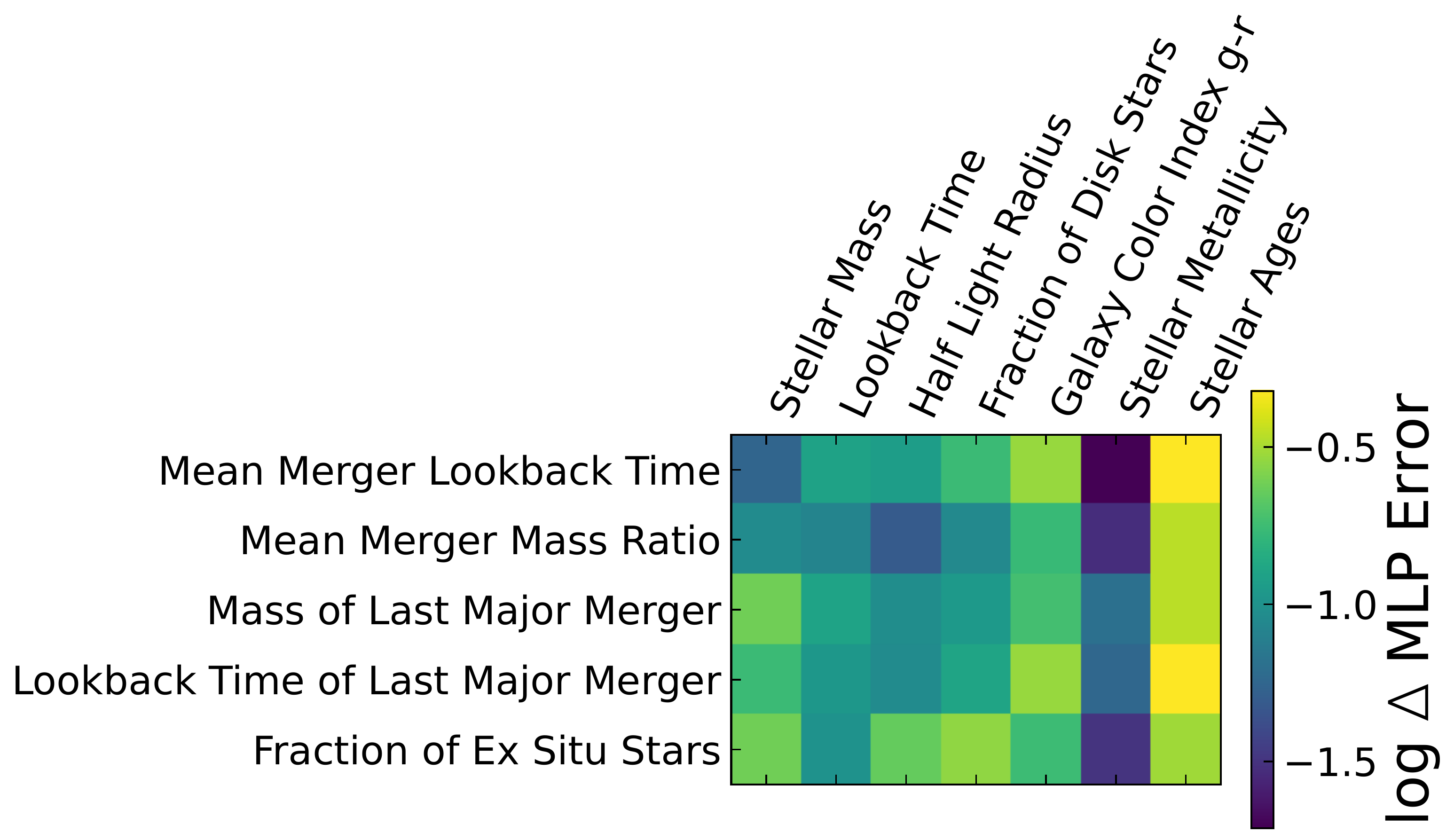}
	\caption{{\bf What inputs are the most informative?} Sensitivity analysis from the MLP model. Each field and color denote the change in the mean prediction error (in absolute value) for the target properties (left) if an observable input (top) is randomly shuffled across the test sample. As the shuffling breaks any information given by the respective input, a shuffled input is considered to be important if the error between the original prediction and the shuffled prediction rises accordingly.
	As the change in error is evaluated target-wise, it is only comparable for each output property separately but not across  targets (i.e. not across rows).
	We see for example a strong dependence of the ex-situ fraction on the stellar mass and on the morphological/structural parameters such as the disk-to-total ratio and the half light radius. Very prominent is also the influence of the (luminosity-weighted) stellar age on the prediction of most merger statistics.}
	\label{fig:mlpsensitivity}
\end{figure}

\subsection{Sensitivity analysis: what are the most informative input galaxy properties?}
\label{sec:sensitivity}

We conclude this analysis by looking into the connections that the networks actually learned, i.e. which input is most important to successfully infer which output. 

As we have shown in Section \ref{sec:mlp} that there are qualitatively no big differences between the MLP point predictions and the cINN MAPs, we here utilize the simpler MLP network to identify to what degree the predicted outputs are sensitive to input variations in the model. In practice, for each observable input parameter, we perform the following steps:
\begin{enumerate}
	\item randomly shuffle the values of the observable across the whole set of test galaxies -- in this way, we prevent this parameter from adding any information to the model without changing the overall prior distribution;
	\item perform an MLP prediction on this modified sample, while all other observables are kept untouched;
	\item measure the mean prediction error (in absolute values) of this modified prediction with respect to the ground truth;
	\item subtract the mean error of the original prediction from the mean error of the modified prediction, hence producing a scalar that summarizes the importance of the observable under scrutiny.
\end{enumerate}
 
The resulting quantity measures how the prediction absolute error changes, on average, if we randomly shuffle one of the inputs. As this shuffling breaks any learned connections, a large increase in error means that the input is important for the prediction capability of the model. In Fig.~\ref{fig:mlpsensitivity}, we summarize our findings by evaluating the importance of each observable input, one at the time, and by color-coding the change in prediction errors for each unobservable output. As the changes in prediction errors are evaluated target-wise, the colors of Fig.~\ref{fig:mlpsensitivity} are comparable only for each target quantity separately but not across rows. 

For example, we see that there is a substantial increase in the average prediction error for the stellar ex-situ fraction not only if the stellar mass is shuffled but also (and perhaps even slightly more) if the fraction of disc stars input is shuffled. The strong dependence on galaxy stellar mass is expected, given the correlation between ex-situ fraction and stellar mass already commented upon and shown in Fig.~\ref{fig:methods/observables}: mergers and satellites accretion play a progressively more important role in the stellar mass growth of progressively more massive galaxies \citep{Pillepich_2014, Pillepich_2018}. However, Fig.~\ref{fig:mlpsensitivity} shows that inputting a galaxy's morphology further improves the predictions. This is consistent with previous non-ML analyses of cosmological simulations, which have shown that, even at fixed stellar mass, galaxies with a higher fraction of disc stars (thus disc galaxies) have a lower ex-situ fraction i.e. their stellar inventory is dominated by in-situ star formation \citep{rodriguezgomez2016exsitu, rodriguezgomez2016role}. Finally, the prediction of the ex-situ fraction is also sensitive to the galaxy stellar size, consistent with previous simulation-based findings whereby a large stellar half mass radius is also an indicator for a large ex-situ fraction, as ex-situ stars tend to populate larger galactocentric distances \citep{Davison_2020, Zhu2021}.

The prediction of the time since the last major merger is strongly affected by the stellar age.

A similar dependence on stellar age is in place for the prediction of the mean merger lookback time. This in truth could also be caused by the fact that galaxies close to $z = 1$ have a more recent merger history and a higher stellar formation rate (thus younger stars). However, in general, as galaxies' average stellar age increase monotonically with galaxy stellar mass above $10^{10}\, \MSUN$ both in the Universe and in the TNG100 simulation \citep{nelson2017results}, the merging with less massive objects is destined to reduce the stellar age of a galaxy. Furthermore, the merger of gas rich galaxies may also cause a burst in star formation, hence causing the (luminosity-weighted) stellar age to decrease.

Interestingly, the stellar metallicity of galaxies appears not to be important for the prediction of the chosen stellar assembly and merger statistics. This is at first glance in contradiction with results from simulations, where it has been seen that the merger history also affects the stellar metallicity \citep{10.1093/mnras/stz538}. We speculate that the same information provided by the metallicity is in fact also contained in the galaxy color and stellar age and we warn the reader that this may be different if focus is placed on the stellar haloes at low-surface brightnesses rather than on the central region of galaxies and on the the metallicity gradients rather than the average stellar metallicity (\textcolor{blue}{Zessner et al in prep.}).

%%%%%%%%%%%%%%%%% DISCUSSION %%%%%%%%%%%%%%%%%%

\section{Discussion and outlook}
\label{sec:discussion}

In this paper, we have shown that it is in principle possible to infer important information about the past assembly and merger history of galaxies by solely relying on a handful of galaxy features as inputs, such as galaxy stellar mass and redshift, morphology, stellar size and average stellar age. However, we have also shown that not all chosen unobservable properties can be predicted by supervised deep neural networks to a good level of accuracy. Moreover, we have for now trained and tested the method on the data from one cosmological simulation only. In this Section, we hence discuss in more detail the more subtle and complex findings of our work and provide considerations for the future application of these ML models to real galaxy data.

\subsection{Understanding the cINN results}

\begin{figure*}
	\centering
		\includegraphics[width=0.27\linewidth]{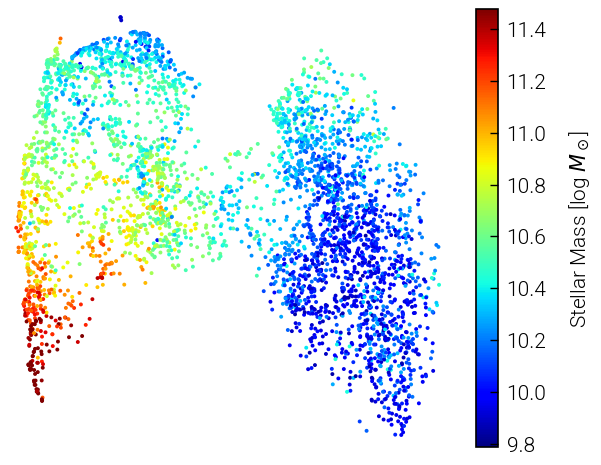}
		\includegraphics[width=0.27\linewidth]{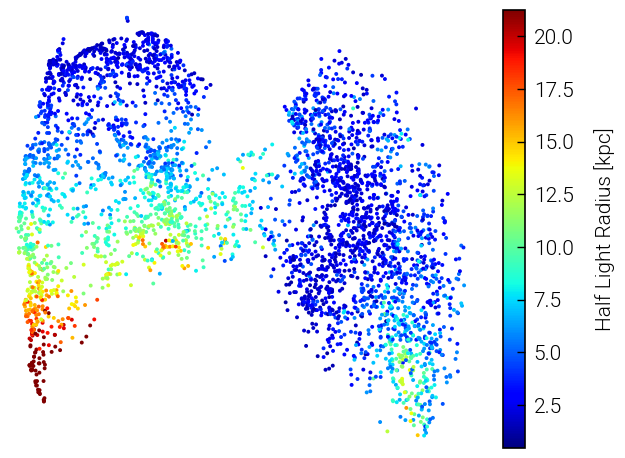}
		\includegraphics[width=0.27\linewidth]{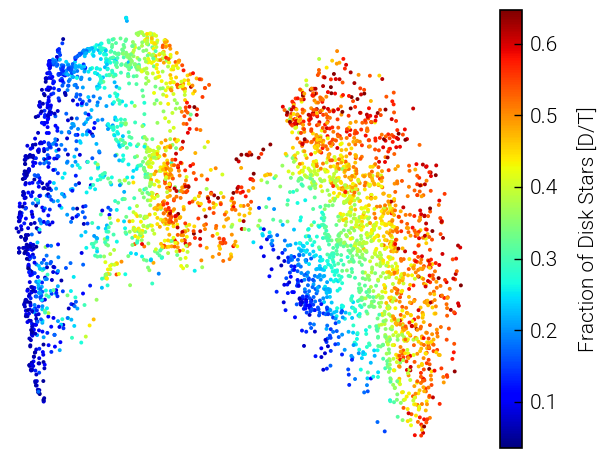}
		\includegraphics[width=0.27\linewidth]{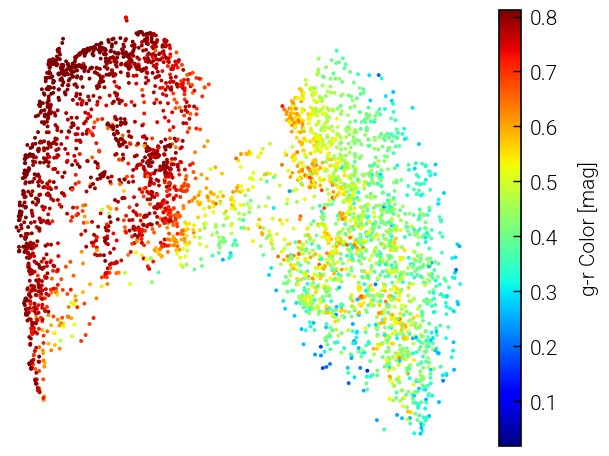}
		\includegraphics[width=0.27\linewidth]{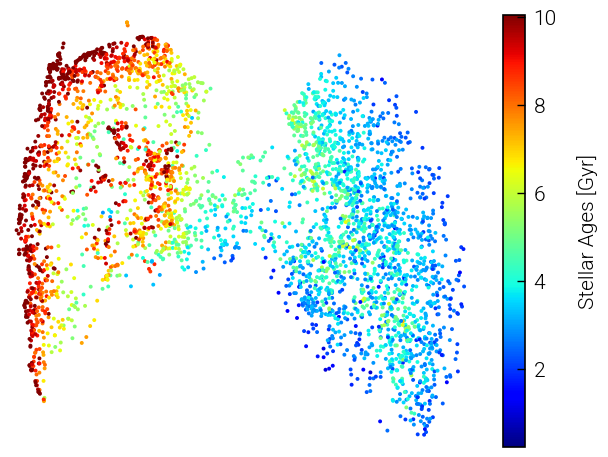}\\
		\vspace{1.5cm}
		\includegraphics[width=0.27\linewidth]{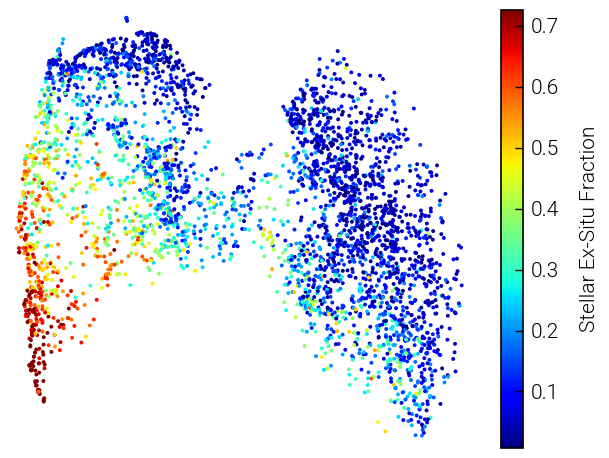}
		\includegraphics[width=0.27\linewidth]{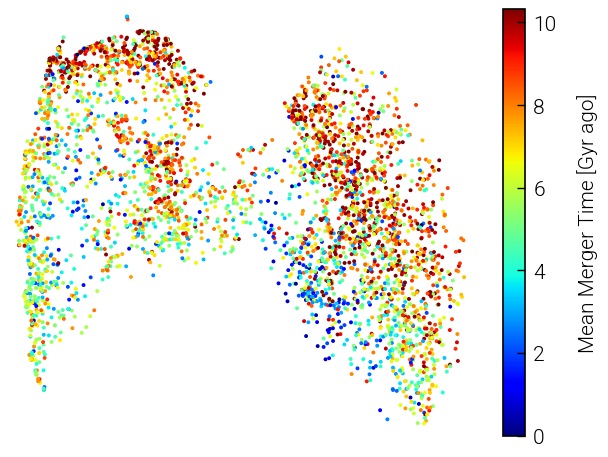}
		\includegraphics[width=0.27\linewidth]{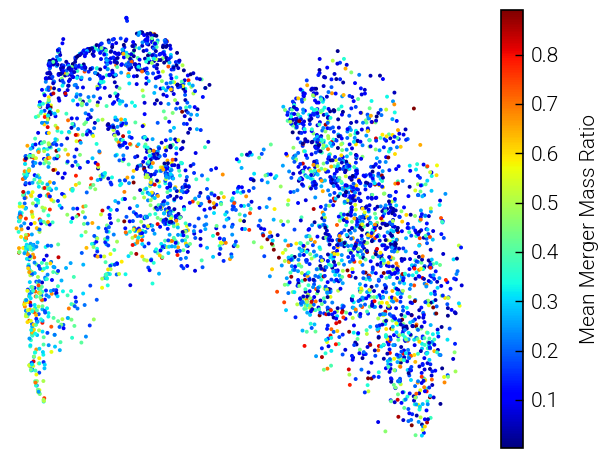}
		\includegraphics[width=0.27\linewidth]{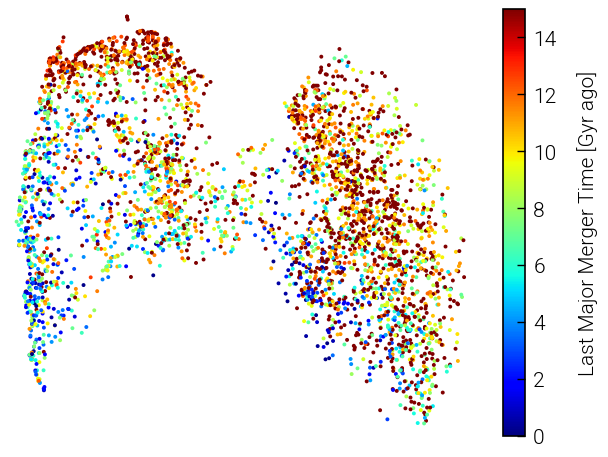}
		\includegraphics[width=0.27\linewidth]{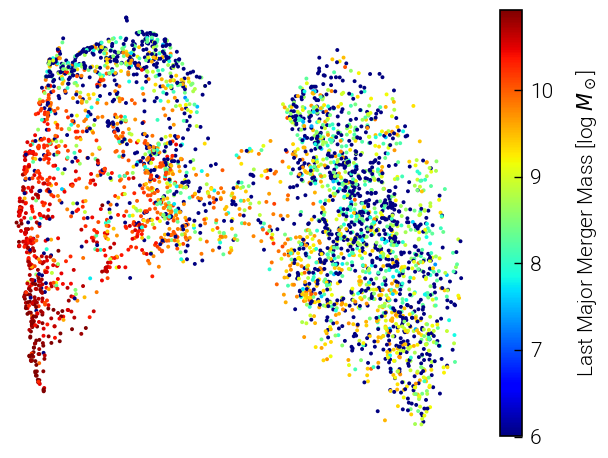}
	\caption{{\bf UMAP representation of TNG100 galaxies at $z=0$}, that embeds the six-dimensional space of the observable statistics (excluding the galaxies' redshift) into a 2D representation space. We deliberately omit axes to the panels as they have no immediate physical meaning. One dot corresponds to one galaxy. The colors denote, in each panel, the ground truth values from the simulation of the observable inputs (top) and the unobservable statistics of the stellar assembly and merger history (bottom). From top-left to bottom-right: stellar mass, stellar size, fraction of disc stars, g-r color, stellar age,  ex-situ fraction, mean merger lookback time and mean mass ratios, and the lookback time and mass of the last major merger. For the last two quantities, galaxies with no major merger in their history are plotted with fixed (unphysical) masses and lookbacktimes according to Section \ref{sec:properties}. To obtain the UMAP representation, here we have used the whole TNG100 sample, including train, validation and test sets. In this representation space, the sample clearly decomposes into star-forming (right grouping) and quenched (left grouping) galaxies. It is remarkable how smoothly and coherently the ex-situ fraction is represented across this 2D space, even if only observable features have been used to create this representation.}
	\label{fig:UMAP}
\end{figure*}

\subsubsection{The case of the mean merger mass ratio}
\label{sec:mass_ratio}

From the analysis in the previous sections, we have seen that the mean merger mass ratio of galaxies is not predicted well at all by the MLP nor the cINN. More specifically, we have seen that the cINN posterior of the mean merger mass ratio is often similar to the prior distribution.
Are these results just caused by an insufficient training or modeling of the networks? Or is the information simply not contained in the list of observables we decided to use as inputs?

To get a rough idea about the nature of this problem, we use UMAP \citep[Uniform Manifold Approximation and Projection for Dimension Reduction,][]{mcinnes2020umap} to visualize the 6-dimensional observational space of galaxies at redshift $z=0$. With this, we get a non-linear embedding into an easily-visualizable 2D space, which is supposed to preserve the overall topology of the dataset as best as possible. We gauge the parameters of this method in such a way that the representation shows a clear bi-modality and also represents a smooth function in stellar mass. The first criterion is reasoned by the expected prominent difference in observables between star-forming and quenched galaxies; the second is justified by our fiducial demand that the representation should be a simple function for a quantity as basic and fundamental as the stellar mass of galaxies.

The UMAP results are shown in Fig.~\ref{fig:UMAP}, where every data point corresponds to one $z=0$ galaxy in our overall TNG100 sample. We deliberately omit axes to the panels as they have no immediate physical meaning. To show the connection with the quantities studied throughout, we color code each galaxy (= datapoint) according to one of the observable (top) and unobservable (bottom) properties, one property per panel. For example, looking at the galaxy g-r colors and the stellar ages (panels in the second row from the top), we can see that the 6D space reduces to two sub-regions: the red and old (quenched) galaxies on the left and the blue and young (star-forming) ones on the right. The galaxies with a dominant accretion and merger history (and hence high ex-situ fractions) are mainly located in the lower edge of the quenched population, with the highest stellar masses. Yet, non-negligible ex-situ fractions can be found also on the left edge of the blue and young galaxies ``wing''.

The very good MLP/cINN predictions of the ex-situ fraction of galaxies, for example, can be understood with the fact that the ex-situ fraction changes smoothly and coherently across this representation space  -- even if the ex-situ fraction was not used to create it! Similarly, the changes of stellar mass across the UMAP space align well with the ex-situ fraction, and indeed the stellar mass is an important predictor of the ex-situ fraction, as also shown in e.g. Fig.~\ref{fig:mlpsensitivity}. However, in addition to this and as already demonstrated in the previous sections, the cINN is able to capture also more subtle and secondary dependencies. For example, the enhanced ex-situ galaxies in the blue and young regime coincide with lower fractions of disc stars, consistently with the findings of Fig.~\ref{fig:mlpsensitivity}: the morphology of a galaxy, i.e. here the disc to total mass ratio, is especially important to distinguish low and high ex-situ galaxies in the blue regime. 

Returning now to the original question of this section, we can see that galaxies with very different mean merger mass ratios can occupy drastically-different locations in the UMAP representation space, with little coherence: for example, low merger mass ratios can be found throughout the UMAP. 
In our view, this is a strong hint that the set of observables used in this work is not sufficient to fully determine the mean merger mass ratio across galaxies' histories. As also previously pointed out by \cite{rodriguezgomez2016role} in the Illustris simulation, a weak correlation with the ex-situ fraction is in place; however there are still many galaxies with low merger mass ratios in the high ex-situ mass regime as well as many galaxies with high merger mass ratios in low ex-situ mass regime. Our cINN model describes all this by returning rather broad posteriors regarding this parameter, which is often similar to the overall prior.

In conclusion, the overall massiveness of all the merger events in a galaxy's history seems to leave a rather ambiguous imprint on the set of observables we have investigated. However, previous studies on stellar haloes in e.g. Illustris \citep{Pop_2018} have shown that the appearance of stellar halo features such as shells (as opposed to streams) heavily depends on the massiveness of the mergers. We are therefore confident that including light images of galaxies, and in particular images of their dim stellar haloes beyond the central luminous bodies, can further help to constrain the average mass ratio of a galaxy's mergers.

\begin{figure}
	\centering
	\includegraphics[width=\columnwidth]{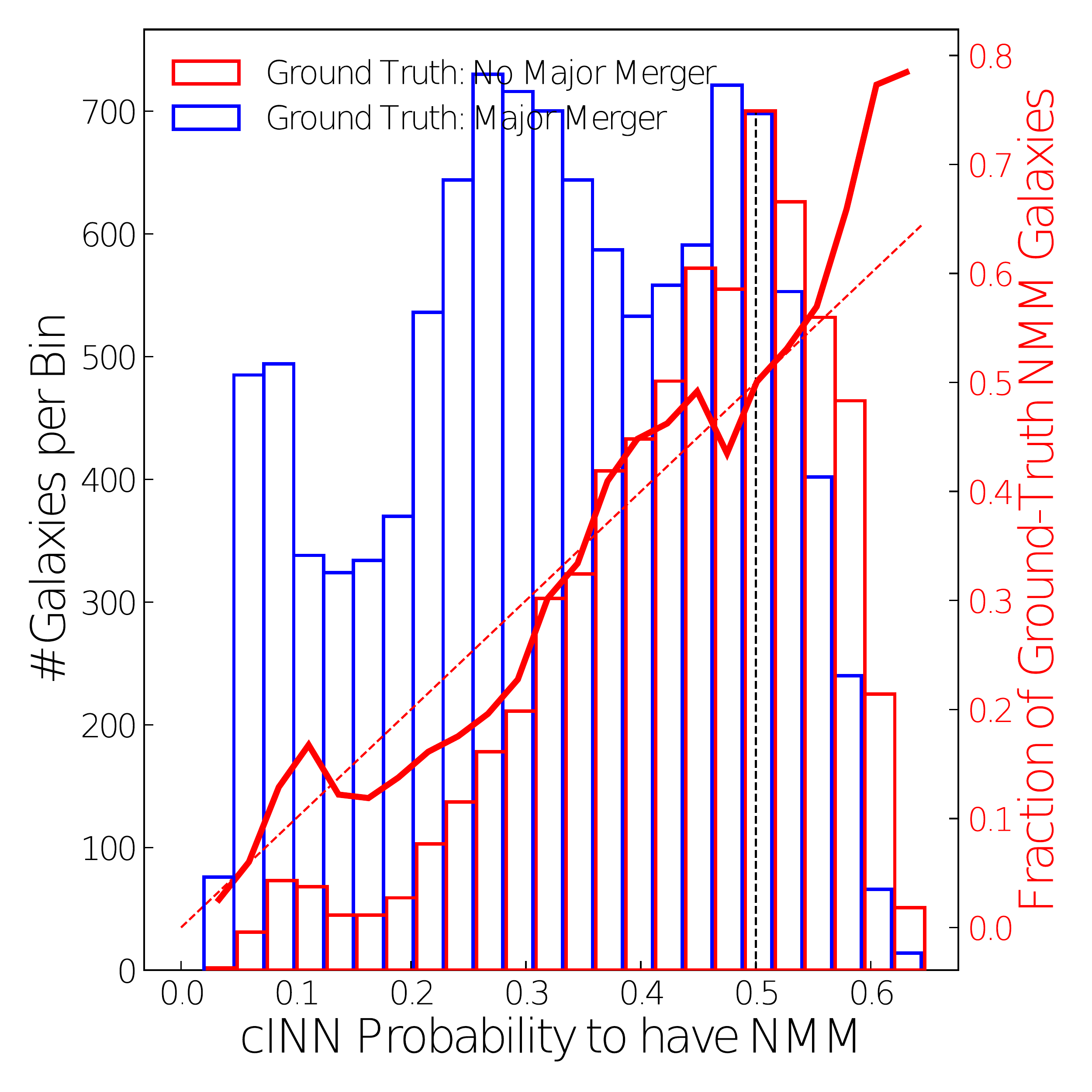}
	\caption{{\bf Capability of the cINN to deal with galaxies with and without a major merger.} We investigate how the fraction of cINN samples that fall into the NMM (No Major Merger) bin, i.e. the probability, relates to the ground truth from the TNG100 simulation. We plot the histograms of these NMM probabilities for test galaxies with no major merger in their ground truth (red) and with at least one major merger in their ground truth (blue). A cINN probability of $0.5$ (vertical dotted line) has been chosen as the boundary above which galaxies are {\it classified} by the cINN to have had no major merger, i.e. we set the MAP estimate to the NMM bin (unphysical time or mass of the major merger) only if more than $50$ per cent of the posterior samples fall into this bin. Our expectation is that the fraction of cINN posterior samples in the NMM bin is a good proxy for the fraction of ground-truth galaxies that had no major merger. This is the case: the red thick curve denotes the number of ground-truth NMM galaxies divided by the total number of galaxies in bins of cINN probability. We see that this quantity (red curve) indeed follows the fraction of cINN NMM samples (red dotted line).}
	\label{fig:non_valid}
\end{figure}

\subsubsection{Stellar mass and time of the last major merger}
\label{sec:nmm}

In this work, we have aimed at inferring the stellar mass and time of the last {\it major} merger. These are of course very interesting statistics of the past of a galaxy, as a massive merger is expected to affect its evolution substantially. However, this choice adds great complexity to the analysis due to the fact that, as we have seen, not all galaxies have ever experienced a major merger (i.e. with stellar mass ratio $\ge0.25$) throughout their lifetime. And so we have seen that the cINN posteriors of these merger summary statistics can be very complex (Fig.~\ref{fig:results/TNG100/posterior_example}), often with wide plateaus or with multiple probability peaks. 

This complexity could have been easily avoided should we have chosen to determine, instead, the mass and time since the last merger of whatever type, i.e. with no constraints on the mass ratio or mass. However, we think this may have not been fully interesting from a scientific perspective and might have also revealed ambiguous without an imposition on the minimum stellar mass of the secondary, as a massive merger might be concealed by a more recent minor merger. 

Throughout the analysis we have dealt with the fraction of galaxies that never had a major merger (NMM) by de facto including a classification scheme: firstly, we have set the ground-truth parameters of the mass and time of the ``last major merger'' to unphysical values in the case with no ground-truth major merger (NMM bin); secondly, we have set the cINN MAP estimate to the NMM bin only if more than 50 per cent of the posterior samples fall into this bin. But there are several open questions:
\begin{itemize}
    \item Have the cINN posteriors successfully learnt the classification between galaxies with and without major mergers?
    \item For which galaxies is this ambiguity the largest?
    \item Are the MAP results more accurate if we use the second cINN peak for the galaxies wrongly classified to have never had a major merger?
\end{itemize}

For the first question, we look in Fig.~\ref{fig:non_valid} at the histogram of the probability inferred by the cINN that a galaxy is a NMM (No-Major-Merger) case: we show these probabilities for all TNG100 test galaxies divided in those that according to the ground truth had no major merger (red) and those with at least one major merger in their lifetime (blue).  Firstly, it is noticeable that only a small fraction of the test galaxies have been predicted by the cINN to have a probability $\ge50$ per cent to have had no major merger in their history; in fact there is no galaxy with such NMM probability $>70$ per cent. Therefore there are no galaxies that had been identified by the cINN to unambiguously be NMM. 

We can also see that roughly half of the galaxies with a predicted NMM probability of $50$ per cent had indeed no major merger in their lifetime according to the ground truth: red vs. blue histograms at the location of the black dotted line. On the other hand, the galaxies with a low assigned NMM probability are overwhelmingly those with major mergers according to the ground truth, which hence the cINN seems to correctly identify. To further quantify this, we show as red thick curve the ground-truth fraction of the galaxies with no major merger with respect to the total number of galaxies (for each histogram bin). We clearly see that this ground-truth fraction is related to the cINN posterior probability, as the red dotted line denotes the 'ideal' 1:1 relation between cINN probability and the ground truth fraction.

\begin{figure}
	\centering
	\includegraphics[width=9cm]{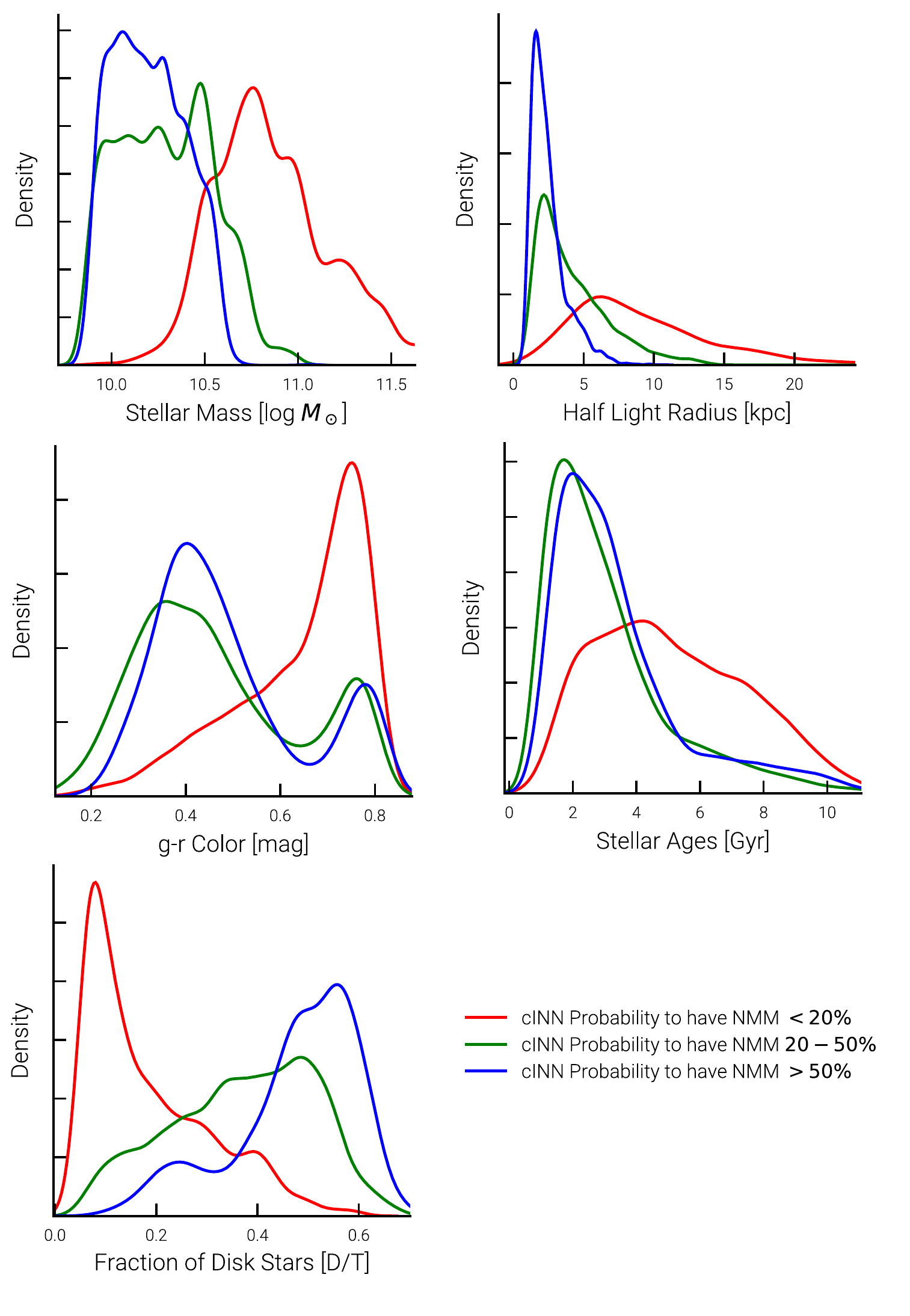}
	\caption{{\bf Which galaxies (in terms of observable properties) have the least ambiguous prediction in relation to having had or not a major merger?} We plot how the TNG100 test galaxies populate the space of the observables when divided in three classes, based on the probability assigned by the cINN to be a No-Major-Merger (NMM) case: $<20$ (red), $20 - 50$ (green), and $>50$ (blue) per cent, respectively. The galaxies with higher probability to be classified as NMM (blue curves) are typically lower-mass, bluer and diskier. Galaxies with masses $\gtrsim 10^{11}\,\MSUN$ are in general assigned by the cINN a very low probability that there was never a major merger in their history (red curves). However, the galaxies with ambiguous cINN probability (green curves) share similar physical properties with those with higher ones. The redshift and the stellar metallicity of galaxies do not exhibit significant biases across these three classes of galaxies, and so they are omitted.}
	\label{fig:non_valid_2}
\end{figure}

Still, a non-negligible fraction of galaxies are wrongly classified by the cINN.  In Fig.~\ref{fig:non_valid_2} we investigate which galaxies (in terms of observable features) are preferred to have no major mergers and which are instead ambiguous regarding this prediction. We split the test galaxies into three classes based on the NMM probabilities given by the cINN: namely, probability to have no major merger $<20$ per cent (red),  $20 - 50$ per cent (green), $> 50$ per cent (blue). The latter are those that in the MAP point prediction are classified as NMM, whereas the sum of the low- and intermediate-probability ones (red + green) gives the population that is predicted to have experienced a major merger. Following the comparison to the ground truth in Fig.~\ref{fig:non_valid} and given that there is no NMM probability $\ge70$ per cent, we consider all galaxies with NMM probability in the $20-70$ per cent range to have a rather ambiguous prediction. 

Fig.~\ref{fig:non_valid_2} shows that the observable properties of the galaxy populations with low vs. intermediate/high  NMM probability are quite different: galaxies with a more ambiguous and higher NMM probability (green and blue curves) are in general lower-mass, disc-like, bluer and somewhat younger than those with very low NMM probability. We interpret that the cINN cannot distinguish the ambiguity between a) isolated disc-like and blue galaxies that never had any/recent major merger events vs. b) disc galaxies that had major mergers but were able to retain their disc shape thanks to the continuing star formation, the latter being sustained or even enhanced by the gas available during the merger (wet merger scenario). In contrast to that, we find that more massive, redder, and more elliptical galaxies are identified by the cINN to have overwhelmingly experienced major mergers, in agreement with the expectations from cosmological simulations. 

In relation to the third aforementioned question, we notice that more accurate and less ambiguous results for the last major merger would have been obtained if we had chosen as cINN point prediction the best posterior peak, namely the one closest to the ground-truth value, instead of the highest peak of the posterior distribution (the MAP, used throughout). This alternative approach is justified by the fact that the relevant posteriors are often bimodal or with multiple peaks. In such a case, the picture of Fig.~\ref{fig:results/TNG100/map} turns into better fits. Namely, for the mass of the last major merger, we get a good median prediction down to progenitor masses $ \gtrsim10^{8.5}\,\MSUN$ instead of the limit of $\gtrsim 10^{9.5}\,\MSUN$ from the MAPs. For the time since the last major merger, the scatter in prediction accuracy is smaller, particularly for galaxies with more ancient (ground-truth) major merger. 
Clearly, when applying this model to observational data, there is no possibility to choose the ``best peak''. However, this analysis shows that also the complex major merger posterior distributions and the resulting major merger classification is feasible and physically interpretable.

\subsubsection{Dependence on the shape of the TNG100 prior}

Finally, a few points of discussion are needed in relation to the influence of the TNG100 prior distribution on the posteriors of the cINN. In a number of occasions, we have seen that the underlying prior strongly biases the cINN predictions. For example,  firstly, the posteriors of the mean merger mass ratio are mostly equal to the prior itself (Figs.~\ref{fig:results/TNG100/posterior_example}) and, as expected, the mean merger mass ratio is unconstrained. Moreover, the ex-situ fractions tend to be under-predicted and to be biased towards the prior peak, which peaks at lower ex-situ fractions. Finally, for most unobservable target quantities, the MAPs are in general shifted towards the prior peaks (Figs.~\ref{fig:cinn_prior_lowmass} and \ref{fig:cinn_prior_highmass}). 

Whether these behaviours are generally desirable is not clear: on the one hand, the cINN uses the knowledge of the unobservable parameters and their inter-relationships from the overall TNG100 distributions (e.g. that low ex-situ galaxies are much more common than high ex-situ galaxies). On the other hand, however, this could be a problem when applying the resulting network to other cosmological simulations or even observations, i.e. when the shape of the underlying prior is not necessarily identical to the one used for the training. In this work, we have chosen to not re-normalize the resulting cINN posteriors to a flat prior; i.e. by dividing the cINN posteriors by the TNG100 prior. We leave the assessment of e.g. such an alternative approach as an open point for future work, where we will also compare the predictions from different cosmological simulations and therefore bracket the uncertainties associated to the galaxy-formation model and the numerical resolution.

\begin{figure*}
	\centering
		\includegraphics[width=0.49\linewidth]{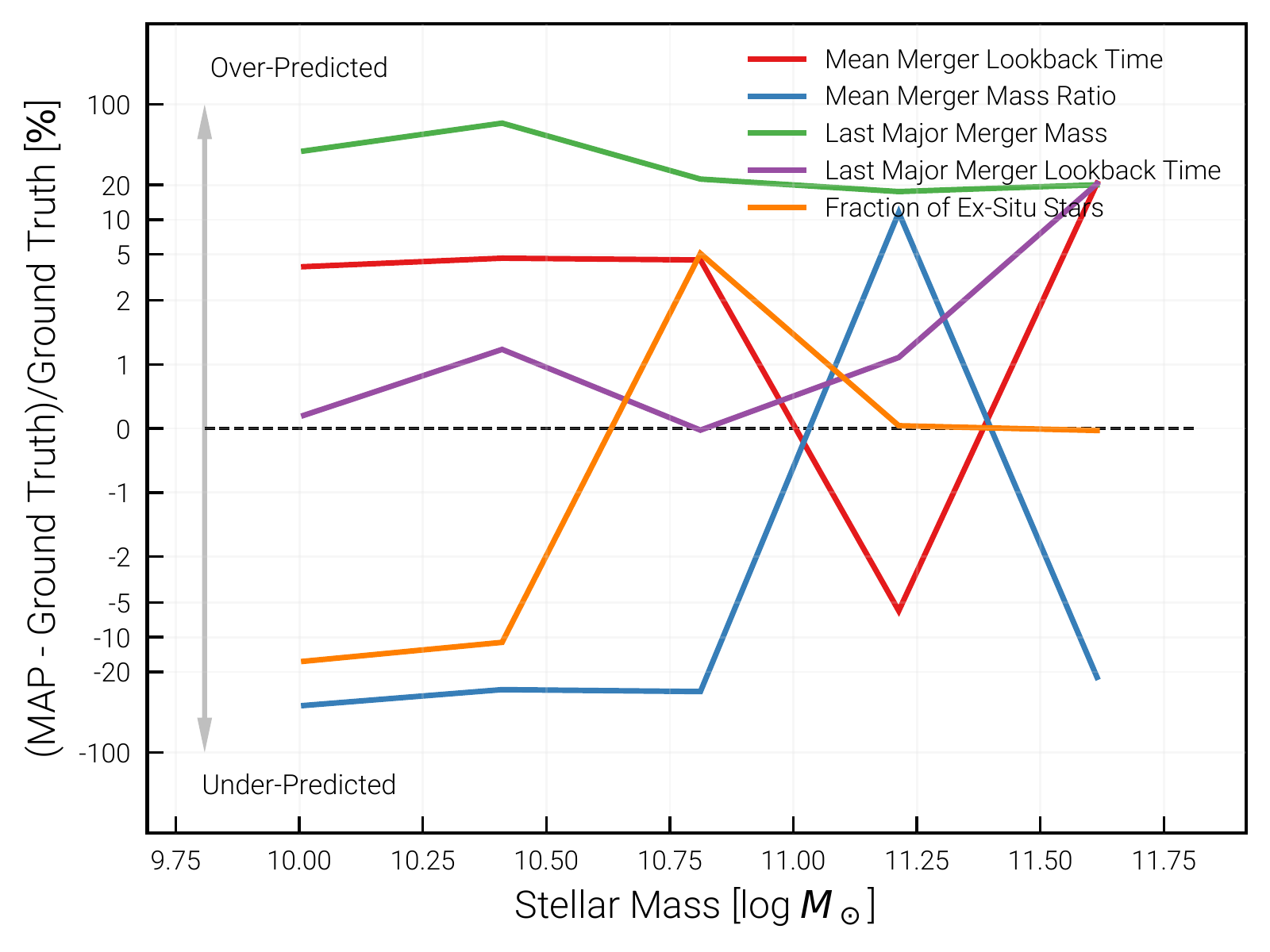}
		\includegraphics[width=0.49\linewidth]{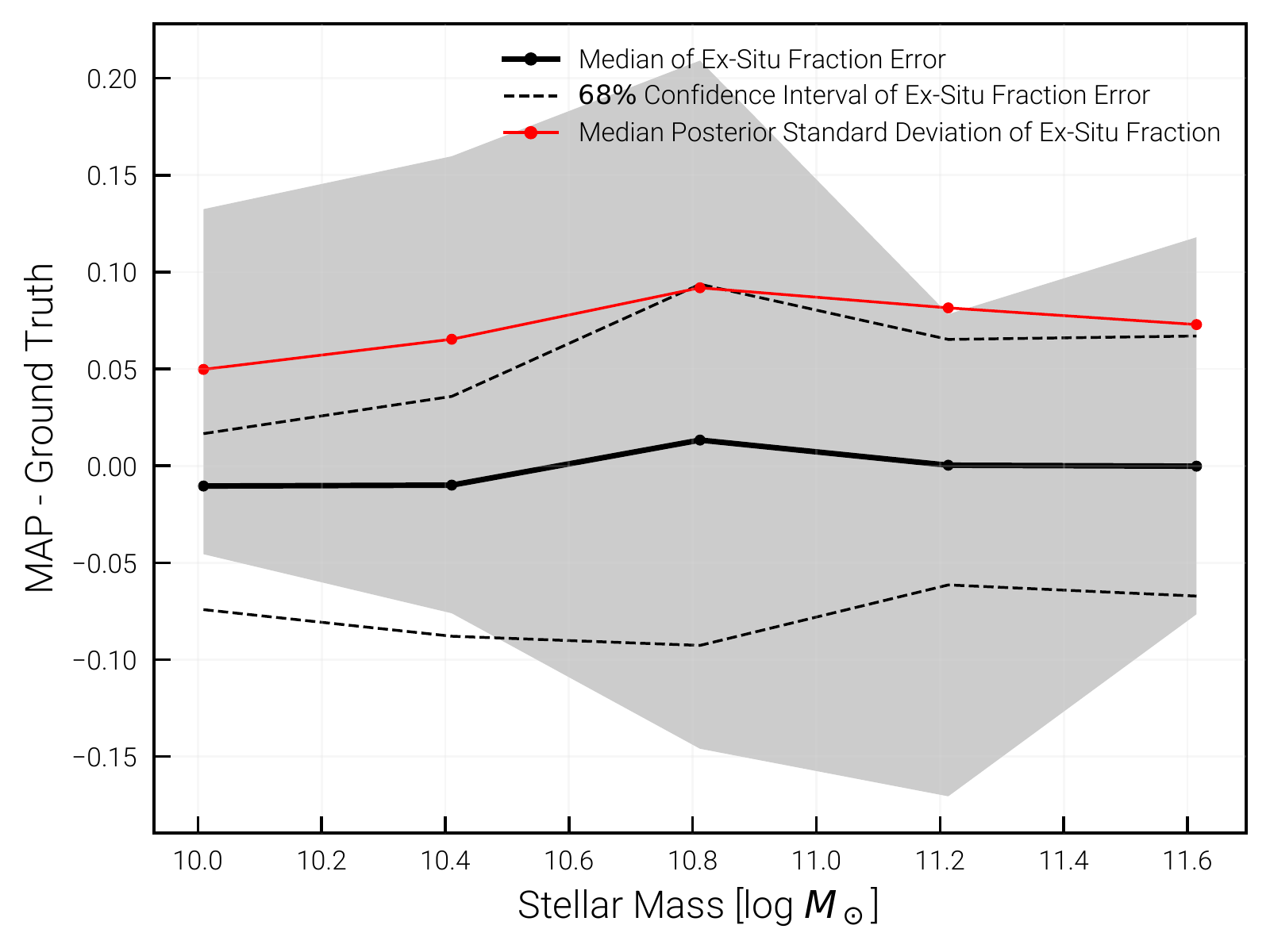}
	\caption{{\bf Prediction performance as a function of galaxy mass.} We investigate how the MAP error of our cINN model (MAP - Ground Truth) depends on galaxy stellar mass, by using all galaxies in our TNG100 test set in 5 uniformly-spaced mass bins. {\bf Left}: Median relative errors (i.e. normalized by the ground truth values on a galaxy-by-galaxy basis) for the five inferred unobservable statistics of the assembly and merger history of galaxies. For the time and mass of the last major merger, we neglect galaxies without a major merger in their history.
	{\bf Right}: Errors (MAP - Ground Truth) for the ex-situ fraction as a function of galaxy stellar mass. The median error is drawn as solid black line while the area containing $68 \%$ of the galaxies ($1 \sigma$) is included within dashed lines. For comparison, we also plot the median of the posterior standard deviations given by our cINN model as a solid red line. To quantify how our results improve upon the bare relationship between galaxy mass and ex-situ fraction that could be extracted directly from the TNG100 simulation, we plot as grey shaded area the $68$ per cent confidence interval of the prior distribution in the ex-situ vs. mass plane (i.e. the 16th-84th percentiles in bins of stellar mass of Fig.~\ref{fig:methods/unobservables}, rightmost panel, having subtracted the median value).}
	\label{fig:errors}
\end{figure*}

\subsection{For which galaxies can we get the best predictions on their past history?}
\label{sec:best_prediction}

We have already quantified the overall performance and predictive capability of the cINN model in Section~\ref{sec:performance}. However, we have not yet explicitly discussed whether these inferences are more promising for certain classes of galaxies than others. This is of relevance in view of the future application of this framework to actual galaxy observations, which may be magnitude (i.e. mass) limited or which may reach different galaxies at different redshifts. 

In terms of redshift dependence, Fig.~\ref{fig:mlpsensitivity} of Section~\ref{sec:sensitivity} shows that the redshift of a galaxy (its lookback time) is not a crucially-informative input for the accuracy of the predictions. This statement applies to galaxies up to $z=1$, as we have focused exclusively on the $z=0-1$ range. In turn, this implies that we expect to be able to predict similarly well the past assembly and merger history of galaxies in the low-redshift Universe, where data from e.g. SDSS, DES, and HSC-SSP abounds, as well as up to 8 billion years ago.

In terms of mass dependence, we have seen by comparing Figs.~\ref{fig:cinn_prior_lowmass} and \ref{fig:cinn_prior_highmass} that the cINN is capable of recovering the overall cross-dependencies among predicted features to similar degrees for the low and high-mass end of the studied sample. We build upon this and quantify the errors on the predictions as a function of galaxy stellar mass in Fig.~\ref{fig:errors}. As we could not find any significant variations with redshift (see above), here we show average trends for all galaxies at $z=0-1$ combined. We show the errors (MAP - Ground Truth) normalized by the ground-truth values for all inferred statistics in the left panel, and the non-normalized (MAP - Ground Truth) for the ex-situ fraction only on the right. The medians are given as solid curves in bins of galaxy stellar mass; the area containing $68$ per cent of the galaxies around the median is between dashed lines.

Firstly, Fig.~\ref{fig:errors} explicitly demonstrates that -- except for the mean merger mass ratio and for the mass of the last major merger for low-mass galaxies -- the cINN predicts the properties of the past assembly and merger histories of the average galaxies with a relative prediction error (compared to the ground truth) that is better than $10-20$ per cent throughout the $10^{10-12}\, \MSUN$ range. The accuracy of the prediction depends differently on galaxy stellar mass for different unobservable quantities, and e.g. weakly for the mass and time of the last major merger, which are typically overpredicted -- see for comparison the absolute values of the prediction errors in Fig.~\ref{fig:results/TNG100/map}.

The ex-situ fraction -- the unobservable that is best predicted by the cINN and MLP based on the analysis thus far -- in fact performs significantly worse for galaxies with stellar mass $\lesssim10^{10.7}\,\MSUN$ and essentially perfectly (relative prediction errors smaller than a few percent) for galaxies with stellar mass  $\gtrsim10^{11}\, \MSUN$. This is not in contradiction with what we have seen so far: in terms of non-normalized prediction errors (right panel of Fig.~\ref{fig:errors} and Fig.~\ref{fig:results/TNG100/map}), the cINN predicts on average the ex-situ fraction within $\pm$ 2.5 percentage points (solid black curve) independently of galaxy stellar mass. However, because of the strong correlation of ex-situ fraction with stellar mass, the median prediction error is much larger relatively to the ground truth for low-mass galaxies, where the ex-situ fraction is low and typically under predicted (see also Section~\ref{sec:cinn_vs_gt}, where we found the MAP to be smaller than the ground truth) . 

\begin{figure*}
	\centering
	\includegraphics[width=4.4cm]{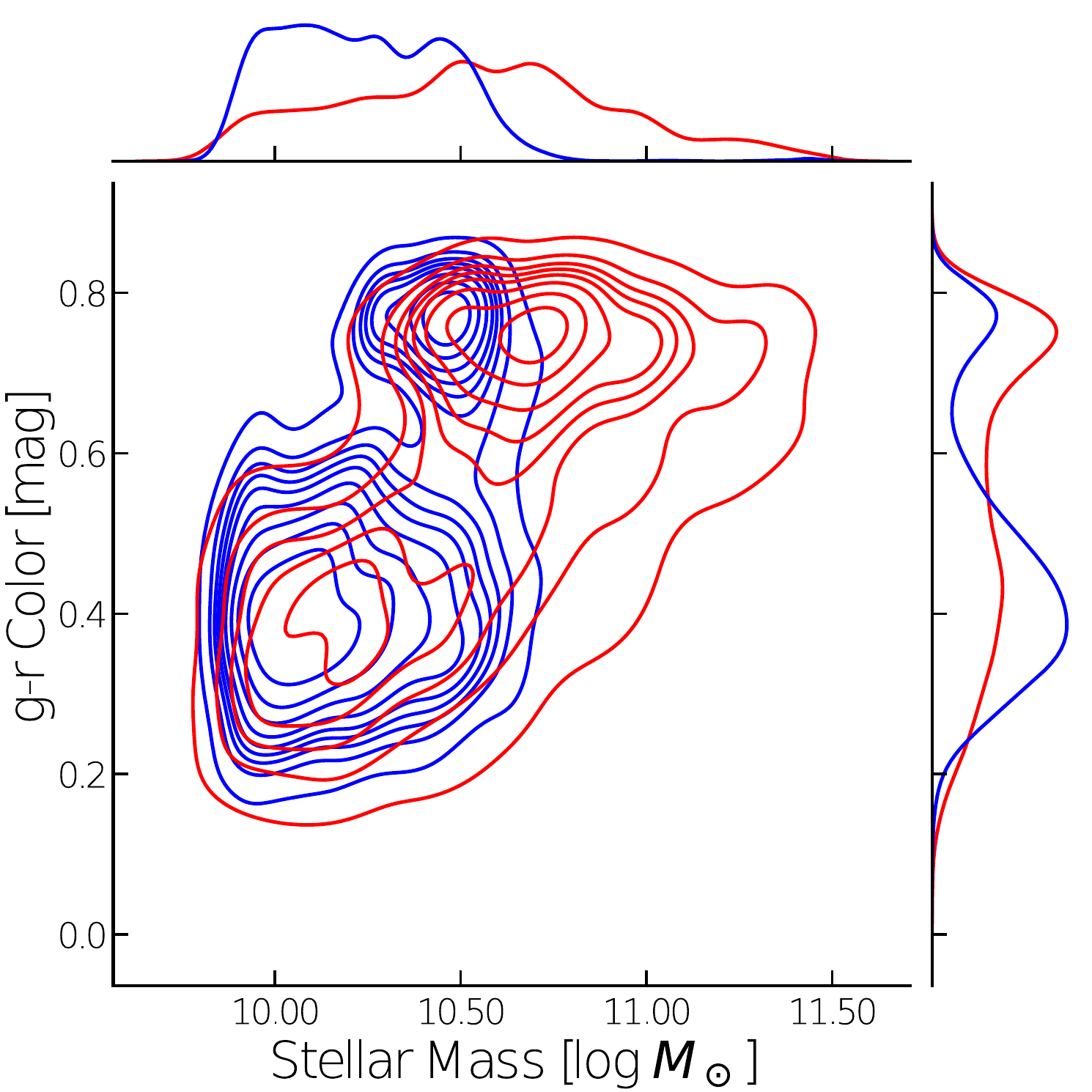}
	\includegraphics[width=4.4cm]{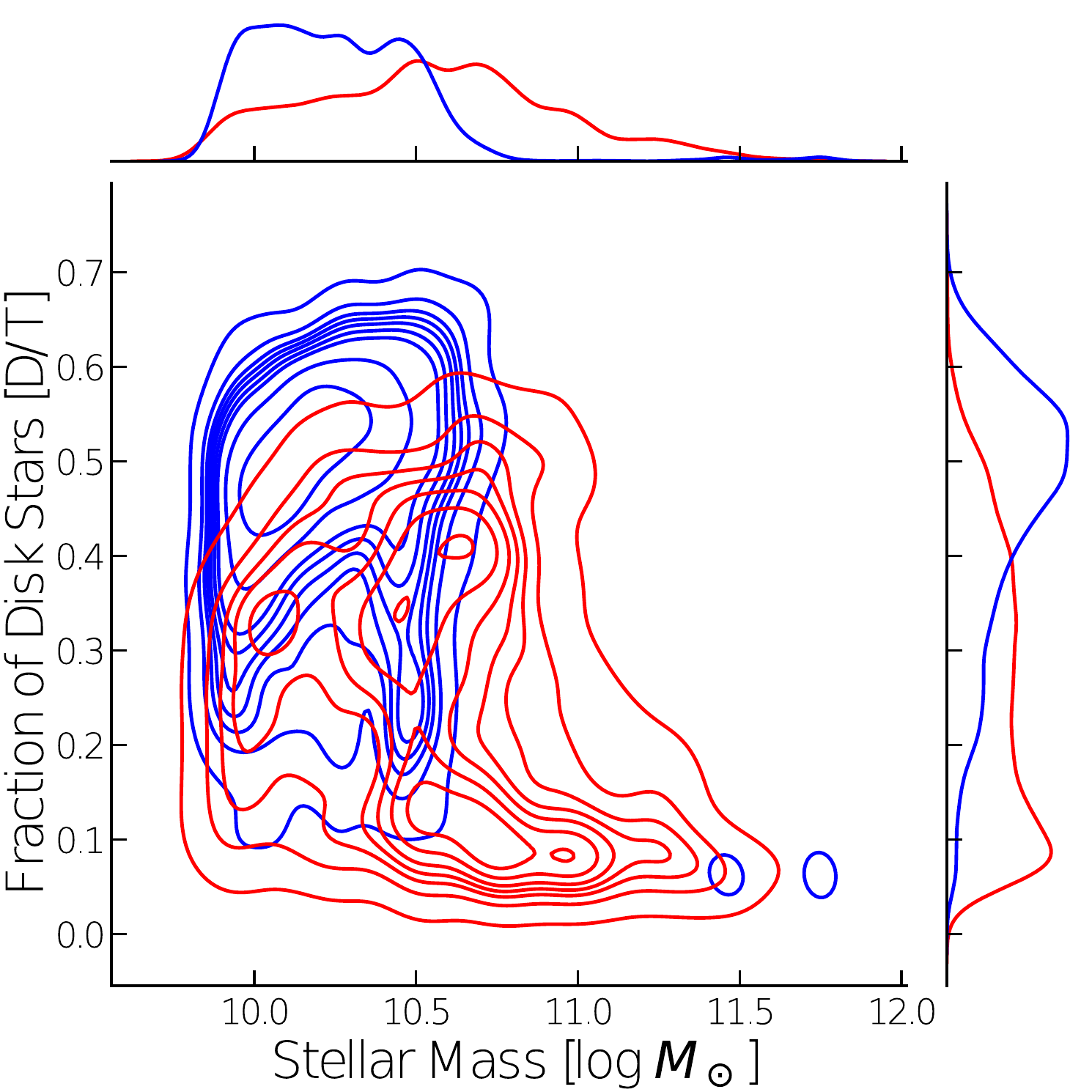}
	\includegraphics[width=4.4cm]{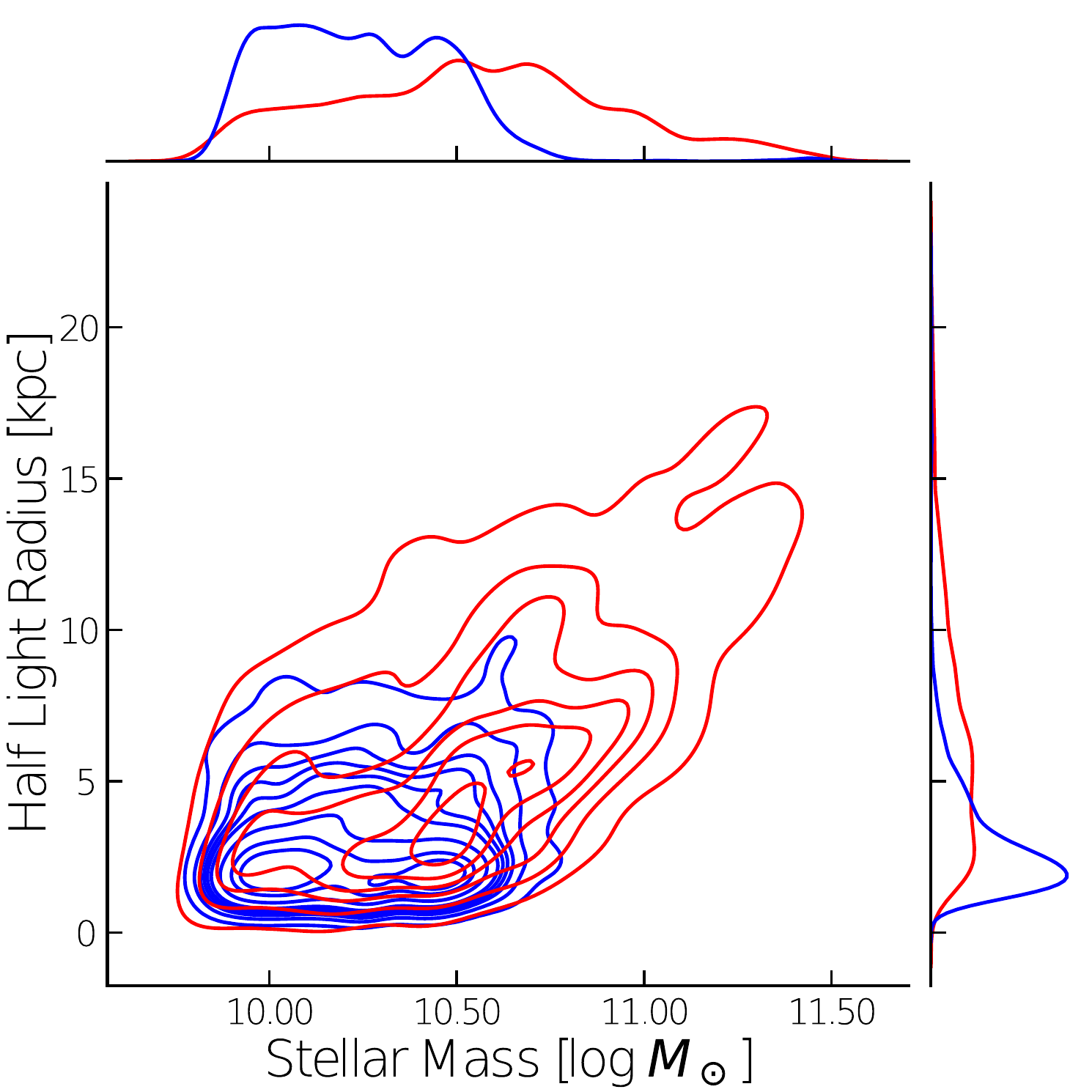}
	\includegraphics[width=4.4cm]{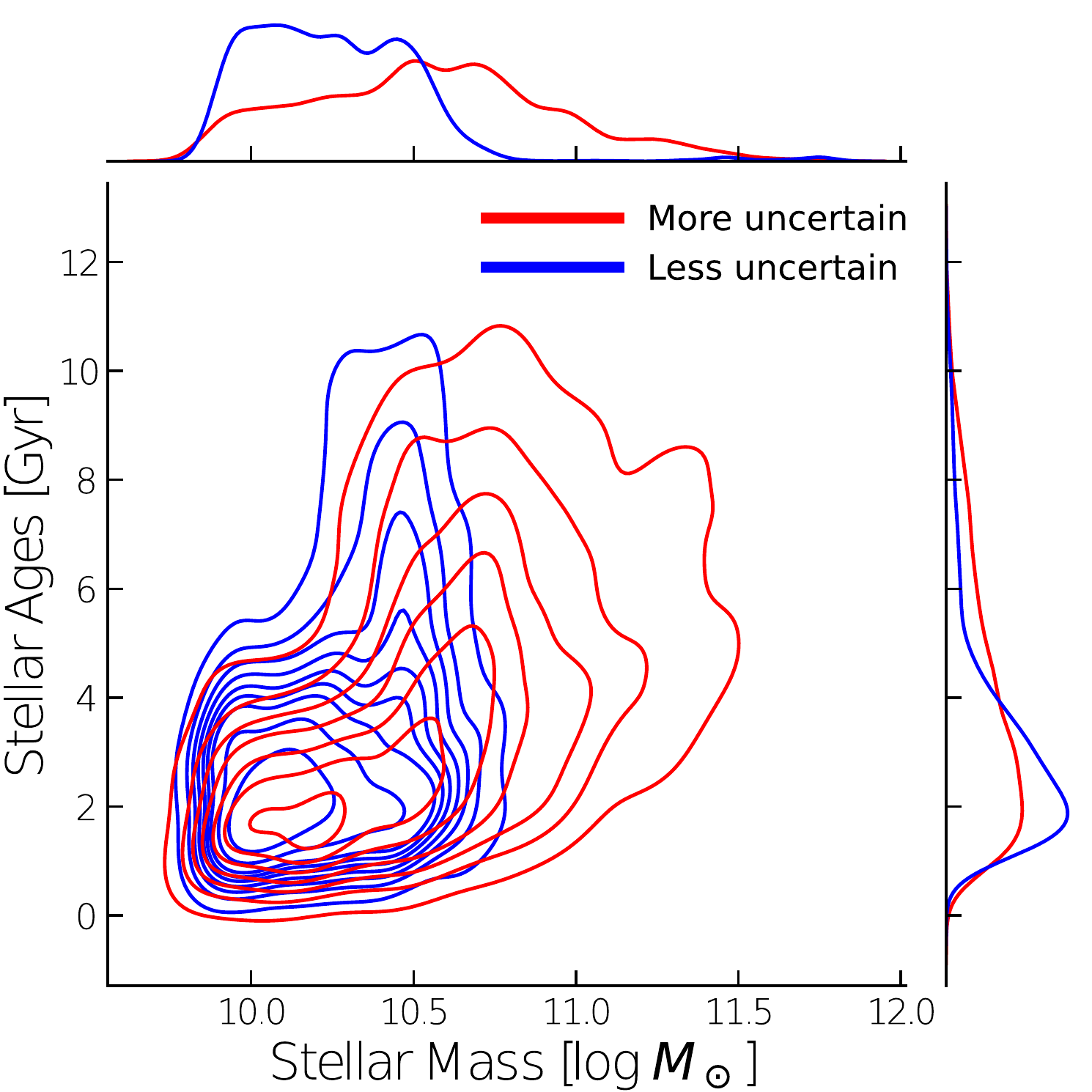}
	\caption{{\bf Which galaxies (in terms of observable properties) have the least ambiguous assembly and merger history?} We answer this question by electing the ex-situ fraction as example. We separate the TNG100 test set in two subsets (stacking all redshifts): galaxies with a predicted ex-situ cINN posterior standard deviation that is below (blue) and above (red) the median of the whole test sample. The blue subset hence represents the galaxies whose predictions are less uncertain than those in the red subset. We show how these two sets populate the space of stellar mass vs. (from left to right) g-r colour, fraction of disc stars, half-light radius and average stellar age. In the plot edges, we show the projections of the two distributions on the x and y axis. We see that, for the ex-situ fraction, the cINN returns especially small error bars in the regime of low-mass, blue, and disc-like galaxies.}
	\label{fig:results/TNG100/best_prediction}
\end{figure*}

Moreover, below $\lesssim 10^{10.7}\, \MSUN$, where typically in-situ star formation is the dominant stellar mass growth channel, the strong dependence between ex-situ fraction and galaxy mass is missing (see e.g. Fig.~\ref{fig:methods/unobservables}): although the cINN is still able to distinguish low- from high-ex-situ galaxies, in the case of low-mass galaxies the ambiguity hampers the ability to make an accurate point prediction and the MAP is influenced by the prior, which overall favours low values of ex-situ fractions. This is also expressed in the confidence interval on the right panel of Fig.~\ref{fig:errors} (dashed curves), which is clearly biased towards an under-prediction of the ex-situ parameter at lower masses. 

As there is a strong relationship between galaxy mass and ex-situ fraction visible in the rightmost panel of Fig.~\ref{fig:methods/unobservables}, we also investigate to what degree the cINN (using all 7 observables) performs better in comparison to what could be inferred from this bare connection. To to so, we also plot as grey shaded area the $68$ per cent confidence interval of the prior distribution in the ex-situ vs. mass plane (i.e. the 16th-84th percentiles in bins of stellar mass). From this we see that the cINN posterior MAP error is significantly smaller than the scatter in the prior data.

To see for which galaxies, distinguished based on their observable properties, we get predictions with a low uncertainty and therefore low ambiguity, we move from quantifying the prediction error (i.e. the accuracy) to assessing the width or standard deviations of the cINN posterior distributions (i.e. the precision). We do this again focusing on the ex-situ fraction and plot in Fig.~\ref{fig:results/TNG100/best_prediction} how two subsets of the TNG100 test galaxies populate the observable space: i) the ones whose cINN posterior standard deviation for the ex-situ fraction is above the median of the test sample and whose predictions are hence affected by larger uncertainties (red) and ii) those with below-median standard deviation and whose prediction is hence less uncertain (blue). 

It is manifest that the errorbars on the ex-situ fraction predictions tend to be indeed smaller for low-mass, blue, disc-like, and compact galaxies, which corresponds to the low ex-situ regime (see also Fig.~\ref{fig:results/TNG100/sigma} and our discussion in Section \ref{sec:uncertainties}). Namely, it is easier for the cINN to identify the galaxies in the sample with a rather quiet merger history whereas the ambiguity rises with the level of past interactions.  

These results would be qualitatively similar, albeit not identical, if we were to replace the posterior standard deviation (i.e. the prediction uncertainty) by the prediction error (MAP - Ground Truth). This aligns well with our findings in Fig.\ref{fig:results/TNG100/sigma} where we have shown that the standard deviation of the posterior is small for low ex-situ galaxies and that the standard deviation correlates well with the prediction error.

Finally, the distributions of Fig.~\ref{fig:results/TNG100/best_prediction}, which focus on the ex-situ fraction, are qualitatively similar for the case of the mass of the last major merger: less ambiguous and uncertain predictions occur for lower-mass, bluer, diskier, more compact and younger galaxies. For the mean merger lookback time and the lookback time of the last major merger, on the other hand, we find a strong dependence on the redshift of the galaxy under scrutiny: for galaxies at earlier redshifts, the physical available timespan for merging is naturally shorter than for galaxies observed close to $z=0$ and therefore the prediction errorbars are much smaller.

\subsection{Towards the application to real observations}
\label{sec:results/low_redshift}

Having demonstrated with the TNG100 simulation the feasibility of inferring unobservable properties of galaxies' past history from a handful of observable galaxy features, we plan in future work to apply these or similar types of ML models to real observational data. This will require quantifying and marginalising over the systematic uncertainties due to adopting different cosmological galaxy simulations for the predictions, such Illustris, EAGLE, and others.

With the scalar-to-scalar models developed in this work, we are especially aiming for large-scale photometric galaxy surveys, such as the aforementioned SDSS, GAMA, or HSC-SSP, which can provide multi-band images and integral properties of large samples of galaxies. In fact, as discussed in Section~\ref{sec:best_prediction}, the framework can be used not only at low redshift, but similarly successfully also up to $z\sim1$. 

We have already designed our input observable parameters (stellar mass, g-r color, half-light radius, r-band weighted average age, etc) in such a way that they are (in principle) derivable from existing  galaxy datasets, either via conventional modelling based on the images or via  ML approaches as done by e.g. \cite{Pasquet_2018}. The only observable that we have used and whose operational definition is in fact far from what is possible with photometric surveys is the disk-to-total ratio, which proves to have important predictive power, especially for the stellar ex-situ fraction. Here we have obtained this ``morphological estimator'' from the kinematics of the stellar particles, whereas in reality either IFU data is needed or Sersic-index based morphological estimators shall be used. 

We also plan to completely bypass the issue of deriving and choosing summary statistics of galaxies by directly using the full images, possibly in multiple bands. Within our framework, this will require introducing an additional encoder head (e.g. a convolutional neural network) that translates the maps into an intermediate lower-dimensional representation space and which hence can then be used as conditional input to the cINN. In turn, this will allow us to exploit the full information content of the images, including those reaching low surface brightness levels and hence revealing stellar halo features (see e.g. deep surveys with the Dragonfly Array, with HSC, and LSST). Even more so, realistic mock images of the simulated galaxies will be of the essence, including the effects induced by the survey-specific resolution, point-spread function, redshift-effects, and artifacts. Albeit demanding, such stellar light mocks from simulated galaxies are now achievable \citep[e.g.][]{Snyder_2015, Bottrell_2017, Trayford_2017, Rodriguez-Gomez_2019, Merritt_2020}, and have already been used for diverse ML-based analyses \citep[e.g.][]{Huertas_Company_2019, Bottrell_2019, Zanisi_2020}.

%%%%%%%%%%%%%%%%% SUMMARY %%%%%%%%%%%%%%%%%%

\section{Summary and Conclusions}
\label{sec:conclusions}

In this work, we have used the cosmological magnetohydrodynamical galaxy formation simulation TNG100 of the IllustrisTNG project to investigate the connection between integral properties of galaxies at one point in time, on the one side, and their past stellar assembly and merger history on the other. In particular, we have trained, tested, and compared a Multilayer Perceptron (MLP) and a conditional Invertible Neural Network (cINN) to determine to what degree summary statistics of the unobservable past of galaxies can be predicted from a handful of galaxy features that are commonly available in large photometric surveys. 

We have focused on TNG100 central galaxies in the $z=0-1$ range with stellar masses of $10^{10-12}\, \MSUN$, for a total of 182'625 galaxies. We have made sure that the progenitors and descendants of each test and validation galaxy are not contained in the training set, to reduce the additional information that the tree-like structure of the simulation data may provide.

As observable inputs, we have chosen seven integral galaxy properties, namely stellar mass, redshift, stellar half-light radius, fraction of disc stars,  integrated dust-attenuated galaxy color (SDSS g-r color index), luminosity-weighted stellar average metallicity and luminosity-weighted average stellar age (Fig.~\ref{fig:methods/observables}).
To quantify the stellar assembly and merger history of galaxies, we have elected five summary statistics, namely the stellar ex-situ fraction (i.e. the fraction of accreted stars via mergers and satellite stripping), the mass-weighted average merger mass ratio and the mass-weighted average merger lookback time throughout a galaxy's history, and the stellar mass and time of the last major merger (if any; Fig.~\ref{fig:methods/unobservables}).

Thanks to the cINN, we have been able to evaluate not only the accuracy of the predictions (Figs.~\ref{fig:results/TNG100/posterior_example}, \ref{fig:results/TNG100/map}, and \ref{fig:errors}) but also to estimate their uncertainty, via the width (standard deviation) of the posterior probability distributions (Fig.~\ref{fig:results/TNG100/sigma}). We have dealt with galaxies that never had a major merger in their past by de facto including in the modeling a classification scheme. And we have interpreted our findings by exploring the relationships between inputs and outputs via a UMAP representation of the simulated galaxies (Fig.~\ref{fig:UMAP}).

We have shown that it is in principle possible to infer important information about the past stellar assembly and merger history of galaxies by using solely scalar galaxy properties as inputs. In particular, we have shown that -- except for the mean merger mass ratio -- it is possible to reconstruct the studied unobservable properties well, i.e. with a median prediction error relative to the ground truth that is better than $10-20$ per cent throughout the considered mass and redshift range (Fig.~\ref{fig:errors}; for the mass of the last major merger this statement applies only for galaxies with $\gtrsim 10^{10.7}\,\MSUN$ ). The fraction of ex-situ stars is recovered particularly successfully throughout the galaxy population: the cINN posterior maxima (Maximum A-Posteriori estimates or MAPs) align well around the ground truth values (Fig.~\ref{fig:results/TNG100/map}); furthermore, the standard deviation of the ex-situ posterior behaves like a normal-distributed error (Fig.~\ref{fig:results/TNG100/sigma}). In contrast, we have found that, given the adopted set of observable inputs, it is not possible to predict the mean merger mass ratio (Fig.~\ref{fig:results/TNG100/map}): the cINN posteriors are often close to the prior distributions of this parameter (Figs.~\ref{fig:results/TNG100/posterior_example}, \ref{fig:cinn_prior_lowmass} and \ref{fig:cinn_prior_highmass}). 
Similarly as for the ex-situ fraction, the predicted average merger lookback time aligns well with the ground truth for the bulk of the galaxies. On the other hand, for the mass and lookback time of the last major merger, some caution is needed due to the shape of the posteriors, which are often complex or exhibit strong bimodalities -- this being especially driven by the ambiguity with galaxies that never had any major mergers in their history. Yet, for those galaxies that the cINN correctly identifies as having experienced a major merger, the predicted time and stellar mass of the merger are within $\pm1$ billion years and $\pm0.1$dex from the true values for the average galaxy (but only if the merger is more massive than about $10^{9.5}\,\MSUN$ in stars; Fig.~\ref{fig:results/TNG100/map}). 

Importantly, the cINN is also able to learn the cross-correlations among the chosen target quantities (Figs.~\ref{fig:cinn_prior_lowmass} and \ref{fig:cinn_prior_highmass}), with e.g. larger ex-situ fractions in galaxies with more massive and more recent last major mergers, also at fixed galaxy mass. Moreover, when comparing the cINN MAPs with the MLP point predictions (Fig.\ref{fig:results/MLP_cINN_comparison}), we find that, in terms of prediction performance (i.e. distance from the ground truth), both ML algorithms yield equivalent results.

Thanks to the MLP, we can easily uncover which of the adopted galaxy properties are most informative for the prediction of the various target quantities (Fig.~\ref{fig:mlpsensitivity}). We find that well-known correlations are learned by the networks, e.g. that the stellar ex situ fraction is larger for larger galaxy masses and hence stellar mass is a fundamental input. However, the networks also learn more subtle and residual dependencies, so that the stellar size and morphology of galaxies are also important predictors of the ex-situ fraction. The luminosity-weighted stellar age plays in general an important role for the inference of all merger statistics, whereas the luminosity-weighted average stellar metallicity is not essential. 

Finally, our framework is able to predict similarly well the past assembly and merger history of galaxies both in the low-redshift Universe as well as up to $z\sim1$. And so our tools have the potential to open up a powerful new avenue to quantify the stellar assembly and merger history of thousands of {\it observed} galaxies across cosmic epochs. 

\section*{Acknowledgements}

LE and AP thank Victor Ksoll for useful inputs. The authors thank Ullrich Koethe, Lynton Ardizzone and their team members for suggesting using cINNs and for making their implementation publicly available. LE and this work are supported by the Deutsche Forschungsgemeinschaft (DFG, German Research Foundation) under Germany’s Excellence Strategy EXC 2181/1-390900948, Exploratory project EP 3.4 (the Heidelberg STRUCTURES Excellence Cluster). DN acknowledges funding from the Deutsche Forschungsgemeinschaft (DFG) through an Emmy Noether Research Group (grant number NE 2441/1-1). The primary TNG simulations were carried out with compute time granted by the Gauss Centre for Supercomputing (GCS) under Large-Scale Projects GCS-ILLU and GCS-DWAR on the GCS share of the supercomputer Hazel Hen at the High Performance Computing Center Stuttgart (HLRS). Part of this research was carried out using the High Performance Computing resources at the Max Planck Computing and Data Facility (MPCDF) in Garching, operated by the Max Planck Society.

\section*{Data Availability}

Data directly related to this publication and its figures will be made available on request from the corresponding author. In fact, the IllustrisTNG simulations are already publicly available and accessible in their entirety at \href{www.tng-project.org/data}{www.tng-project.org/data} \citep{nelson2019illustristng}. Upon acceptance, we plan to release the associated code and ML models developed here on GitHub.

%%%%%%%%%%%%%%%%%%%% REFERENCES %%%%%%%%%%%%%%%%%%

\bibliographystyle{mnras}
\bibliography{references}

\begin{thebibliography}{}
\makeatletter
\relax
\def\mn@urlcharsother{\let\do\@makeother \do\$\do\&\do\#\do\^\do\_\do\%\do\~}
\def\mn@doi{\begingroup\mn@urlcharsother \@ifnextchar [ {\mn@doi@}
  {\mn@doi@[]}}
\def\mn@doi@[#1]#2{\def\@tempa{#1}\ifx\@tempa\@empty \href
  {http://dx.doi.org/#2} {doi:#2}\else \href {http://dx.doi.org/#2} {#1}\fi
  \endgroup}
\def\mn@eprint#1#2{\mn@eprint@#1:#2::\@nil}
\def\mn@eprint@arXiv#1{\href {http://arxiv.org/abs/#1} {{\tt arXiv:#1}}}
\def\mn@eprint@dblp#1{\href {http://dblp.uni-trier.de/rec/bibtex/#1.xml}
  {dblp:#1}}
\def\mn@eprint@#1:#2:#3:#4\@nil{\def\@tempa {#1}\def\@tempb {#2}\def\@tempc
  {#3}\ifx \@tempc \@empty \let \@tempc \@tempb \let \@tempb \@tempa \fi \ifx
  \@tempb \@empty \def\@tempb {arXiv}\fi \@ifundefined
  {mn@eprint@\@tempb}{\@tempb:\@tempc}{\expandafter \expandafter \csname
  mn@eprint@\@tempb\endcsname \expandafter{\@tempc}}}

\bibitem[\protect\citeauthoryear{{Ardizzone} et~al.,}{{Ardizzone}
  et~al.}{2018}]{2018arXiv180804730A}
{Ardizzone} L.,  et~al., 2018, arXiv e-prints, \href
  {https://ui.adsabs.harvard.edu/abs/2018arXiv180804730A} {p. arXiv:1808.04730}

\bibitem[\protect\citeauthoryear{{Ardizzone}, {L{\"u}th}, {Kruse}, {Rother}  \&
  {K{\"o}the}}{{Ardizzone} et~al.}{2019a}]{2019arXiv190702392A}
{Ardizzone} L.,  {L{\"u}th} C.,  {Kruse} J.,  {Rother} C.,   {K{\"o}the} U.,
  2019a, arXiv e-prints, \href
  {https://ui.adsabs.harvard.edu/abs/2019arXiv190702392A} {p. arXiv:1907.02392}

\bibitem[\protect\citeauthoryear{{Ardizzone}, {L{\"u}th}, {Kruse}, {Rother}  \&
  {K{\"o}the}}{{Ardizzone} et~al.}{2019b}]{FrEIA}
{Ardizzone} L.,  {L{\"u}th} C.,  {Kruse} J.,  {Rother} C.,   {K{\"o}the} U.,
  2019b, arXiv e-prints, \href
  {https://ui.adsabs.harvard.edu/abs/2019arXiv190702392A} {p. arXiv:1907.02392}

\bibitem[\protect\citeauthoryear{{Belokurov}, {Erkal}, {Evans}, {Koposov}  \&
  {Deason}}{{Belokurov} et~al.}{2018}]{Belokurov2018}
{Belokurov} V.,  {Erkal} D.,  {Evans} N.~W.,  {Koposov} S.~E.,   {Deason}
  A.~J.,  2018, \mn@doi [\mnras] {10.1093/mnras/sty982}, \href
  {https://ui.adsabs.harvard.edu/abs/2018MNRAS.478..611B} {478, 611}

\bibitem[\protect\citeauthoryear{{Blanton} \& {Moustakas}}{{Blanton} \&
  {Moustakas}}{2009}]{Blanton&Moustakas_2009}
{Blanton} M.~R.,  {Moustakas} J.,  2009, \mn@doi [\araa]
  {10.1146/annurev-astro-082708-101734}, \href
  {https://ui.adsabs.harvard.edu/abs/2009ARA&A..47..159B} {47, 159}

\bibitem[\protect\citeauthoryear{{Bonaca} et~al.,}{{Bonaca}
  et~al.}{2020}]{Bonaca2020}
{Bonaca} A.,  et~al., 2020, \mn@doi [\apjl] {10.3847/2041-8213/ab9caa}, \href
  {https://ui.adsabs.harvard.edu/abs/2020ApJ...897L..18B} {897, L18}

\bibitem[\protect\citeauthoryear{{Bottrell}, {Torrey}, {Simard}  \&
  {Ellison}}{{Bottrell} et~al.}{2017}]{Bottrell_2017}
{Bottrell} C.,  {Torrey} P.,  {Simard} L.,   {Ellison} S.~L.,  2017, \mn@doi
  [\mnras] {10.1093/mnras/stx276}, \href
  {https://ui.adsabs.harvard.edu/abs/2017MNRAS.467.2879B} {467, 2879}

\bibitem[\protect\citeauthoryear{{Bottrell} et~al.,}{{Bottrell}
  et~al.}{2019}]{Bottrell_2019}
{Bottrell} C.,  et~al., 2019, \mn@doi [\mnras] {10.1093/mnras/stz2934}, \href
  {https://ui.adsabs.harvard.edu/abs/2019MNRAS.490.5390B} {490, 5390}

\bibitem[\protect\citeauthoryear{{Bottrell}, {Hani}, {Teimoorinia}, {Patton}
  \& {Ellison}}{{Bottrell} et~al.}{2021}]{Bottrell_2020}
{Bottrell} C.,  {Hani} M.~H.,  {Teimoorinia} H.,  {Patton} D.~R.,   {Ellison}
  S.~L.,  2021, \mn@doi [\mnras] {10.1093/mnras/stab3717}, \href
  {https://ui.adsabs.harvard.edu/abs/2021MNRAS.tmp.3422B} {}

\bibitem[\protect\citeauthoryear{{Boylan-Kolchin}, {Springel}, {White},
  {Jenkins}  \& {Lemson}}{{Boylan-Kolchin} et~al.}{2009}]{BoylanKolchin_2009}
{Boylan-Kolchin} M.,  {Springel} V.,  {White} S. D.~M.,  {Jenkins} A.,
  {Lemson} G.,  2009, \mn@doi [\mnras] {10.1111/j.1365-2966.2009.15191.x},
  \href {https://ui.adsabs.harvard.edu/abs/2009MNRAS.398.1150B} {398, 1150}

\bibitem[\protect\citeauthoryear{Bundy et~al.,}{Bundy
  et~al.}{2014}]{Bundy_2014}
Bundy K.,  et~al., 2014, \mn@doi [The Astrophysical Journal]
  {10.1088/0004-637x/798/1/7}, 798, 7

\bibitem[\protect\citeauthoryear{Chollet et~al.}{Chollet
  et~al.}{2015}]{chollet2015keras}
Chollet F.,  et~al., 2015, Keras, \url{https://keras.io}

\bibitem[\protect\citeauthoryear{{{\'C}iprijanovi{\'c}}
  et~al.,}{{{\'C}iprijanovi{\'c}} et~al.}{2021}]{Ciprianovic_2021}
{{\'C}iprijanovi{\'c}} A.,  et~al., 2021, \mn@doi [\mnras]
  {10.1093/mnras/stab1677}, \href
  {https://ui.adsabs.harvard.edu/abs/2021MNRAS.506..677C} {506, 677}

\bibitem[\protect\citeauthoryear{Cook, Conroy, Pillepich, Rodriguez-Gomez  \&
  Hernquist}{Cook et~al.}{2016}]{Cook_2016}
Cook B.~A.,  Conroy C.,  Pillepich A.,  Rodriguez-Gomez V.,   Hernquist L.,
  2016, \mn@doi [The Astrophysical Journal] {10.3847/1538-4357/833/2/158}, 833,
  158

\bibitem[\protect\citeauthoryear{{Davison}, {Norris}, {Pfeffer}, {Davies}  \&
  {Crain}}{{Davison} et~al.}{2020}]{Davison_2020}
{Davison} T.~A.,  {Norris} M.~A.,  {Pfeffer} J.~L.,  {Davies} J.~J.,   {Crain}
  R.~A.,  2020, \mn@doi [\mnras] {10.1093/mnras/staa1816}, \href
  {https://ui.adsabs.harvard.edu/abs/2020MNRAS.497...81D} {497, 81}

\bibitem[\protect\citeauthoryear{{De Silva} et~al.,}{{De Silva}
  et~al.}{2015}]{DeSilva2015}
{De Silva} G.~M.,  et~al., 2015, \mn@doi [\mnras] {10.1093/mnras/stv327}, \href
  {https://ui.adsabs.harvard.edu/abs/2015MNRAS.449.2604D} {449, 2604}

\bibitem[\protect\citeauthoryear{{Deng} et~al.,}{{Deng}
  et~al.}{2012}]{Deng2012}
{Deng} L.-C.,  et~al., 2012, \mn@doi [Research in Astronomy and Astrophysics]
  {10.1088/1674-4527/12/7/003}, \href
  {https://ui.adsabs.harvard.edu/abs/2012RAA....12..735D} {12, 735}

\bibitem[\protect\citeauthoryear{{Dinh}, {Sohl-Dickstein}  \& {Bengio}}{{Dinh}
  et~al.}{2016}]{2016arXiv160508803D}
{Dinh} L.,  {Sohl-Dickstein} J.,   {Bengio} S.,  2016, arXiv e-prints, \href
  {https://ui.adsabs.harvard.edu/abs/2016arXiv160508803D} {p. arXiv:1605.08803}

\bibitem[\protect\citeauthoryear{Dolag, Borgani, Murante  \& Springel}{Dolag
  et~al.}{2009}]{doi:10.1111/j.1365-2966.2009.15034.x}
Dolag K.,  Borgani S.,  Murante G.,   Springel V.,  2009, \mn@doi [MNRAS]
  {10.1111/j.1365-2966.2009.15034.x}, 399, 497

\bibitem[\protect\citeauthoryear{{Ellison}, {Mendel}, {Patton}  \&
  {Scudder}}{{Ellison} et~al.}{2013}]{Ellison_2013}
{Ellison} S.~L.,  {Mendel} J.~T.,  {Patton} D.~R.,   {Scudder} J.~M.,  2013,
  \mn@doi [\mnras] {10.1093/mnras/stt1562}, \href
  {https://ui.adsabs.harvard.edu/abs/2013MNRAS.435.3627E} {435, 3627}

\bibitem[\protect\citeauthoryear{{Fakhouri} \& {Ma}}{{Fakhouri} \&
  {Ma}}{2008}]{Fakhouri&Ma_2008}
{Fakhouri} O.,  {Ma} C.-P.,  2008, \mn@doi [\mnras]
  {10.1111/j.1365-2966.2008.13075.x}, \href
  {https://ui.adsabs.harvard.edu/abs/2008MNRAS.386..577F} {386, 577}

\bibitem[\protect\citeauthoryear{Ferreira, Conselice, Duncan, Cheng, Griffiths
  \& Whitney}{Ferreira et~al.}{2020}]{Ferreiras_2020}
Ferreira L.,  Conselice C.~J.,  Duncan K.,  Cheng T.-Y.,  Griffiths A.,
  Whitney A.,  2020, \mn@doi [The Astrophysical Journal]
  {10.3847/1538-4357/ab8f9b}, 895, 115

\bibitem[\protect\citeauthoryear{{Gaia Collaboration} et~al.,}{{Gaia
  Collaboration} et~al.}{2018}]{GaiaDR22018}
{Gaia Collaboration} et~al., 2018, \mn@doi [\aap]
  {10.1051/0004-6361/201833051}, \href
  {https://ui.adsabs.harvard.edu/abs/2018A&A...616A...1G} {616, A1}

\bibitem[\protect\citeauthoryear{{Genel}, {Genzel}, {Bouch{\'e}}, {Naab}  \&
  {Sternberg}}{{Genel} et~al.}{2009}]{Genel_2009}
{Genel} S.,  {Genzel} R.,  {Bouch{\'e}} N.,  {Naab} T.,   {Sternberg} A.,
  2009, \mn@doi [\apj] {10.1088/0004-637X/701/2/2002}, \href
  {https://ui.adsabs.harvard.edu/abs/2009ApJ...701.2002G} {701, 2002}

\bibitem[\protect\citeauthoryear{{Genel}, {Fall}, {Hernquist}, {Vogelsberger},
  {Snyder}, {Rodriguez-Gomez}, {Sijacki}  \& {Springel}}{{Genel}
  et~al.}{2015}]{2015ApJ...804L..40G}
{Genel} S.,  {Fall} S.~M.,  {Hernquist} L.,  {Vogelsberger} M.,  {Snyder}
  G.~F.,  {Rodriguez-Gomez} V.,  {Sijacki} D.,   {Springel} V.,  2015, \mn@doi
  [The Astrophysical Journal Letters] {10.1088/2041-8205/804/2/L40}, \href
  {https://ui.adsabs.harvard.edu/abs/2015ApJ...804L..40G} {804, L40}

\bibitem[\protect\citeauthoryear{HST}{HST}{2008}]{HST_Merger}
HST 2008, A Hubble Atlas of Interacting Galaxies, \url
  {https://sci.esa.int/web/hubble/-/42690-a-hubble-atlas-of-interacting-galaxies}

\bibitem[\protect\citeauthoryear{{Helmi}, {Babusiaux}, {Koppelman}, {Massari},
  {Veljanoski}  \& {Brown}}{{Helmi} et~al.}{2018}]{Helmi2018}
{Helmi} A.,  {Babusiaux} C.,  {Koppelman} H.~H.,  {Massari} D.,  {Veljanoski}
  J.,   {Brown} A. G.~A.,  2018, \mn@doi [\nat] {10.1038/s41586-018-0625-x},
  \href {https://ui.adsabs.harvard.edu/abs/2018Natur.563...85H} {563, 85}

\bibitem[\protect\citeauthoryear{Huertas-Company et~al.,}{Huertas-Company
  et~al.}{2019}]{Huertas_Company_2019}
Huertas-Company M.,  et~al., 2019, \mn@doi [MNRAS] {10.1093/mnras/stz2191},
  489, 1859–1879

\bibitem[\protect\citeauthoryear{Ioffe \& Szegedy}{Ioffe \&
  Szegedy}{2015}]{DBLP:journals/corr/IoffeS15}
Ioffe S.,  Szegedy C.,  2015, Computing Research Repository, abs/1502.03167

\bibitem[\protect\citeauthoryear{{Ishiyama} et~al.,}{{Ishiyama}
  et~al.}{2021}]{Ishiyama_2021}
{Ishiyama} T.,  et~al., 2021, \mn@doi [\mnras] {10.1093/mnras/stab1755}, \href
  {https://ui.adsabs.harvard.edu/abs/2021MNRAS.506.4210I} {506, 4210}

\bibitem[\protect\citeauthoryear{Kingma \& Ba}{Kingma \&
  Ba}{2017}]{kingma2017adam}
Kingma D.~P.,  Ba J.,  2017, Adam: A Method for Stochastic Optimization
  (\mn@eprint {arXiv} {1412.6980})

\bibitem[\protect\citeauthoryear{{Kingma} \& {Dhariwal}}{{Kingma} \&
  {Dhariwal}}{2018}]{2018arXiv180703039K}
{Kingma} D.~P.,  {Dhariwal} P.,  2018, arXiv e-prints, \href
  {https://ui.adsabs.harvard.edu/abs/2018arXiv180703039K} {p. arXiv:1807.03039}

\bibitem[\protect\citeauthoryear{Ksoll et~al.,}{Ksoll
  et~al.}{2020}]{10.1093/mnras/staa2931}
Ksoll V.~F.,  et~al., 2020, \mn@doi [MNRAS] {10.1093/mnras/staa2931}, 499, 5447

\bibitem[\protect\citeauthoryear{{Lacey} \& {Cole}}{{Lacey} \&
  {Cole}}{1993}]{Lacey&Cole_1997}
{Lacey} C.,  {Cole} S.,  1993, \mn@doi [\mnras] {10.1093/mnras/262.3.627},
  \href {https://ui.adsabs.harvard.edu/abs/1993MNRAS.262..627L} {262, 627}

\bibitem[\protect\citeauthoryear{{Majewski} et~al.,}{{Majewski}
  et~al.}{2017}]{Majewski2017}
{Majewski} S.~R.,  et~al., 2017, \mn@doi [\aj] {10.3847/1538-3881/aa784d},
  \href {https://ui.adsabs.harvard.edu/abs/2017AJ....154...94M} {154, 94}

\bibitem[\protect\citeauthoryear{Marinacci, Pakmor  \& Springel}{Marinacci
  et~al.}{2013}]{Marinacci_2013}
Marinacci F.,  Pakmor R.,   Springel V.,  2013, \mn@doi [MNRAS]
  {10.1093/mnras/stt2003}, 437, 1750–1775

\bibitem[\protect\citeauthoryear{Marinacci et~al.,}{Marinacci
  et~al.}{2018}]{Marinacci_2018}
Marinacci F.,  et~al., 2018, \mn@doi [MNRAS] {10.1093/mnras/sty2206}

\bibitem[\protect\citeauthoryear{{Mart{\'\i}nez-Delgado}
  et~al.,}{{Mart{\'\i}nez-Delgado} et~al.}{2010}]{martinezdelgado2010stellar}
{Mart{\'\i}nez-Delgado} D.,  et~al., 2010, \mn@doi [The Astronomical Journal]
  {10.1088/0004-6256/140/4/962}, \href
  {https://ui.adsabs.harvard.edu/abs/2010AJ....140..962M} {140, 962}

\bibitem[\protect\citeauthoryear{McInnes, Healy  \& Melville}{McInnes
  et~al.}{2020}]{mcinnes2020umap}
McInnes L.,  Healy J.,   Melville J.,  2020, UMAP: Uniform Manifold
  Approximation and Projection for Dimension Reduction (\mn@eprint {arXiv}
  {1802.03426})

\bibitem[\protect\citeauthoryear{{Merritt}, {Pillepich}, {van Dokkum},
  {Nelson}, {Hernquist}, {Marinacci}  \& {Vogelsberger}}{{Merritt}
  et~al.}{2020}]{Merritt_2020}
{Merritt} A.,  {Pillepich} A.,  {van Dokkum} P.,  {Nelson} D.,  {Hernquist} L.,
   {Marinacci} F.,   {Vogelsberger} M.,  2020, \mn@doi [\mnras]
  {10.1093/mnras/staa1164}, \href
  {https://ui.adsabs.harvard.edu/abs/2020MNRAS.495.4570M} {495, 4570}

\bibitem[\protect\citeauthoryear{Monachesi et~al.,}{Monachesi
  et~al.}{2019}]{10.1093/mnras/stz538}
Monachesi A.,  et~al., 2019, \mn@doi [Monthly Notices of the Royal Astronomical
  Society] {10.1093/mnras/stz538}, 485, 2589

\bibitem[\protect\citeauthoryear{{Naidu} et~al.,}{{Naidu}
  et~al.}{2021}]{Naidu2021}
{Naidu} R.~P.,  et~al., 2021, arXiv e-prints, \href
  {https://ui.adsabs.harvard.edu/abs/2021arXiv210303251N} {p. arXiv:2103.03251}

\bibitem[\protect\citeauthoryear{{Naiman} et~al.,}{{Naiman}
  et~al.}{2018}]{naiman2017results}
{Naiman} J.~P.,  et~al., 2018, \mn@doi [MNRAS] {10.1093/mnras/sty618}, \href
  {https://ui.adsabs.harvard.edu/abs/2018MNRAS.477.1206N} {477, 1206}

\bibitem[\protect\citeauthoryear{{Nelson} et~al.,}{{Nelson}
  et~al.}{2015}]{nelson2015illustris}
{Nelson} D.,  et~al., 2015, \mn@doi [Astronomy and Computing]
  {10.1016/j.ascom.2015.09.003}, \href
  {https://ui.adsabs.harvard.edu/abs/2015A&C....13...12N} {13, 12}

\bibitem[\protect\citeauthoryear{{Nelson} et~al.,}{{Nelson}
  et~al.}{2018}]{nelson2017results}
{Nelson} D.,  et~al., 2018, \mn@doi [MNRAS] {10.1093/mnras/stx3040}, \href
  {https://ui.adsabs.harvard.edu/abs/2018MNRAS.475..624N} {475, 624}

\bibitem[\protect\citeauthoryear{{Nelson} et~al.,}{{Nelson}
  et~al.}{2019a}]{nelson2019illustristng}
{Nelson} D.,  et~al., 2019a, \mn@doi [Computational Astrophysics and Cosmology]
  {10.1186/s40668-019-0028-x}, \href
  {https://ui.adsabs.harvard.edu/abs/2019ComAC...6....2N} {6, 2}

\bibitem[\protect\citeauthoryear{Nelson et~al.,}{Nelson
  et~al.}{2019b}]{Nelson_2019}
Nelson D.,  et~al., 2019b, \mn@doi [MNRAS] {10.1093/mnras/stz2306}, 490,
  3234–3261

\bibitem[\protect\citeauthoryear{{Oser}, {Ostriker}, {Naab}, {Johansson}  \&
  {Burkert}}{{Oser} et~al.}{2010}]{Oser_2010}
{Oser} L.,  {Ostriker} J.~P.,  {Naab} T.,  {Johansson} P.~H.,   {Burkert} A.,
  2010, \mn@doi [\apj] {10.1088/0004-637X/725/2/2312}, \href
  {https://ui.adsabs.harvard.edu/abs/2010ApJ...725.2312O} {725, 2312}

\bibitem[\protect\citeauthoryear{Pasquet, Bertin, Treyer, Arnouts  \&
  Fouchez}{Pasquet et~al.}{2018}]{Pasquet_2018}
Pasquet J.,  Bertin E.,  Treyer M.,  Arnouts S.,   Fouchez D.,  2018, \mn@doi
  [Astronomy & Astrophysics] {10.1051/0004-6361/201833617}, 621, A26

\bibitem[\protect\citeauthoryear{Paszke et~al.,}{Paszke
  et~al.}{2019}]{NEURIPS2019_9015}
Paszke A.,  et~al., 2019, in Wallach H.,  Larochelle H.,  Beygelzimer A.,
  d\textquotesingle Alch\'{e}-Buc F.,  Fox E.,   Garnett R.,  eds, , Advances
  in Neural Information Processing Systems 32.
Curran Associates, Inc., pp 8024--8035

\bibitem[\protect\citeauthoryear{Pillepich et~al.,}{Pillepich
  et~al.}{2014}]{Pillepich_2014}
Pillepich A.,  et~al., 2014, \mn@doi [MNRAS] {10.1093/mnras/stu1408}, 444,
  237–249

\bibitem[\protect\citeauthoryear{Pillepich, Madau  \& Mayer}{Pillepich
  et~al.}{2015}]{Pillepich_2015}
Pillepich A.,  Madau P.,   Mayer L.,  2015, \mn@doi [The Astrophysical Journal]
  {10.1088/0004-637x/799/2/184}, 799, 184

\bibitem[\protect\citeauthoryear{Pillepich et~al.,}{Pillepich
  et~al.}{2018a}]{Pillepich_2018a}
Pillepich A.,  et~al., 2018a, \mn@doi [MNRAS] {10.1093/mnras/stx2656}, 473,
  4077–4106

\bibitem[\protect\citeauthoryear{Pillepich et~al.,}{Pillepich
  et~al.}{2018b}]{Pillepich_2018}
Pillepich A.,  et~al., 2018b, \mn@doi [MNRAS] {10.1093/mnras/stx3112}, 475,
  648–675

\bibitem[\protect\citeauthoryear{Pillepich et~al.,}{Pillepich
  et~al.}{2019}]{Pillepich_2019}
Pillepich A.,  et~al., 2019, \mn@doi [MNRAS] {10.1093/mnras/stz2338}, 490,
  3196–3233

\bibitem[\protect\citeauthoryear{{Planck Collaboration} et~al.,}{{Planck
  Collaboration} et~al.}{2016}]{Planck_2015}
{Planck Collaboration} et~al., 2016, \mn@doi [\aap]
  {10.1051/0004-6361/201525830}, \href
  {https://ui.adsabs.harvard.edu/abs/2016A&A...594A..13P} {594, A13}

\bibitem[\protect\citeauthoryear{Pop, Pillepich, Amorisco  \& Hernquist}{Pop
  et~al.}{2018}]{Pop_2018}
Pop A.-R.,  Pillepich A.,  Amorisco N.~C.,   Hernquist L.,  2018, \mn@doi
  [MNRAS] {10.1093/mnras/sty1932}, 480, 1715–1739

\bibitem[\protect\citeauthoryear{{Potter}, {Stadel}  \& {Teyssier}}{{Potter}
  et~al.}{2017}]{Potter_2017}
{Potter} D.,  {Stadel} J.,   {Teyssier} R.,  2017, \mn@doi [Computational
  Astrophysics and Cosmology] {10.1186/s40668-017-0021-1}, \href
  {https://ui.adsabs.harvard.edu/abs/2017ComAC...4....2P} {4, 2}

\bibitem[\protect\citeauthoryear{{Quinn}, {Hernquist}  \& {Fullagar}}{{Quinn}
  et~al.}{1993}]{Quinn_1993}
{Quinn} P.~J.,  {Hernquist} L.,   {Fullagar} D.~P.,  1993, \mn@doi [\apj]
  {10.1086/172184}, \href
  {https://ui.adsabs.harvard.edu/abs/1993ApJ...403...74Q} {403, 74}

\bibitem[\protect\citeauthoryear{Rodriguez-Gomez et~al.,}{Rodriguez-Gomez
  et~al.}{2015}]{Rodriguez_Gomez_2015}
Rodriguez-Gomez V.,  et~al., 2015, \mn@doi [MNRAS] {10.1093/mnras/stv264}, 449,
  49–64

\bibitem[\protect\citeauthoryear{{Rodriguez-Gomez} et~al.,}{{Rodriguez-Gomez}
  et~al.}{2016}]{rodriguezgomez2016exsitu}
{Rodriguez-Gomez} V.,  et~al., 2016, \mn@doi [\mnras] {10.1093/mnras/stw456},
  \href {https://ui.adsabs.harvard.edu/abs/2016MNRAS.458.2371R} {458, 2371}

\bibitem[\protect\citeauthoryear{{Rodriguez-Gomez} et~al.,}{{Rodriguez-Gomez}
  et~al.}{2017}]{rodriguezgomez2016role}
{Rodriguez-Gomez} V.,  et~al., 2017, \mn@doi [MNRAS] {10.1093/mnras/stx305},
  \href {https://ui.adsabs.harvard.edu/abs/2017MNRAS.467.3083R} {467, 3083}

\bibitem[\protect\citeauthoryear{{Rodriguez-Gomez} et~al.,}{{Rodriguez-Gomez}
  et~al.}{2019}]{Rodriguez-Gomez_2019}
{Rodriguez-Gomez} V.,  et~al., 2019, \mn@doi [\mnras] {10.1093/mnras/sty3345},
  \href {https://ui.adsabs.harvard.edu/abs/2019MNRAS.483.4140R} {483, 4140}

\bibitem[\protect\citeauthoryear{{Sarzi} et~al.,}{{Sarzi}
  et~al.}{2018}]{Sarzi2018}
{Sarzi} M.,  et~al., 2018, \mn@doi [\aap] {10.1051/0004-6361/201833137}, \href
  {https://ui.adsabs.harvard.edu/abs/2018A&A...616A.121S} {616, A121}

\bibitem[\protect\citeauthoryear{{Schaye} et~al.,}{{Schaye}
  et~al.}{2015}]{Schaye_2015}
{Schaye} J.,  et~al., 2015, \mn@doi [\mnras] {10.1093/mnras/stu2058}, \href
  {https://ui.adsabs.harvard.edu/abs/2015MNRAS.446..521S} {446, 521}

\bibitem[\protect\citeauthoryear{{Schechter}}{{Schechter}}{1976}]{Schechter_1976}
{Schechter} P.,  1976, \mn@doi [\apj] {10.1086/154079}, \href
  {https://ui.adsabs.harvard.edu/abs/1976ApJ...203..297S} {203, 297}

\bibitem[\protect\citeauthoryear{{Shi}, {Wang}, {Li}, {Han}, {Shi},
  {Rodriguez-Gomez}  \& {Peng}}{{Shi} et~al.}{2021}]{Rui_2021}
{Shi} R.,  {Wang} W.,  {Li} Z.,  {Han} J.,  {Shi} J.,  {Rodriguez-Gomez} V.,
  {Peng} Y.,  2021, arXiv e-prints, \href
  {https://ui.adsabs.harvard.edu/abs/2021arXiv211207203S} {p. arXiv:2112.07203}

\bibitem[\protect\citeauthoryear{{Snyder} et~al.,}{{Snyder}
  et~al.}{2015}]{Snyder_2015}
{Snyder} G.~F.,  et~al., 2015, \mn@doi [\mnras] {10.1093/mnras/stv2078}, \href
  {https://ui.adsabs.harvard.edu/abs/2015MNRAS.454.1886S} {454, 1886}

\bibitem[\protect\citeauthoryear{Springel}{Springel}{2005}]{Springel_2005}
Springel V.,  2005, \mn@doi [MNRAS] {10.1111/j.1365-2966.2005.09655.x}, 364,
  1105–1134

\bibitem[\protect\citeauthoryear{Springel}{Springel}{2010}]{Springel_2010}
Springel V.,  2010, \mn@doi [MNRAS] {10.1111/j.1365-2966.2009.15715.x}, 401,
  791–851

\bibitem[\protect\citeauthoryear{{Springel}, {White}, {Tormen}  \&
  {Kauffmann}}{{Springel} et~al.}{2001}]{2001MNRAS.328..726S}
{Springel} V.,  {White} S. D.~M.,  {Tormen} G.,   {Kauffmann} G.,  2001,
  \mn@doi [MNRAS] {10.1046/j.1365-8711.2001.04912.x}, \href
  {https://ui.adsabs.harvard.edu/abs/2001MNRAS.328..726S} {328, 726}

\bibitem[\protect\citeauthoryear{{Springel} et~al.,}{{Springel}
  et~al.}{2005}]{2005Natur.435..629S}
{Springel} V.,  et~al., 2005, \mn@doi [Nature] {10.1038/nature03597}, \href
  {https://ui.adsabs.harvard.edu/abs/2005Natur.435..629S} {435, 629}

\bibitem[\protect\citeauthoryear{Springel et~al.,}{Springel
  et~al.}{2018}]{Springel_2017}
Springel V.,  et~al., 2018, \mn@doi [MNRAS] {10.1093/mnras/stx3304}, 475,
  676–698

\bibitem[\protect\citeauthoryear{{Trayford} et~al.,}{{Trayford}
  et~al.}{2017}]{Trayford_2017}
{Trayford} J.~W.,  et~al., 2017, \mn@doi [\mnras] {10.1093/mnras/stx1051},
  \href {https://ui.adsabs.harvard.edu/abs/2017MNRAS.470..771T} {470, 771}

\bibitem[\protect\citeauthoryear{{Turnbull}, {Bridges}  \& {Carter}}{{Turnbull}
  et~al.}{1999}]{Turnbull_1999}
{Turnbull} A.~J.,  {Bridges} T.~J.,   {Carter} D.,  1999, \mn@doi [\mnras]
  {10.1046/j.1365-8711.1999.02724.x}, \href
  {https://ui.adsabs.harvard.edu/abs/1999MNRAS.307..967T} {307, 967}

\bibitem[\protect\citeauthoryear{{Vogelsberger} et~al.,}{{Vogelsberger}
  et~al.}{2014}]{Vogelsberger.2014Nature}
{Vogelsberger} M.,  et~al., 2014, \mn@doi [Nature] {10.1038/nature13316}, \href
  {https://ui.adsabs.harvard.edu/abs/2014Natur.509..177V} {509, 177}

\bibitem[\protect\citeauthoryear{{Vogelsberger}, {Marinacci}, {Torrey}  \&
  {Puchwein}}{{Vogelsberger} et~al.}{2020}]{Vogelberger_2020}
{Vogelsberger} M.,  {Marinacci} F.,  {Torrey} P.,   {Puchwein} E.,  2020,
  \mn@doi [Nature Reviews Physics] {10.1038/s42254-019-0127-2}, \href
  {https://ui.adsabs.harvard.edu/abs/2020NatRP...2...42V} {2, 42}

\bibitem[\protect\citeauthoryear{{Wechsler} \& {Tinker}}{{Wechsler} \&
  {Tinker}}{2018}]{Wechsler&Tinker_2018}
{Wechsler} R.~H.,  {Tinker} J.~L.,  2018, \mn@doi [\araa]
  {10.1146/annurev-astro-081817-051756}, \href
  {https://ui.adsabs.harvard.edu/abs/2018ARA&A..56..435W} {56, 435}

\bibitem[\protect\citeauthoryear{Weinberger et~al.,}{Weinberger
  et~al.}{2017}]{Weinberger_2016}
Weinberger R.,  et~al., 2017, \mn@doi [MNRAS] {10.1093/mnras/stw2944}, 465,
  3291–3308

\bibitem[\protect\citeauthoryear{{Zanisi} et~al.,}{{Zanisi}
  et~al.}{2021}]{Zanisi_2020}
{Zanisi} L.,  et~al., 2021, \mn@doi [\mnras] {10.1093/mnras/staa3864}, \href
  {https://ui.adsabs.harvard.edu/abs/2021MNRAS.501.4359Z} {501, 4359}

\bibitem[\protect\citeauthoryear{{Zhu} et~al.,}{{Zhu} et~al.}{2020}]{Zhu2020}
{Zhu} L.,  et~al., 2020, \mn@doi [\mnras] {10.1093/mnras/staa1584}, \href
  {https://ui.adsabs.harvard.edu/abs/2020MNRAS.496.1579Z} {496, 1579}

\bibitem[\protect\citeauthoryear{{Zhu} et~al.,}{{Zhu} et~al.}{2021}]{Zhu2021}
{Zhu} L.,  et~al., 2021, arXiv e-prints, \href
  {https://ui.adsabs.harvard.edu/abs/2021arXiv211013172Z} {p. arXiv:2110.13172}

\bibitem[\protect\citeauthoryear{{de los Rios}, {Peta{\v{c}}}, {Zaldivar},
  {Bonaventura}, {Calore}  \& {Iocco}}{{de los Rios} et~al.}{2021}]{Rios_2021}
{de los Rios} M.~E.,  {Peta{\v{c}}} M.,  {Zaldivar} B.,  {Bonaventura} N.~R.,
  {Calore} F.,   {Iocco} F.,  2021, arXiv e-prints, \href
  {https://ui.adsabs.harvard.edu/abs/2021arXiv211108725D} {p. arXiv:2111.08725}

\bibitem[\protect\citeauthoryear{{von Marttens} et~al.,}{{von Marttens}
  et~al.}{2021}]{Marttens_2021}
{von Marttens} R.,  et~al., 2021, arXiv e-prints, \href
  {https://ui.adsabs.harvard.edu/abs/2021arXiv211101185V} {p. arXiv:2111.01185}

\makeatother
\end{thebibliography}

%%%%%%%%%%%%%%%%% APPENDICES %%%%%%%%%%%%%%%%%%%%%

\appendix

\section{Calibration errors of the cINN model}
\label{sec:calibrationerrors}

While our analysis in the main text of this paper has mainly focused on the performance of the cINN in terms of MAPs and posterior standard deviations, we show here more quantitatively to what degree the posterior distribution itself is related to the distribution of ground truth galaxies. To do so, we calculate the calibration error for each output: this is the deviation between the ground truth {\it distribution} of the TNG100 galaxy test sample and the respectively-predicted cINN posterior distribution. To obtain these calibration errors, we measure for the whole sample how often the ground truth falls into a certain confidence interval of the cINN posteriors, for each target quantity.

For example, we expect that we find within the $50$ per cent interval of the sampled posterior $p(x|c)$ also $50$ per cent of the ground truth test galaxies. A positive value therefore denotes an under-confident posterior (the interval contains more test galaxies than displayed by the posterior) and a negative value therefore denotes an overconfident posterior (the interval contains less test galaxies than displayed by the posterior).

\begin{figure}
	\centering
	\includegraphics[width=8cm]{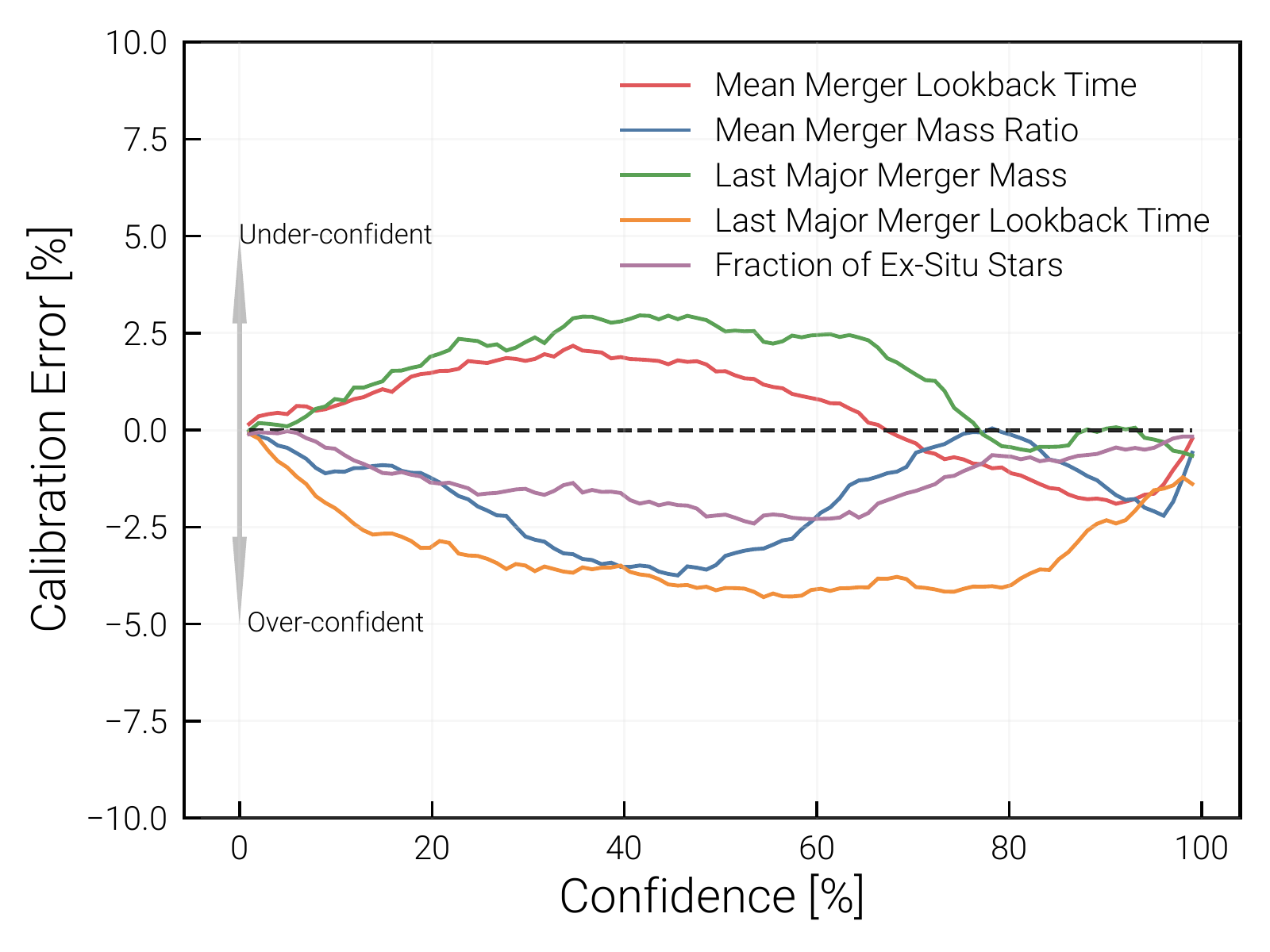}
	\caption{{\bf Calibration error of the MLP model as function of the confidence level for each of the $5$ target quantities.} The calibration error is the deviation between the fraction of test galaxy ground truths contained within a certain posterior confidence interval and the expected fraction according to the respective confidence level. A positive calibration error therefore denotes an under-confident posterior while a negative value denotes an over-confident posterior. We see that especially the ex-situ fraction posterior are well within $\pm 2 \%$, whereas the mean merger mass ratio and the lookback time of the last major merger tends to be slightly over-confident. We exclude galaxies with no major merger in their history for the calculation of the last major merger calibration error.}
	\label{fig:results/TNG100/calibration_error}
\end{figure}

We show the calibration errors as function of the posterior confidence level for each of the $5$ outputs in Fig.~\ref{fig:results/TNG100/calibration_error}. Note that we perform these calculations on the raw posteriors i.e. without using the previously-utilized ``No-Major-Merger'' bin. Especially for the ex-situ fraction, our networks are characterized by a low calibration error: $\lesssim 2$ per cent. The over-confidence for all 5 statistics is smaller than $\approx 2.5$ per cent. We can conclude that the cINN posteriors therefore follow the ground truth distribution quite well, as we have also already verified in Figs.~\ref{fig:cinn_prior_lowmass} and \ref{fig:cinn_prior_highmass}.

%%%%%%%%%%%%%%%%%%%%%%%%%%%%%%%%%%%%%%%%%%%%%%%%%%

\label{lastpage}
\end{document}